\documentclass[tosem, manuscript]{acmart}
\AtBeginDocument{%
  }

\setcopyright{acmlicensed}

\copyrightyear{2026}
\acmYear{2026}
\acmJournal{TOSEM}
\acmVolume{XX}
\acmNumber{X}
\acmArticle{XXX}
\acmDOI{XXXXXXX.XXXXXXX}
\usepackage{longtable}
\usepackage{svg}
\usepackage{subcaption} 
\usepackage{booktabs}
\usepackage{multirow}
\usepackage{makecell} \usepackage{xcolor}
\usepackage[none]{hyphenat}
\usepackage[most]{tcolorbox}
\usepackage{array}
\usepackage{ragged2e}
\usepackage{adjustbox}
\usepackage{enumitem}
\usepackage{placeins} 
\usepackage{threeparttable}
\usepackage{tabularx}

\newtcolorbox{rqanswer}[1]{
    enhanced,
    breakable,
    center,
    width=0.95\linewidth,
    colback=white,
    colframe=black,
    boxrule=0.4pt,
    arc=6pt,
    left=6pt,
    right=6pt,
    top=5pt,
    bottom=5pt,
    before skip=8pt,
    after skip=8pt,
    fonttitle=\bfseries,
    title={#1}
}
\newtcolorbox{myframe}[1][]{
  enhanced,
  arc=0pt,
  outer arc=0pt,
  colback=white,
  boxrule=0.8pt,
  #1
}

\newcolumntype{L}[1]{>{\RaggedRight\arraybackslash}p{#1}}

\usepackage[framemethod=tikz]{mdframed}

\newmdenv[
    backgroundcolor=gray!5,
    linecolor=black!70,
    linewidth=0.8pt,
    roundcorner=8pt,
    innertopmargin=10pt,
    innerbottommargin=10pt,
    innerleftmargin=12pt,
    innerrightmargin=12pt,
    skipabove=8pt,
    skipbelow=8pt
]{searchbox}
\acmISBN{978-1-4503-XXXX-X/2018/06}

\begin{document}

\title{Monitoring Machine Learning Systems in Production: A Multivocal Literature Review}

\author{Hira Naveed}
\email{hira.naveed@monash.edu}
\orcid{1234-5678-9012}
\affiliation{%
  \institution{Monash University}
  \country{Australia}
}
\author{Aleksandar Pasquini}
\email{aleksandar.pasquini@deakin.edu.au}
\orcid{1234-5678-9012}
\affiliation{%
  \institution{Deakin University}
  \country{Australia}
}

\author{Scott Barnett}
\affiliation{%
  \institution{Deakin University}
  \country{Australia}}
  \email{scott.barnett@deakin.edu.au}

\author{Chetan Arora}
\orcid{0000-0003-1466-7386}
\affiliation{%
  \institution{Monash University}
  \country{Australia}
}
  \email{chetan.arora@monash.edu}

\author{John Grundy}
\orcid{0000-0003-4928-7076}
\affiliation{%
 \institution{Monash University}
 \country{Australia}}
  \email{john.grundy@monash.edu}

\author{Hourieh Khalajzadeh}
\affiliation{%
  \institution{Deakin University}
  \country{Australia}}
  \email{hkhalajzadeh@deakin.edu.au}

\author{Omar Haggag}
\orcid{0000-0003-2346-3131}
\affiliation{%
  \institution{Monash University}
  \country{Australia}}
\email{omar.haggag@monash.edu}

\renewcommand{\shortauthors}{Naveed et al.}

\begin{abstract}
Monitoring is essential for maintaining reliable machine learning (ML) systems after deployment. However, knowledge about its practical implementation and effectiveness remains fragmented across academic and gray literature. We present a multivocal literature review of 94 sources: 37 academic and 57 gray. We develop taxonomies covering monitored objects and properties, techniques, input signals, tools, automation, frequency, evaluation, and challenges. Our findings reveal an imbalanced monitoring process. Existing approaches provide broad and largely automated support for observing system behavior, detecting problems, and communicating findings, while monitor configuration, investigation, and response retain substantial human involvement. Monitoring is concentrated on inference, with no source covering the entire production ML system, limiting end-to-end diagnosis. Evidence of effectiveness is also limited: among 78 sources presenting an approach or tool, 45 provide no formal evaluation, 24 report offline or controlled evaluations, and only nine evaluate monitoring in live production. This gap is particularly evident for generative AI, where evidence is dominated by gray literature and vendor claims. We propose six directions for improving monitor design, cross-layer diagnosis, proxy validation, corrective responses, production-oriented evaluation, and organizational practices. Our review provides a structured basis for improving the detection, diagnosis, and resolution of production ML problems.

\end{abstract}

\ccsdesc[500]{Software and its engineering~Maintaining software}
\ccsdesc[500]{Computing methodologies~Machine learning}

\keywords{Software Engineering, Runtime Monitoring, Machine Learning, Survey}

\received{20 September 2009}
\received[revised]{12 September 2009}
\received[accepted]{5 December 2009}

\maketitle
\section{Introduction}
Machine learning (ML) systems are increasingly deployed in real-world settings, where their outputs influence users, software behavior, and organizational decisions \cite{paleyes2022challenges, measureimproved2025}. An ML system combines one or more learned models with the data pipelines, software components, serving infrastructure, downstream applications, and operational processes required to use those models in practice \cite{sculley2015hidden, lwakatare2020large, huyen2022designing}. However, the system behavior observed during development does not necessarily persist after deployment \cite{leest2025tea}. Production data and operating conditions may change, data quality may deteriorate, upstream dependencies may fail, and the relationships learned by a model may no longer hold \cite{leest2025tea, haveidea2024}. Changes to models, data pipelines, infrastructure, and downstream applications can also introduce unexpected interactions and feedback loops \cite{sculley2015hidden, haveidea2024, paleyes2022challenges}. These factors make production ML systems difficult and costly to maintain \cite{sculley2015hidden}. Their reliability therefore cannot be ensured only through quality assurance activities performed before deployment; it requires ongoing assessment of the system throughout its operation \cite{mltestscore2017, grieser2020assuring, martinez2021developing, sculley2015hidden}.

Monitoring is a key maintenance activity that involves collecting and analyzing runtime data to determine whether a production ML system continues to operate as expected \cite{leest2025tea, haveidea2024, sculley2015hidden}. Its scope varies across different types of ML systems. In traditional ML systems, monitoring commonly addresses data quality, drift, predictive performance, safety, fairness, and operational reliability \cite{amazonsagemaker2021,alertiger2023, ghosh2022faircanary}. Generative AI systems introduce additional concerns related to correctness, relevance, faithfulness, toxicity, and policy compliance, while retrieval-augmented generation (RAG) systems require retrieval and generation quality to be assessed separately \cite{dong2024agentops, es2024ragas, bommasani2023holistic, llamaguard2023}. Agentic systems further extend monitoring beyond individual inputs and outputs to multi-step execution trajectories involving planning, memory, tool selection, and tool execution \cite{dong2024agentops}.

\subsection{Motivation}
A substantial body of research has examined ML monitoring, including data quality, drift, outlier detection, performance degradation, robustness, and fairness \cite{sculley2015hidden, paleyes2022challenges, daqlmonitor2019, amazonsagemaker2021, ghosh2022faircanary}. However, comparatively fewer studies examine monitoring in production, where approaches must operate in real-world contexts across several data pipelines, models, infrastructure, and applications \cite{measureimproved2025}. The available production focused research and practitioner knowledge is also dispersed across software engineering, machine learning, data engineering, MLOps, and responsible AI. Studies differ in their monitoring objectives, the parts of the system they monitor, the runtime data they collect, and how they evaluate monitoring effectiveness. Therefore, despite the volume of work on ML monitoring, knowledge about monitoring ML systems in production remains fragmented \cite{measureimproved2025}.

This fragmentation makes it difficult to determine which monitoring approaches are suitable for different production contexts, how rigorously they have been evaluated, and whether they address the problems encountered by practitioners \cite{measureimproved2025}. Evidence on emerging monitoring approaches and tools is also available in arXiv preprints, such as AgentOps \cite{dong2024agentops}, and practitioner resources, such as Arize Phoenix \cite{arizephoenixnd}. Existing reviews cover particular aspects of monitoring, including challenges and methods \cite{schroder2022monitoring}, contextual information \cite{leest2025tea}, and robustness in production \cite{bayram2025towards}. Although there have been multivocal reviews that integrate academic and gray literature, their focus is on monitoring and maintenance \cite{nguyen2023machine} or on monitoring and mitigation steps in production \cite{measureimproved2025}. A multivocal synthesis is therefore needed to provide a holistic overview of academic and gray literature on monitoring production ML systems.

To address this need, we conduct a multivocal literature review (MLR) on monitoring production ML systems. We focus specifically on production ML systems because monitoring approaches that perform well in controlled academic settings may not satisfy the operational, integration, scalability, and reliability requirements of real-world systems \cite{paleyes2022challenges}. Combining academic and gray literature offers several benefits. First, it provides a broader and more current view of the field than either body of literature can offer alone. Second, it enables research and practice to be triangulated, helping assess the maturity of existing approaches and revealing where academic evidence aligns with or diverges from production practices and needs. Finally, drawing on evidence from varied contexts improves the practical relevance of the findings and makes broader trends and gaps more visible. 

This review can help researchers identify underexplored problems and strengthen the practical relevance of future studies. It can support practitioners in understanding the available approaches and selecting those suited to their systems and monitoring needs. It can also provide valuable guidance for tool developers to improve the design and effectiveness of monitoring tools for production ML systems. For this MLR, we systematically searched four academic and two gray literature databases, resulting in 94 included studies: 37 academic and 57 gray. Our review characterizes the existing literature and the ML systems studied; identifies the monitored objects and properties, monitoring approaches, metrics, and tools; and examines how monitoring effectiveness has been evaluated. We further synthesize monitoring challenges, assess the extent to which existing approaches address them, and identify the remaining open problems.

\subsection{Contributions}
Our main contributions are summarized as follows:

\begin{itemize}
    \item \textbf{Comprehensive characterization of monitoring production ML systems.} We consolidate a rapidly growing and fragmented body of academic and practitioner knowledge. Our analysis characterizes the ML systems, the monitored properties, techniques, and tools, providing an overview of how the field has evolved and where research and practice are concentrated.

    \item \textbf{A comprehensive set of taxonomies for production ML monitoring.} We develop taxonomies that organize the findings across all research questions, covering ML system characteristics, monitored objects and properties, monitoring techniques and input signals, evaluation methods, and monitoring challenges. Together, these taxonomies provide a structured, end-to-end view of production ML monitoring.

    \item \textbf{Synthesis of evaluation practices and evidence.} We examine how monitoring approaches have been evaluated, including the evaluation context, scale, domain, and metrics. This analysis reveals the extent of the empirical support available for existing approaches and identifies limitations in current evaluation practices.

    \item \textbf{Connection between existing approaches and reported challenges.} We synthesize the challenges reported in the literature and assess the extent to which they are addressed by existing production monitoring approaches. This comparison highlights areas in which available research and tools align with production needs, as well as challenges for which support remains limited.

    \item \textbf{Open problems and implications for research and practice.} We consolidate the unresolved problems identified across academic and gray literature and derive directions for future work. For researchers, these findings identify opportunities to address underexplored production challenges. For practitioners, they provide a structured basis for understanding and selecting monitoring approaches for different production contexts.
\end{itemize}
\subsection{Organization}
The remainder of this paper is organized as follows: Section~\ref{RW} presents the background and related work. Section~\ref{method} describes the review methodology. Section~\ref{results} presents the review results. Section~\ref{discussion} discusses the findings and their implications, and Section \ref{future} presents recommendations for future work. Section~\ref{threats} presents the threats to validity. Lastly, Section~\ref{conclusion} concludes the paper.

\section{Background and Related Work}\label{RW}
\subsection{ML Systems}
A machine learning (ML) model is a parameterized function learned from data, whereas an \textbf{ML system} is the larger software and organizational arrangement that makes one or more learned models useful in practice \cite{huyen2022designing}. It includes the data sources and transformations that produce features, the model artifact and inference code, the serving and storage infrastructure, the application that consumes predictions, and the people and processes that define, operate, and act on the system \cite{jordan2015machine, sculley2015hidden, lwakatare2020large, kuwajima2020engineering}. This system-level distinction is important because the learned component is commonly only a small part of the production implementation, while data dependencies, feedback loops, configuration, glue code, and downstream consumers account for much of its operational risk~\cite{sculley2015hidden, kuwajima2020engineering, jordan2015machine, mltestscore2017, huyen2022designing}. We use production ML system to mean an ML system that processes live or operational data and whose outputs affect users, software behavior, and organizational decisions \cite{lwakatare2019taxonomy, lwakatare2020large, huyen2022designing}. The definition therefore covers batch and online inference; cloud, on-premise, and edge deployments; and traditional ML, deep learning, and generative AI.

\subsection{Monitoring ML Systems}
\textbf{Monitoring} is a maintenance activity that collects and analyzes runtime data from a production system to observe system health, detect faults, and support timely mitigation \cite{delgado2004taxonomy, swebok2026}. Traditional software systems are monitored primarily for operational and safety properties such as latency, availability, throughput, resource utilization, and error rates \cite{delgado2004taxonomy, vierhauser2023grum, swebok2026}. Although these properties remain important, monitoring them alone may leave degradation in ML behavior undetected \cite{george2025silent, cummaudo2020beware, ghosh2022faircanary}. An ML system may remain available and return technically valid responses while its predictive performance, fairness, relevance, or safety deteriorates \cite{ghosh2022faircanary, cummaudo2020beware}. This is because the behavior of an ML system depends not only on its code and infrastructure but also on learned model parameters \cite{mltestscore2017}, evolving production data \cite{moreno2012unifying, daqlmonitor2019}, its operating environment \cite{moreno2012unifying, monitoringobservability2025}, downstream interactions \cite{monitoringobservability2025}, and feedback generated through system use. These dependencies can produce silent failures that do not cause explicit software errors but still negatively affect users and organizational outcomes \cite{george2025silent, cummaudo2020beware, roland2022domain}.

To maintain reliable ML systems, monitoring must extend beyond traditional operational properties to capture changes in data patterns \cite{moreno2012unifying, daqlmonitor2019}, degradation in model quality \cite{lu2018learning}, infrastructure failures \cite{mltestscore2017}, harmful outcomes, and violations of business, safety, fairness, or compliance requirements \cite{successfulmachine2019, monitoringobservability2025, ghosh2022faircanary}. Consistent with this broader scope, monitoring ML systems has been characterized as continuous verification and validation under actual operating conditions \cite{leest2025tea}. Monitoring ML systems begins by defining the expected behavior of the system, relevant quality requirements, comparison baselines \cite{moreno2012unifying, lu2018learning}, and acceptable operating ranges \cite{lu2018learning, daqlmonitor2019}. Runtime evidence is then collected through metrics, logs, traces, model inputs and outputs, and ground truth outcomes. This evidence may be analyzed using predefined rules \cite{daqlmonitor2019, mltestscore2017}, statistical tests \cite{moreno2012unifying, lu2018learning, mltestscore2017}, performance or business metrics \cite{moreno2012unifying, lu2018learning, mltestscore2017}, learned evaluators, human-in-the-loop \cite{ginart2022mldemon}, or hybrid monitoring approaches \cite{arizeaind}. Detecting production faults can support corrective actions such as fixing data, changing system configurations, rolling back or retraining a model \cite{lewis2022augur}, restricting system functionality \cite{haveidea2024}, or modifying code. ML monitoring therefore provides the runtime evidence needed to determine whether a deployed system continues to behave as expected and when intervention may be required.

Monitoring is also a socio-technical activity \cite{humancentricperspective2022, understandingpractitioners2025}. Decisions about what to monitor, which thresholds indicate unacceptable behavior, how alerts should be prioritized, and when intervention is required depend on the system's goals and operational context \cite{humancentricperspective2022, haveidea2024, understandingpractitioners2025}. These decisions require coordination among multidisciplinary teams comprising data scientists, software engineers, operations engineers, domain experts, system owners, and managers \cite{mltestscore2017, humancentricperspective2022, haveidea2024, collaborationchallenges2022}. Consequently, the effectiveness of monitoring depends not only on the accuracy of detection techniques but also on clearly assigned responsibilities, access to contextual information, appropriate escalation procedures, and the organization’s ability to act on detected problems \cite{haveidea2024, challengesmonitoring2026, collaborationchallenges2022}.

\subsection{Monitored Properties Across ML System Layers}
Monitoring only the model endpoint gives an incomplete account of system health because a visible degradation can originate upstream, within the model, in the serving path, or downstream in the application. Such faults can propagate across components, making their root causes difficult to identify. We organize monitored objects into five layers to support analysis across a production ML system. The structure draws on ML system architectures reported in prior literature and the included studies \cite{mltestscore2017, sculley2015hidden, monitoringobservability2025, feuerriegel2024generative, baylor2017tfx}; its definitions and boundaries were refined during analysis, as described in Section~\ref{method}. We present the layers here to establish the system-level terminology used throughout the review. 


\begin{enumerate}
    \item The \textbf{preprocessing layer} contains data ingestion, validation, joining, transformation, and feature engineering before inference~\cite{baylor2017tfx,sculley2015hidden}. Relevant monitored properties include schema conformance, completeness, freshness, consistency, integrity, training--serving skew, and feature drift~\cite{daqlmonitor2019,sagemakermodel2022}. Errors at this layer can silently propagate into otherwise healthy models~\cite{sculley2015hidden,datavalidation2019}. Systems such as TensorFlow Data Validation illustrate mature, production-scale support for learning schemas, computing statistics, and detecting anomalies in continuous ML pipelines~\cite{tensorflowdata2020}. 

    \item The \textbf{inference layer} covers the path from a prepared input to a model output, including model selection and versioning~\cite{mltestscore2017,michelangelo2017}. For predictive models, monitored properties include predictive performance, calibration, prediction drift, changes in feature attribution, and performance or fairness within important slices~\cite{sagemakermodel2022,amazonsagemaker2021,monitoringfeature2021}. Generative systems add faithfulness, relevance, correctness, coherence, toxicity, policy compliance, and other safety properties~\cite{bommasani2023holistic,llamaguard2023}. RAG systems additionally require separation of retrieval quality from generation quality; otherwise, a poor answer cannot be attributed to missing context, irrelevant retrieval, or generation over the available evidence~\cite{es2024ragas,arizephoenixnd}. Operational properties at this layer include model latency, throughput, error rates, token consumption, and inference cost \cite{coralogix2026, dynatracend, llmobservability2nd}.

    \item The \textbf{agentic layer} is only present in generative AI-based agentic systems; it covers orchestration built on top of a model, including planning, memory, tool selection, tool execution, and multi-step control flow~\cite{dong2024agentops,agentobservability2024}. Monitoring the final response alone can conceal looping, unnecessary tool calls, invalid arguments, partial task completion, or a correct answer reached through an unsafe action~\cite{beyondblackbox2025,arizephoenixnd}. The monitored properties are therefore the execution trajectory as well as the outcome: task correctness and completion, step efficiency, tool-use validity, safety, and alignment with the assigned goal~\cite{llamafirewall2025,chainmonitorability2025}. Operational properties include agent-run duration, tool call latency and errors, and the cost of multi-step execution \cite{opikcomet2026, superwise2026, galileond}. Recent developer studies similarly argue that LLM observability must connect application-level outcomes with prompts, responses, intermediate operations, and operational context~\cite{designprinciples2025}.

    \item The \textbf{serving infrastructure} comprises the compute, networking, storage, and platform services that host the deployed system~\cite{mltestscore2017,michelangelo2017}. Operational properties at this layer include availability, saturation, resource utilization, infrastructure failures, and the latency and reliability of the underlying serving services~\cite{delgado2004taxonomy,vierhauser2023grum}. These measurements provide important context for interpreting problems observed at other layers ~\cite{monitoringobservability2025,haveidea2024}
    

    \item The \textbf{application and business layer} contains the downstream workflow in which model outputs are consumed~\cite{sculley2015hidden,monitoringobservability2025}. Monitored properties include user experience, task success, conversion or other domain key performance indicators, harmful impact, compliance, auditability, and accountability~\cite{successfulmachine2019,haveidea2024,loggingrequirement2025}. This layer determines whether a statistically detectable change is materially important~\cite{measureimproved2025}. It also exposes failures that model-centered metrics miss, such as a recommendation model whose offline accuracy is stable while user satisfaction or a protected group's outcomes deteriorate~\cite{ghosh2022faircanary,amazonsagemaker2021}. Operational properties at this layer include end-to-end application response time, application request volume and errors, and the cost of delivering application-level tasks \cite{scalingml2026, deployingmachine2018}.
\end{enumerate}

\subsection{Related Surveys}
Several secondary studies have been conducted to synthesize the literature on monitoring ML systems, including MLRs, systematic
literature reviews (SLRs), systematic mapping studies (SMS), scoping reviews, and surveys. An MLR combined with an interview study \cite{measureimproved2025} examined the characteristics of production ML systems and identified five steps for monitoring and mitigating production issues: define, measure, assess, act, and control. Another MLR \cite{nguyen2023machine} investigates the methods, tools, challenges, and future directions for ML monitoring and maintenance. Schroder et al. conducted an SLR \cite{schroder2022monitoring} to identify and categorize challenges and methods for monitoring production ML systems. They proposed a taxonomy of ML monitoring methods comprising three categories (data, models, and non-technical), and listed real-world incidents corresponding to each category. Another SLR \cite{karval2023catching} reviewed monitoring and explainability methods for ML; monitoring methods were categorized under data drift, outlier detection, and adversarial detection. Leest et al. conducted an SMS \cite{leest2025tea} on the use of contextual information in ML monitoring. They organized contextual information in a framework with three dimensions - system, aspect, and representation - and outlined recurring patterns in ML monitoring context. 

Andersen et al. \cite{andersen2024monitoring} conducted a scoping review of monitoring methods for AI in healthcare, focusing on performance monitoring methods and the rationales for selecting monitoring strategies. Their review highlighted the scarcity of empirical evidence and practical guidance for implementing performance monitoring in healthcare settings. Wong et al. \cite{wongmodel} reviewed ML model monitoring, data drift detection, and efficient model retraining, focusing on monitoring frameworks and metrics, drift detection techniques, and strategies for selective and automated retraining. Bodor et al. \cite{bodor2023machine} reviewed ML monitoring in the MLOps context, presenting monitoring metrics across different stages of the ML lifecycle and comparing existing monitoring tools and their capabilities.

Bayram et al. \cite{bayram2025towards} surveyed robustness aspects of production ML systems, including monitoring, and provided an overview of relevant tools, challenges, and future research directions. Pattan et al. \cite{pattan2020survey} surveyed factors affecting the behavior of production ML models, along with monitoring challenges, metrics, methods, and tools. Jain et al. \cite{jain2023survey} reviewed the importance of model monitoring and the techniques and tools used for monitoring Natural Language Processing (NLP) applications. Rahman et al. \cite{rahman2021run} surveyed ML monitoring for robotic perception, a subset of computer vision, to identify approaches for monitoring the performance and reliability of perception systems. The authors
mapped monitoring approaches to stages in the perception pipeline and categorized them based on past experiences, inconsistencies during inference, and uncertainty estimation and confidence. Another survey on safety monitoring for ML perception functions was presented in \cite{ferreira2025safety}. The authors provided a comprehensive review of the potential threats in perception systems, requirements elicitation for safety monitors, safety monitoring strategies, techniques to recover
from failure, and methods to evaluate monitors. Bacelar \cite{bacelar2021monitoring} reviewed bias and fairness monitoring in ML models, covering sources of bias and fairness measurement and mitigation approaches.

Despite these contributions, the existing secondary studies provide only partial coverage of the production ML monitoring landscape.
Several studies \cite{ferreira2025safety, andersen2024monitoring, jain2023survey, rahman2021run, bacelar2021monitoring} are domain-specific, focusing on healthcare, NLP, or perception systems. Other studies with a broader scope \cite{karval2023catching, leest2025tea, pattan2020survey, measureimproved2025, schroder2022monitoring, nguyen2023machine, bayram2025towards} focus on particular dimensions of monitoring, such as contextual information, monitoring methods, tools, or challenges, rather than providing a comprehensive review of the existing literature. Furthermore, gray literature and production ML monitoring remain underrepresented in existing reviews. Only two studies combine both, one is a thesis \cite{nguyen2023machine} and the other focuses primarily on monitoring and mitigation steps \cite{measureimproved2025}. Two other reviews \cite{bayram2025towards, schroder2022monitoring} examine production ML monitoring but are limited to academic literature. \textbf{None of the existing reviews comprehensively examine production ML monitoring across system characteristics, monitoring approaches and tools, their effectiveness, monitoring challenges, the extent to which they are addressed, and remaining open problems, while also incorporating gray literature.}

\section{Review Methodology}\label{method}
\begin{figure}[!htbp]
    \centering
    \includegraphics[width=\linewidth]{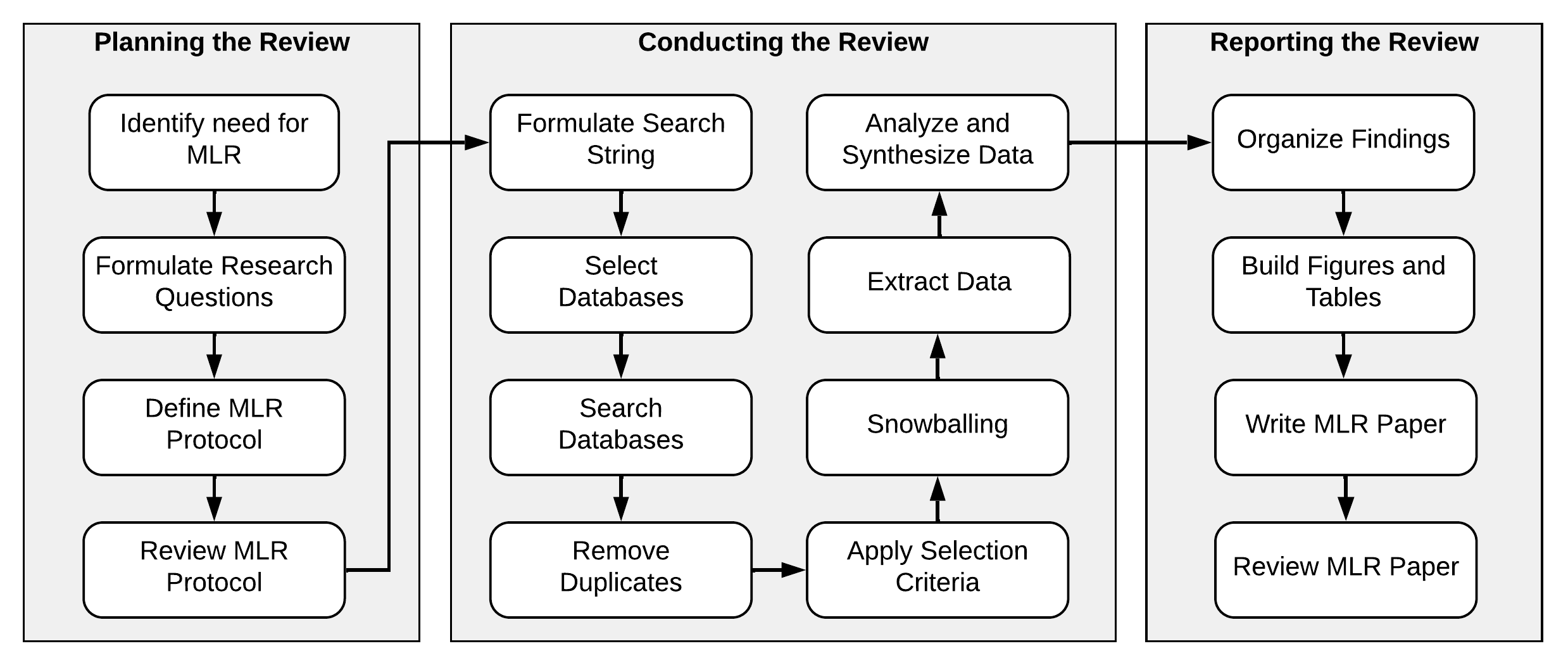}
    \caption{Multivocal Literature Review Process}
    \label{fig:methodology}
\end{figure}

In this section, we describe the methodology used to conduct our MLR. The review followed the guidelines proposed by Garousi et al. \cite{garousi2019guidelines} for MLRs in software engineering and the snowballing guidelines by Wohlin \cite{wohlin2014guidelines}. As illustrated in Figure \ref{fig:methodology}, the review was conducted in three stages: planning the review, conducting the review, and reporting the review. The following sections describe each stage in detail.
\subsection{Planning the Review}
The planning stage established the foundation for the MLR by defining the review protocol and research methodology. This involved identifying the need for the review, formulating the research questions, developing and reviewing the review protocol, and designing the search strategy. These activities helped ensure the review was conducted systematically and transparently, and improved its reproducibility.

\subsubsection{Need for a Review}
We first conducted a preliminary review of the existing secondary literature to understand the current state of knowledge and identify gaps that warranted further investigation. This review showed that existing secondary studies primarily focused on specific monitoring aspects, application domains, or formal literature, providing only a partial view of production ML monitoring. Based on these findings, we determined that a comprehensive MLR was needed to synthesize evidence from both academic and gray literature sources to guide future research and practice.

We deliberately scoped the review to production ML monitoring, as monitoring approaches in experimental or controlled settings differ significantly from those deployed in real-world systems. Production environments introduce unique challenges such as dynamic data, scalability constraints, and operational requirements that are not captured in controlled settings. Additionally, the broader literature on ML monitoring is vast; restricting the scope to production ML monitoring ensured that the review remained focused and feasible.

\subsubsection{Research Questions}
The research questions define the scope of the review and guide the subsequent stages of study selection, data extraction, and evidence synthesis. They were formulated based on the review objectives and the gaps identified during the planning stage to provide a structured understanding of production ML monitoring. The questions progressively investigate the characteristics of monitored production ML systems (RQ1), current monitoring practices (RQ2), evidence of monitoring effectiveness (RQ3), and monitoring challenges and open problems (RQ4), together providing a comprehensive view of the state of production ML monitoring.

\textbf{RQ1. What are the characteristics of production ML systems being monitored?}
This RQ establishes the production context in which monitoring is performed. Understanding the characteristics of monitored ML systems provides the foundation for interpreting the answers of other research questions.

\begin{itemize}
  \item[\textit{RQ1.1.}]\textit{What are the ML families and tasks of production ML systems being monitored?} 
  \item[\textit{RQ1.2.}] \textit{What are the deployment environments and serving methods of production ML systems being monitored?}
\end{itemize}

\textbf{RQ2. How are production ML systems monitored?} This RQ investigates current monitoring practices for production ML systems, examining what is monitored, how monitoring is implemented, which tools are used, and their operational characteristics.

\begin{itemize}
    \item[\textit{RQ2.1.}] \textit{What are the monitored objects and monitored properties of production ML systems?} 
    \item[\textit{RQ2.2.}] \textit{What are the tools, techniques, and input signals used for monitoring production ML systems?}
    \item[\textit{RQ2.3.}] \textit{What are the automation levels and frequencies of monitoring techniques for production ML systems?} 
\end{itemize}

\textbf{RQ3. What evidence exists for the effectiveness of monitoring techniques and tools for production ML systems?} This RQ examines how monitoring techniques are evaluated, providing insights into the maturity and practical applicability of current solutions.
\begin{itemize}
    \item[\textit{RQ3.1.}] \textit{How are monitoring approaches for production ML systems evaluated?} 
    \item[\textit{RQ3.2.}] \textit{What are the evaluation scales and domains of production ML systems being monitored?} 
\end{itemize}

\textbf{RQ4. What are the challenges in monitoring production ML systems, to what extent do existing monitoring approaches address them, and what are the open problems?} This RQ complements the technical findings by synthesizing monitoring challenges reported across both practitioner-oriented and approach/tool-oriented sources. It provides insight into the challenges encountered when monitoring production ML systems, assesses the extent to which existing monitoring approaches address these challenges, and identifies the remaining open problems.
\begin{itemize}
    \item[\textit{RQ4.1.}] \textit{What research methods and practitioner populations underpin these findings?}
    \item[\textit{RQ4.2.}] \textit{What are the challenges in monitoring production ML systems?} 
    \item [\textit{RQ4.3.}] \textit{Which monitoring challenges are addressed by existing approaches and what are the open problems?} 
\end{itemize}

\subsubsection{Review Protocol}
Before commencing the MLR, a review protocol was developed by the first and second authors to define the methodology and ensure that the review was conducted consistently and transparently. The protocol documented the review objectives, research questions, search process, study selection criteria, data extraction fields, and data analysis procedures. It was reviewed by the remaining authors and refined prior to the formal literature search. 

\subsection{Conducting the Review}
The conducting stage implemented the review protocol. We executed the search strategy, applied the predefined study selection criteria to the retrieved studies, performed automated database searches and complemented these with manual snowballing to identify additional relevant studies, extracted data using a structured extraction form, and analyzed the collected evidence to answer the research questions.

\subsubsection{Search Strategy}
The search strategy was developed to maximize the retrieval of relevant studies while maintaining a manageable number of search results. To develop an initial vocabulary for production ML monitoring, we first established a purposive set of seed papers. These seed papers were identified through exploratory searches on Google Scholar \cite{googlescholar}, highly cited publications, and the authors' prior domain knowledge. Ten papers \cite{mltestscore2017, sagemakermodel2022, tensorflowdata2020, haveidea2024, mlenabledsystems2024, alertiger2023, measureimproved2025, domainspecific2023, designprinciples2025, monitoringobservability2025} covering diverse aspects of production ML monitoring, including model performance monitoring, observability, industrial monitoring systems, and LLM monitoring, were selected to span these topics, without establishing that they represented the full literature. These seed papers informed the initial terminology and served as a development check on the search strategy. Keywords were identified by examining the seed papers, with particular emphasis on their titles and abstracts. Similar and synonymous terms were consolidated into four search concept groups representing machine learning, monitoring, production contexts, and practitioners. Each group was further refined by incorporating synonyms, abbreviations, and related terminology surfaced during exploratory and pilot searches. The resulting keywords and their corresponding search terms are presented in Table \ref{tab:search}.

\begin{table*}[!htbp]
\centering
\caption{Search Terms}
\label{tab:search}
\small
\begin{tabularx}{\textwidth}{p{1.8cm}X}
\toprule
\textbf{Term} & \textbf{Keywords} \\
\midrule

Machine Learning &

AI model, machine learning, ML, LLM, large language model, agent, MLOps, LLMOps, deep learning, DL, neural network, generative AI, GenAI \\

Monitoring &

monitor, monitoring, observability, continuous verification, runtime verification, runtime analysis, drift detection, anomaly detection, OOD detection, out of distribution detection \\

Production &

production, operation, deployed, post deployment, live system, industry, industrial, real world, case study, user study, user studies, field study, field studies, production readiness \\

Practitioners &

engineer, developer, practitioner, practitioner experience, engineering team, data scientist, ML engineer, MLOps engineer, product team, postmortem, post-mortem, lessons learned \\

\bottomrule
\end{tabularx}
\end{table*}

The first search round used only machine learning and monitoring terms over full-text fields, and retrieved a substantial number of irrelevant studies, including papers that applied machine learning to monitor non-ML phenomena and studies unrelated to production environments. Introducing production-related terms alone did not sufficiently improve the relevance of the retrieved studies; therefore, practitioner-related terms were added as they are a key aspect of production contexts. Different combinations of search fields and keywords were evaluated, progressively restricting the search to the title and abstract for some terms to improve search results. The refinement process continued until the final search strategy successfully retrieved all representative seed papers while producing a focused and manageable set of candidate studies. The final search string is shown below: 
\begin{searchbox}
\small
(\textit{TITLE} ((\textit{"AI model"} OR \textit{"machine learning"} OR \textit{"ML"} OR
\textit{"LLM"} OR \textit{"large language model"} OR \textit{"agent"} OR
\textit{"MLOps"} OR \textit{"LLMOps"} OR \textit{"deep learning"} OR
\textit{"DL"} OR \textit{"neural network"} OR
\textit{"generative AI"} OR \textit{"GenAI"}))
AND
\textit{TITLE-ABS} ((\textit{"monitor"} OR \textit{"monitoring"} OR
\textit{"observability"} OR \textit{"continuous verification"} OR
\textit{"runtime verification"} OR \textit{"runtime analysis"} OR
\textit{"drift detection"} OR \textit{"anomaly detection"} OR
\textit{"OOD detection"} OR
\textit{"out of distribution detection"}))
AND
\textit{TITLE-ABS} ((\textit{"production"} OR \textit{"operation"} OR
\textit{"deployed"} OR \textit{"post deployment"} OR
\textit{"live system"} OR \textit{"industry"} OR
\textit{"industrial"} OR \textit{"real world"} OR
\textit{"case study"} OR \textit{"user study"} OR
\textit{"user studies"} OR \textit{"field study"} OR
\textit{"field studies"} OR \textit{"production readiness"}))
AND
(\textit{"engineer"} OR \textit{"developer"} OR
\textit{"practitioner"} OR \textit{"practitioner experience"} OR
\textit{"engineering team"} OR \textit{"data scientist"} OR
\textit{"ML engineer"} OR \textit{"MLOps engineer"} OR
\textit{"product team"} OR \textit{"postmortem"} OR
\textit{"post-mortem"} OR \textit{"lessons learned"}))

\end{searchbox}

The search was executed across four digital libraries for academic literature: IEEE Xplore, ACM Digital Library, Scopus, and Web of Science. These databases were selected for their broad coverage of software engineering and machine learning research, with IEEE Xplore and ACM Digital Library focusing on core software engineering venues, and Scopus and Web of Science providing comprehensive indexing across interdisciplinary publications. For gray literature, we selected two widely used sources: arXiv and Google Search Engine. ArXiv was included to capture recent and emerging research not yet indexed in digital libraries, while Google Search Engine was used to identify industry reports, blogs, and tool documentation. For feasibility, Google Search results were limited to the top 200, following the recommendation of Garousi et al. \cite{garousi2019guidelines}. No publication-date restrictions were applied during the search.

\subsubsection{Study Selection}
\begin{table}[!htbp]
\centering
\caption{Study Selection Criteria}
\label{tab:selection_criteria}
\small
\begin{tabular}{p{1.2cm}p{12.5cm}}
\toprule
\textbf{ID} & \textbf{Criterion} \\
\midrule
\multicolumn{2}{l}{\textbf{Inclusion Criteria}} \\[2pt]

IC1 & The source focuses on monitoring ML systems in production environments. \\

IC2 & The source proposes an ML monitoring approach, tool, or platform, or reports empirical evidence of monitoring practices from real-world deployments or practitioners, such as interviews, surveys, user studies, case studies, or organizational deployment reports. \\

IC3 & The source is written in English. \\

\midrule
\multicolumn{2}{l}{\textbf{Exclusion Criteria}} \\[2pt]

EX1 & The source does not focus on ML monitoring. \\

EX2 & The monitoring approach in the source was neither developed nor used by practitioners and provides no evidence of real-world deployment. Benchmark or simulated evaluations alone, including those framed as reproducing production concerns, do not satisfy this requirement. \\

EX3 & The source is a secondary or tertiary study. \\

EX4 & The academic source is shorter than four pages. This criterion does not apply to web pages or online tool documentation. \\

EX5 & An extended version of the source is available. \\

EX6 & The full text of the source is not accessible. \\

EX7 & The source is gray literature with terms that restrict research use or competitor comparisons without permission. \\

\bottomrule
\end{tabular}
\end{table}

The study selection process aimed to identify studies that answer the research questions while excluding studies outside the scope of this review. A study was included only if it satisfied all inclusion criteria and did not meet any exclusion criterion. The complete study selection criteria are presented in Table~\ref{tab:selection_criteria}.

The screening process was conducted in two stages by the first and second authors, who manually applied the selection criteria to each paper. The initial search results from each database were exported to Excel files, and studies were managed using a color-coding scheme: red for excluded studies, yellow for potentially relevant studies, and green for included studies. The first stage involved title and abstract screening, and the second stage involved full-text screening of the remaining studies; at each stage, studies that did not satisfy the inclusion criteria or met any exclusion criterion were removed. For web pages without abstracts, the available title, summary, and webpage content were examined during initial screening. To minimize bias, both authors independently screened the studies, and disagreements were resolved through discussion with the remaining authors until consensus was reached.

\subsubsection{Automated Search and Duplicate Removal}
\begin{figure}[!htbp]
    \centering
    \includegraphics[width=\linewidth]{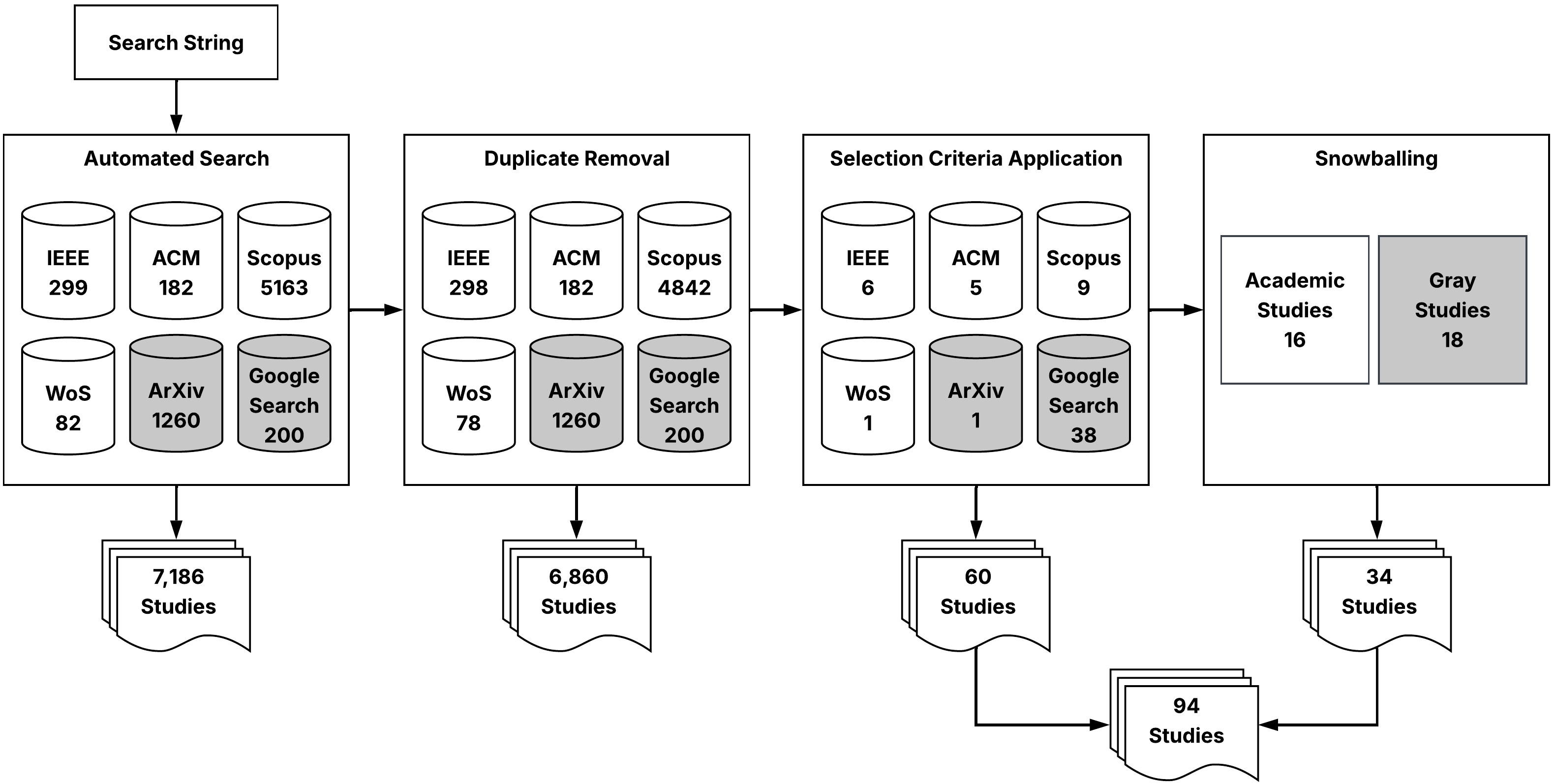}
    \caption{Search Process and Results}
    \label{fig:search_results}
\end{figure}

Figure \ref{fig:search_results} summarizes the automated search, duplicate removal, and screening process, including the number of studies at each stage. The automated search was conducted using the final search strategy across the selected academic and gray literature sources. The initial search retrieved a total of 7,186 studies across all sources. Duplicate removal was performed by matching study titles across sources, resulting in 6,860 unique studies. The study selection criteria were then manually applied to the remaining studies. Title and abstract screening reduced the number of candidate studies to 314, followed by full-text screening, which further refined the set to 60 studies.

\subsubsection{Snowballing}
To complement the automated search, backward and forward snowballing was performed following the guidelines by Wohlin \cite{wohlin2014guidelines}. Backward snowballing examined the reference lists of the included studies, while forward snowballing identified studies citing them. Snowballing was performed manually and iteratively until saturation on the 60 studies selected in the previous stage, following the same two-stage screening process and selection criteria used in the automated search. Each candidate study was screened by the first and second authors, resulting in the inclusion of 34 additional studies (16 academic and 18 gray literature). 

\subsubsection{Data Extraction}

\begin{table*}[!t]
\centering
\small
\caption{Data Extraction Fields}
\label{tab:data-extraction}
\begin{tabularx}{\textwidth}{@{}c L{0.25\textwidth} X L{0.20\textwidth}@{}}
\toprule
\textbf{\#} & \textbf{Data Extracted} & \textbf{Description} & \textbf{RQ(s)} \\
\midrule

1 & Title & Title of the publication. & N/A \\

2 & Publication year & Year the study was published. & N/A \\

3 & Literature type & Type of study (i.e., academic or gray). & N/A \\

4 & Study type & Type of empirical or technical study conducted. & RQ4.1 \\

5 & ML family & Family of the monitored machine learning model. & RQ1.1, RQ2.1, RQ2.2, RQ2.3 \\

6 & ML task & Task performed by the monitored machine learning model. & RQ1.1 \\

7 & Deployment environment & Environment in which the machine learning system is deployed. & RQ1.2 \\

8 & Serving method & Method used to serve the machine learning system for inference. & RQ1.2 \\

9 & Monitored object & System component or layer being monitored. & RQ2.1, RQ2.2 \\

10 & Monitored property & Property or behaviour being monitored. & RQ2.1, RQ2.2 \\

11 & Monitoring tool & Monitoring tool or platform used. & RQ2.2 \\

12 & Monitoring technique & Technique used to perform monitoring. & RQ2.2, RQ2.3 \\

13 & Input signals & Information used as input to the monitoring technique. & RQ2.2 \\

14 & Monitoring frequency & Frequency at which monitoring is performed. & RQ2.3 \\

15 & Automation level & Degree of automation of the monitoring process. & RQ2.3 \\

16 & Evaluation approach & Method used to evaluate the monitoring approach. & RQ3.1 \\

17 & Evaluation metrics & Measures used to evaluate the monitoring approach. & RQ3.1 \\

18 & Evaluation domain & Application domain in which the monitoring approach was evaluated. & RQ3.2 \\

19 & Evaluation scale & Scale at which the monitoring approach was evaluated. & RQ3.2 \\

20 & Practitioner study method & Method used to collect practitioner evidence. & RQ4.1 \\

21 & Number of participants & Number of practitioners participating in the study. & RQ4.1 \\

22 & Participant domain & Industry domains represented by the participants. & RQ4.1 \\

23 & Participant roles & Professional roles of the participants. & RQ4.1 \\

24 & Monitoring challenges & Reported challenges of monitoring production ML systems & RQ4.2 \\

25 & Challenges addressed & Challenges addressed by the monitoring approaches. & RQ4.3 \\

\bottomrule
\end{tabularx}
\end{table*}

Data extraction was performed to systematically collect information required to answer the research questions. The data extraction fields were designed based on the research questions to ensure that all relevant aspects of production ML monitoring were captured. The complete list of these fields is presented in Table~\ref{tab:data-extraction}.

A data extraction form was implemented using Google Forms to standardize the recorded fields and facilitate data collection. The extraction was conducted by the first and second authors. To improve consistency, two pilot extraction rounds were performed on a subset of 10 randomly selected studies. These pilot rounds were used to refine the extraction fields, clarify definitions, and align the understanding of the extraction process between the authors. Following the pilot phase, the first and second authors divided the remaining included studies equally, with each author independently extracting data for 50\% of the studies using the final extraction form. Any uncertainties or disagreements during extraction were resolved through discussion between the authors until consensus was reached.

\subsubsection{Data Analysis and Synthesis}\label{analysis}
Data analysis and synthesis were conducted to interpret the extracted data and answer the research questions. The process combined quantitative analysis, cross-dimensional analysis, taxonomy development, and qualitative synthesis to identify patterns, relationships, and gaps in production ML monitoring literature.

Descriptive quantitative analysis was used to characterize the selected studies. Frequency-based analysis was applied to data extraction fields (e.g., ML families, tasks, deployment environments, monitored properties, monitoring techniques, and evaluation approaches) to establish the distribution of studies across key dimensions. These descriptive summaries provide the foundation for subsequent relational and interpretive analyses.

Beyond descriptive summaries, cross-dimensional analysis was conducted to examine relationships between variables. Given the multi-dimensional nature of the dataset, pairwise and multi-dimensional comparisons were performed to identify co-occurrence patterns and associations across dimensions, for example, which monitoring techniques are applied to specific monitored properties, how monitoring practices vary across ML paradigms, and how operational characteristics relate to monitoring techniques.

To structure the domain, taxonomies were developed inductively using thematic synthesis. We followed the guidelines for taxonomy development in software engineering proposed by Ralph \cite{ralph2018toward}. The extracted data was first open-coded to capture the concepts reported in each study. Codes representing semantically similar or functionally equivalent concepts were then iteratively grouped into high-level categories through comparison across studies. The monitored-object taxonomy was the only exception to this fully inductive process; its five-layer structure was informed by ML system architectures reported in prior literature and the included studies \cite{mltestscore2017, sculley2015hidden, monitoringobservability2025, feuerriegel2024generative, baylor2017tfx}, and its definitions and boundaries were refined during the analysis. 
For all taxonomies, categories were iteratively refined by merging overlapping concepts, separating conceptually distinct categories, clarifying category boundaries, and revisiting earlier coding decisions. This process continued until a stable and coherent classification structure was obtained. The resulting taxonomies support the organization and interpretation of findings across the research questions.

Qualitative synthesis was conducted for open-text fields, including monitoring challenges and challenges addressed. A thematic synthesis approach was used, where extracted observations were initially coded and iteratively grouped into higher-level themes. These themes complement the quantitative analyses by providing contextual and interpretive insights into production ML monitoring practices.

Finally, a gap analysis was conducted to identify underrepresented areas within the literature. This involved examining sparsely populated taxonomy categories, limited or absent combinations across analytical dimensions, and imbalances in coverage across monitoring stages, evaluation contexts, and system characteristics. Together, these analyses supported the identification of research gaps, limitations in the available evidence, and opportunities for future research.

The findings are reported through a combination of descriptive summaries, cross-dimensional analyses, taxonomy-driven organization, and thematic synthesis. All analysis and synthesis steps were conducted by the first and second authors, with iterative discussions used to resolve ambiguities and review consistency across the analysis stages.

\subsection{Reporting the Review}
The reporting stage presents the findings in a structured and interpretable manner. Results are organized around the research questions, providing a clear progression from an overview of the included studies to detailed findings on production ML monitoring. Taxonomies are used as organizing frameworks to structure the presentation of results, while tables and figures are used to summarize key findings and support comparison across studies. Cross-cutting insights are highlighted where relevant to connect findings across different aspects of production ML monitoring.

\section{Results}\label{results}
This section presents an overview of the included studies, the developed taxonomies, and the findings for each research question.
\subsection{Overview of Included Studies}
\begin{figure*}[t]
    \centering
    \begin{subfigure}[t]{0.60\textwidth}
        \centering
        \includegraphics[width=\linewidth]{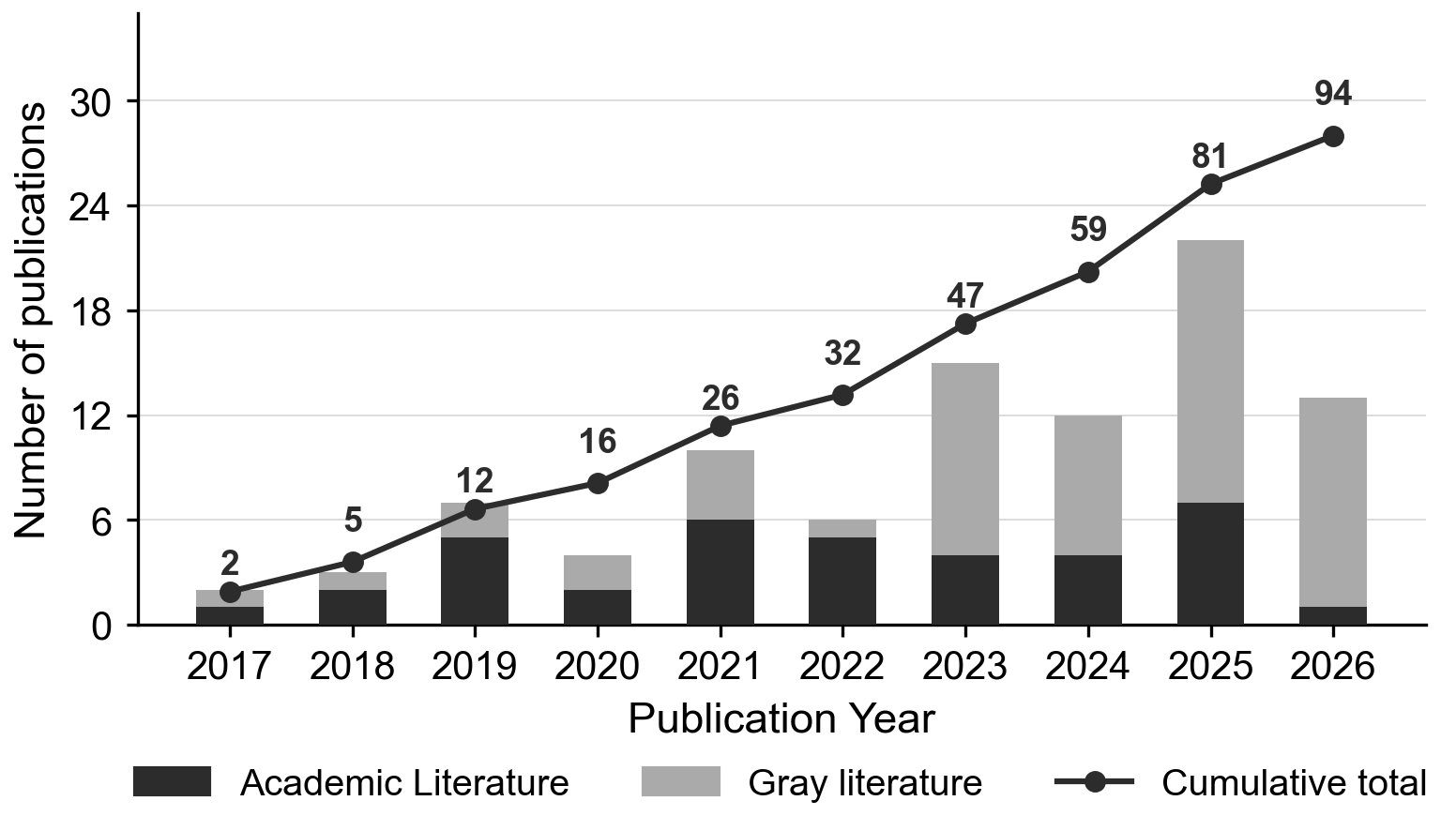}
        \caption{Publications Per Year}
        \label{fig:publications_per_year}
    \end{subfigure}
    \hfill
    \begin{subfigure}[t]{0.38\textwidth}
        \centering
        \includegraphics[width=\linewidth]{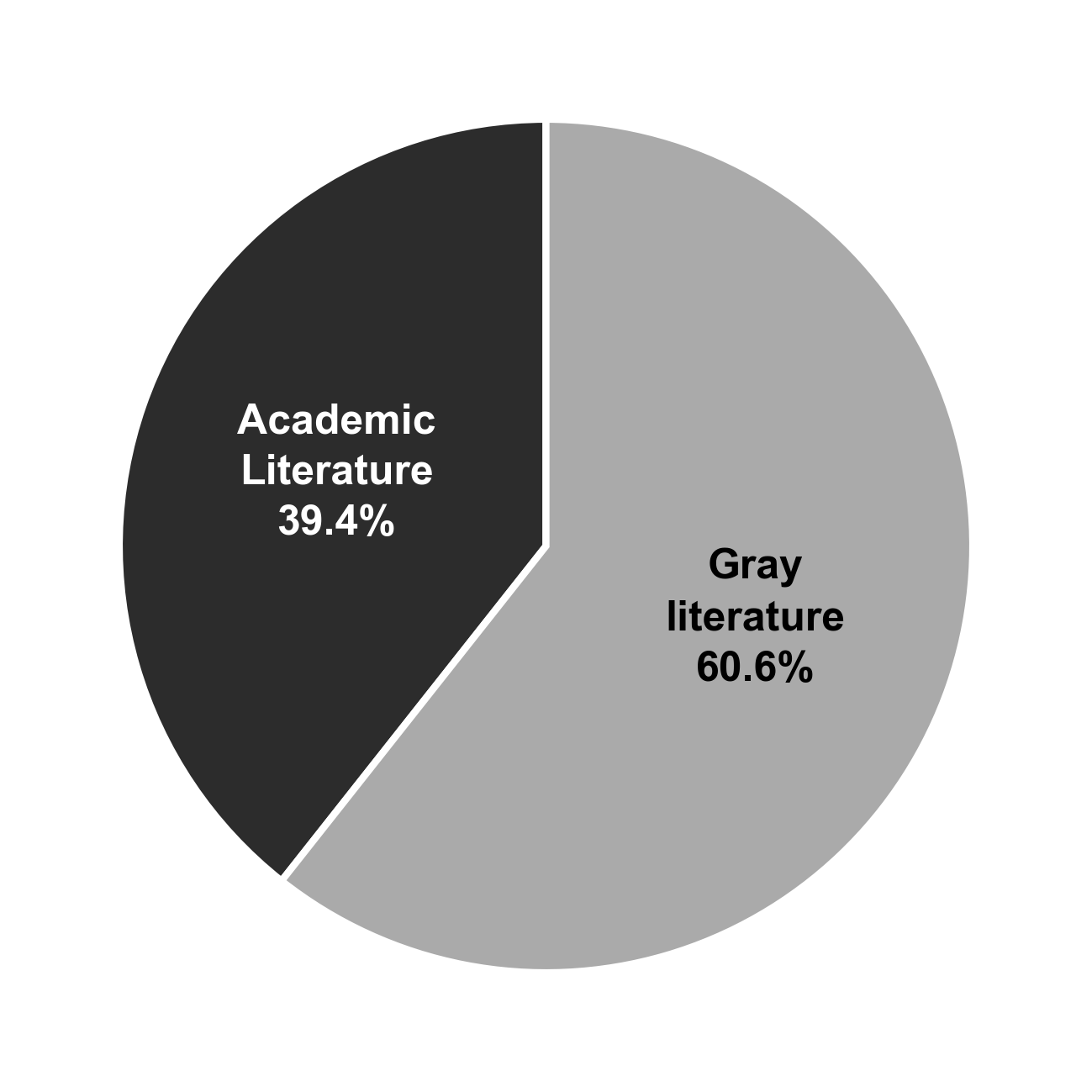}
        \caption{Literature Distribution}
        \label{fig:literature_distribution}
    \end{subfigure}
    \caption{Publication Trends of ML Monitoring Literature}
    \label{fig:pub_trends}
\end{figure*}
\begin{figure}
    \centering
    \includegraphics[width=0.6\linewidth]{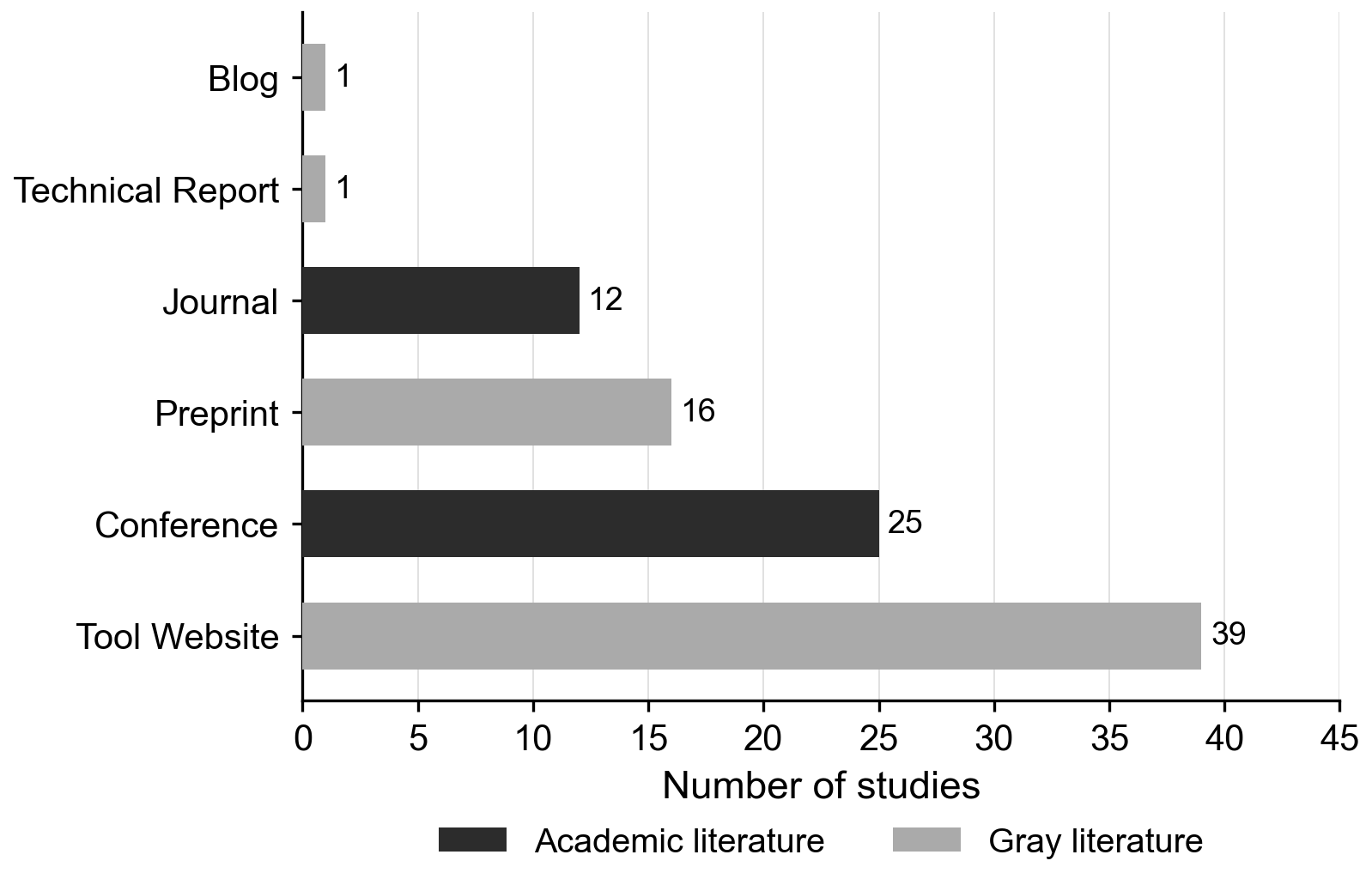}
    \caption{Study Type Distribution}
    \label{fig:study_types}
\end{figure}

The publication trends and study types of the included studies are presented in Figures~\ref{fig:pub_trends} and~\ref{fig:study_types}. Figure \ref{fig:publications_per_year} shows the distribution of the included studies by publication year, distinguishing academic and gray literature. The literature on monitoring production ML systems has grown substantially over the study period, with the cumulative number of included studies increasing from 2 in 2017 to 94 in 2026. Growth increases particularly from 2021 onward, reaching 15 studies in 2023, 12 in 2024, and 22 in 2025. The 2026 count only represents the studies available at the time of the MLR search and should not be considered as the total studies for that year. Overall, gray literature represents 60.6\% of the included studies (57 out of 94), while academic literature represents 39.4\% (37 out of 94), as shown in Figure~\ref{fig:literature_distribution}. The rise in gray literature is particularly evident from 2023 onward, with gray literature constituting the majority of studies in each year.

Figure~\ref{fig:study_types} shows the various types of studies included in the MLR. Tool websites constitute the largest study type, accounting for 39 studies, followed by 25 conference papers, 16 preprints, and 12 journal papers. Technical reports and blogs are the least represented, with one study each. The prevalence of tool websites indicates that a significant portion of current literature about monitoring production ML systems is dominated by industry sources rather than conventional academic publications. The presence of 16 preprints also reflects the rapid evolution of the area, with research findings frequently becoming available prior to formal peer review.

\subsection{RQ1. What are the characteristics of production ML systems being monitored?
}
\subsubsection{RQ1 Taxonomy}
\begin{table}[!htbp]
\centering
\caption{RQ1 Taxonomy}
\label{tab:corpus-taxonomy}
\small
\begin{tabular}{@{}L{0.16\textwidth} L{0.20\textwidth} L{0.56\textwidth}@{}}
\toprule
\textbf{Taxonomy} & \textbf{Category} & \textbf{Description} \\
\midrule

\multirow{6}{=}{ML Family}
 & Traditional ML & Classical statistical or shallow learning models, e.g.\ regression, tree-based, or ensemble methods \\
 & Deep Learning & Neural network models with multiple hidden layers, excluding generative architectures \\
 & Generative AI & Large language models and other generative architectures, including LLM-based agents \\
\midrule

\multirow{10}{=}{ML Task}
 & Classification & Assigning inputs to discrete categories \\
 & Regression & Predicting a continuous numeric value \\
 & Retrieval & Finding relevant items or documents given a query \\
 & Recommendation & Suggesting items to a user based on preferences or behavior \\
 & Ranking & Ordering a set of items by relevance or priority \\
 & Forecasting & Predicting future values from historical time-series data \\
 & Anomaly Detection & Identifying unusual or outlier instances \\
 & Generation & Producing novel text, code, or other content \\
 & Agent Orchestration & Coordinating multi-step or multi-tool agentic workflows \\
 & Task Agnostic & Monitoring approach is not tied to a specific downstream task \\
\midrule

\multirow{3}{=}{Deployment Environment}
 & Cloud & Hosted on public or private cloud infrastructure \\
 & On-premise & Hosted on infrastructure owned and operated by the organization \\
 & Edge & Deployed on edge devices with local inference \\
 
\midrule

\multirow{3}{=}{Serving Method}
 & API / Service & Exposed via a network-accessible API endpoint \\
 & Managed ML Platform & Served through a managed ML platform, e.g.\ SageMaker, Vertex AI \\
 & Embedded in Application & Model runs directly within the client or host application \\
\bottomrule
\end{tabular}
\end{table}

The RQ1 taxonomy characterizes the production ML system that is subject to monitoring. It separates four dimensions---\textbf{ML family}, \textbf{ML task}, \textbf{deployment environment}, and \textbf{serving method}---because each constrains monitoring requirements in a different way. The dimensions are coded independently: for example, a generative model may be hosted on-premise and exposed through an API, while a traditional classifier may be embedded in an edge application. This separation avoids treating a familiar deployment pattern as an intrinsic property of a model family and permits heterogeneous systems to be described through combinations of categories. It also reflects the system-level view that production behavior depends not only on a learned model but on the surrounding data, software, infrastructure, and operational setting~\cite{sculley2015hidden}.

\textbf{ML family} identifies the broad model paradigm. Traditional ML covers classical statistical and shallow-learning models, including regression, tree-based, and ensemble methods. Deep learning covers multilayer neural models used primarily for discriminative or representation-learning purposes, while generative AI covers architectures that produce new content, including large language models and LLM-based agents. The distinction is monitoring-relevant rather than purely historical. Traditional and discriminative deep learning systems commonly expose structured predictions and established task metrics, whereas generative systems produce open-ended outputs whose correctness, faithfulness, safety, and task completion may require semantic or human evaluation. Agents remain within generative AI at the family level, but their orchestration requirements are represented separately by the task dimension.

\textbf{ML task} describes the functional operation performed by the deployed system. Classification and regression capture discrete and continuous prediction; retrieval, recommendation, and ranking cover the selection and ordering of items; forecasting captures temporally ordered prediction; and anomaly detection identifies unusual observations. Generation denotes the production of text, code, or other content, while agent orchestration captures multi-step workflows involving planning, memory, or tool use. Task agnostic is reserved for approaches designed without dependence on a particular downstream task. These categories are not forced to be mutually exclusive at the system level: a retrieval-augmented generation system, for example, can involve both retrieval and generation. Multi-label coding preserves such compositions instead of reducing the system to its final output.

The two operational dimensions locate the model within its runtime setting. \textbf{Deployment environment} distinguishes cloud, on-premise, and edge execution. These settings shape resource constraints, connectivity, instrumentation, data residency, and the feasibility of centralizing monitoring evidence. \textbf{Serving method} distinguishes a network-accessible API or service, a managed ML platform, and a model embedded directly in an application. Serving method is separated from deployment environment because an API can be hosted in cloud or on-premise infrastructure, and managed services differ from self-operated endpoints in the monitoring controls they expose. Together, the four dimensions provide a concise system profile against which monitoring coverage can be interpreted without presuming that one monitoring design transfers unchanged across model families, tasks, or deployment arrangements~\cite{mltestscore2017,haveidea2024}.

\subsubsection{RQ1.1. What are the ML families and tasks of production ML systems being monitored?}
\begin{figure*}[t]
    \centering

    \begin{subfigure}[t]{0.38\textwidth}
        \centering
        \includegraphics[width=\linewidth]{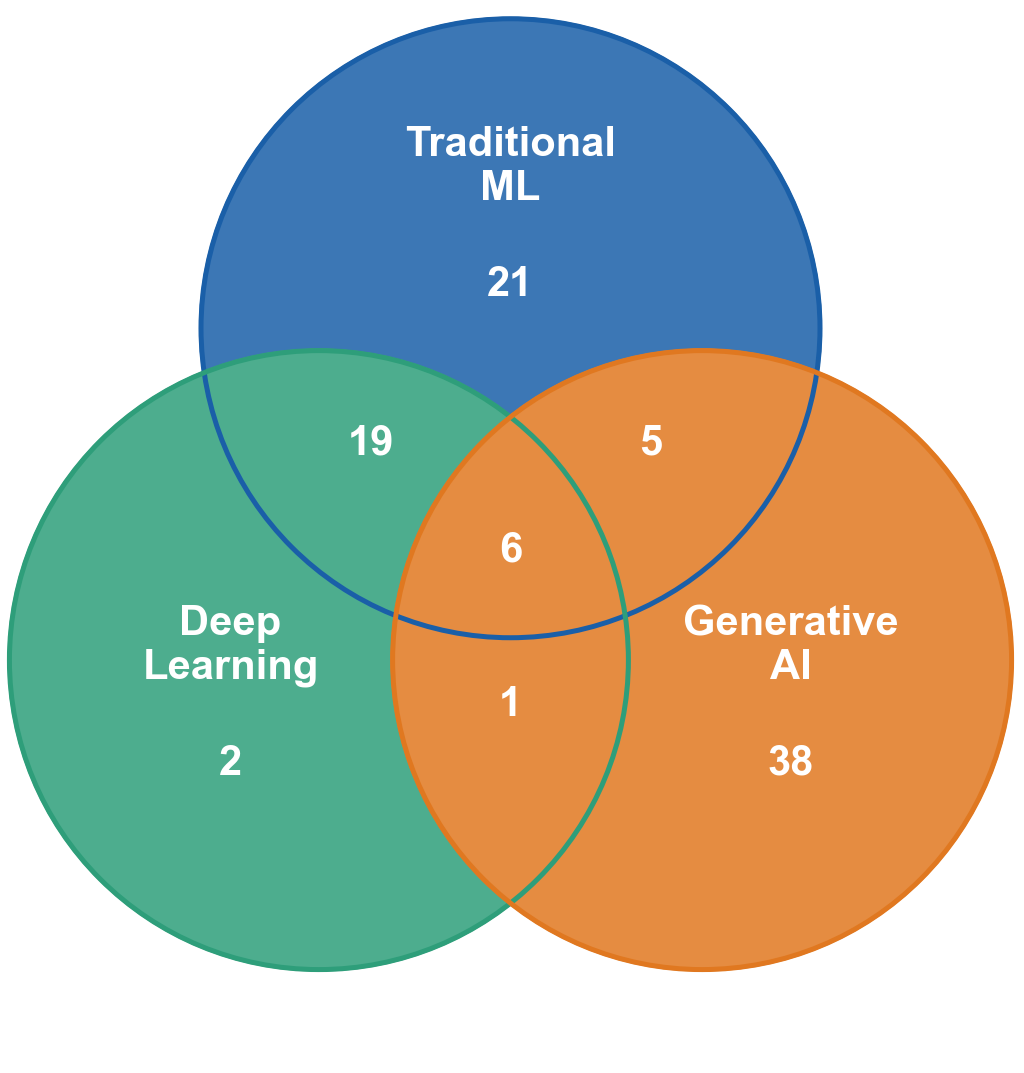}
        \caption{ML Families Monitored}
        \label{fig:ML_families}
    \end{subfigure}
    \hfill
    \begin{subfigure}[t]{0.60\textwidth}
        \centering
        \includegraphics[width=\linewidth]{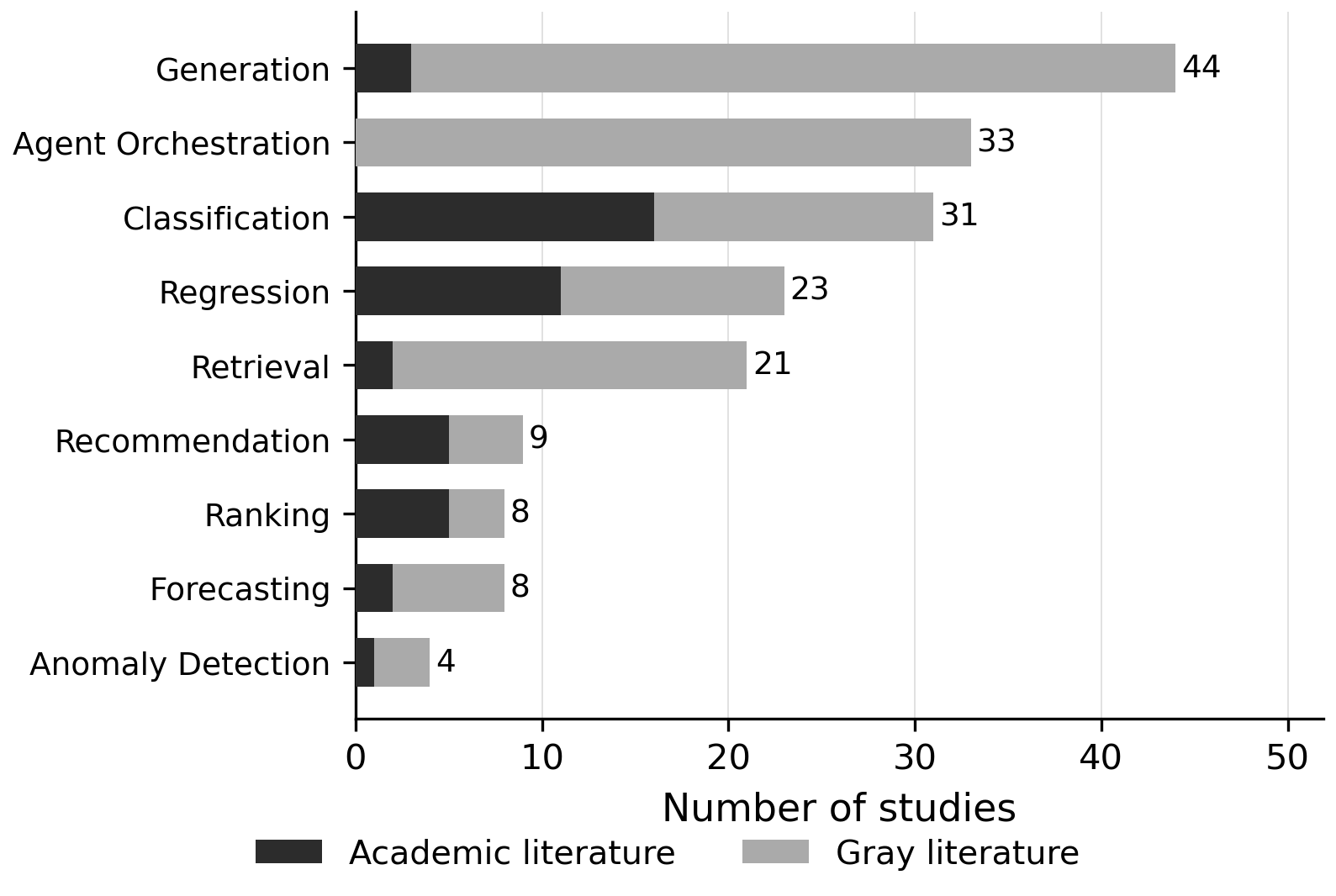}
        \caption{ML Tasks Monitored}
        \label{fig:ML_tasks}
    \end{subfigure}

    \caption{ML Families and Tasks}
    \label{fig:families_and_tasks}
\end{figure*}

\textbf{ML Families. } Figure~\ref{fig:ML_families} shows that production ML monitoring research covers traditional ML, deep learning, and generative AI systems, with substantial overlap between these families. Traditional ML \cite{pipelinemonitoring2023, frameworkmonitor2019, michelangelo2017, alertiger2023, daqlmonitor2019, aitotal2021} and generative AI \cite{arizephoenixnd, beyondblackbox2025, llamafirewall2025, llamaguard2023, chainmonitorability2025, galileond, raindropnd} are the most widely represented, appearing in 51 and 50 sources, respectively, while deep learning appears in 28. However, only two sources focus exclusively on deep learning \cite{overton2019, empiricalinvestigation2024}; most examine it together with traditional ML or generative AI \cite{michelangelo2017, censiusnd, radicalbit2024, fiddlernd, datavalidation2019, timeretrain2025}. This is a property of the coded corpus: the family scheme separates generative architectures from deep learning and allows multiple family labels per source. The small deep learning exclusive count therefore does not independently establish that deep learning lacks a distinct monitoring context in practice. Generative AI, by contrast, has developed into a prominent monitoring area, with 38 sources focusing exclusively on this family. Its coverage is driven primarily by gray literature, within the included corpus of generative AI monitoring sources \cite{honeycomb2026, braintrust2026, shieldgemma2024, arizephoenixnd, llmobservabilitynd, llmobservability2nd, agentobservability2024, dynatracend}.

\textbf{ML Tasks. } The monitored tasks are similarly diverse, as shown in Figure~\ref{fig:ML_tasks}. Generation is the most frequently represented task, appearing in 44 sources, followed by agent orchestration, classification, regression, and retrieval. Comparatively, recommendation, ranking, forecasting, and anomaly detection appear in fewer sources. One source presents a task-agnostic approach, while 20 sources do not report a particular ML task. Established predictive tasks such as classification, regression, forecasting, and recommendation are represented in traditional and deep learning systems \cite{pipelinemonitoring2023, michelangelo2017, alertiger2023, sagemakermodel2022}. In contrast, generation, retrieval, and agent orchestration commonly occur together in generative AI applications, including RAG pipelines and agentic systems \cite{arizephoenixnd, beyondblackbox2025, agentobservability2024, galileond}.  Academic literature is concentrated mainly on established predictive tasks such as classification and regression, whereas gray literature dominates generation, agent orchestration, and retrieval. This division indicates that academic research remains comparatively focused on conventional predictive systems, while practitioner sources provide much of the available coverage of newer generative and agentic systems. For example, \cite{frameworkmonitor2019} demonstrates monitoring for conventional predictive models, whereas \cite{arizephoenixnd} traces retrieval operations, model generations, and tool calls across RAG and agentic application workflows.

\subsubsection{RQ1.2. What are the deployment environments and serving methods of production ML systems being monitored?}

\begin{figure*}[t]
    \centering

    \begin{subfigure}[t]{0.49\textwidth}
        \centering
        \includegraphics[width=\linewidth]{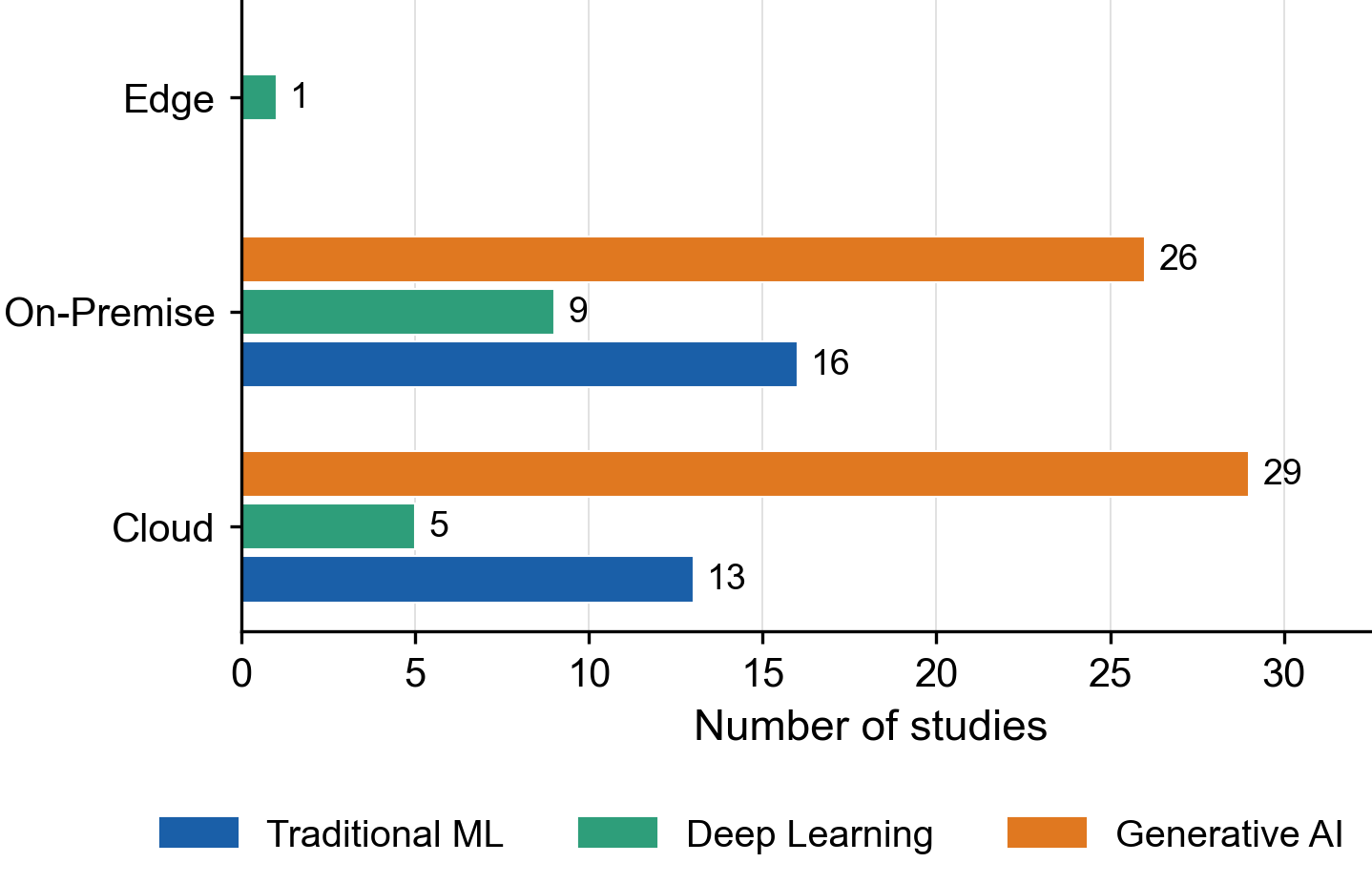}
        \caption{Deployment Environments across ML Families}
        \label{fig:ML_deployment}
    \end{subfigure}
    \hfill
    \begin{subfigure}[t]{0.49\textwidth}
        \centering
        \includegraphics[width=\linewidth]{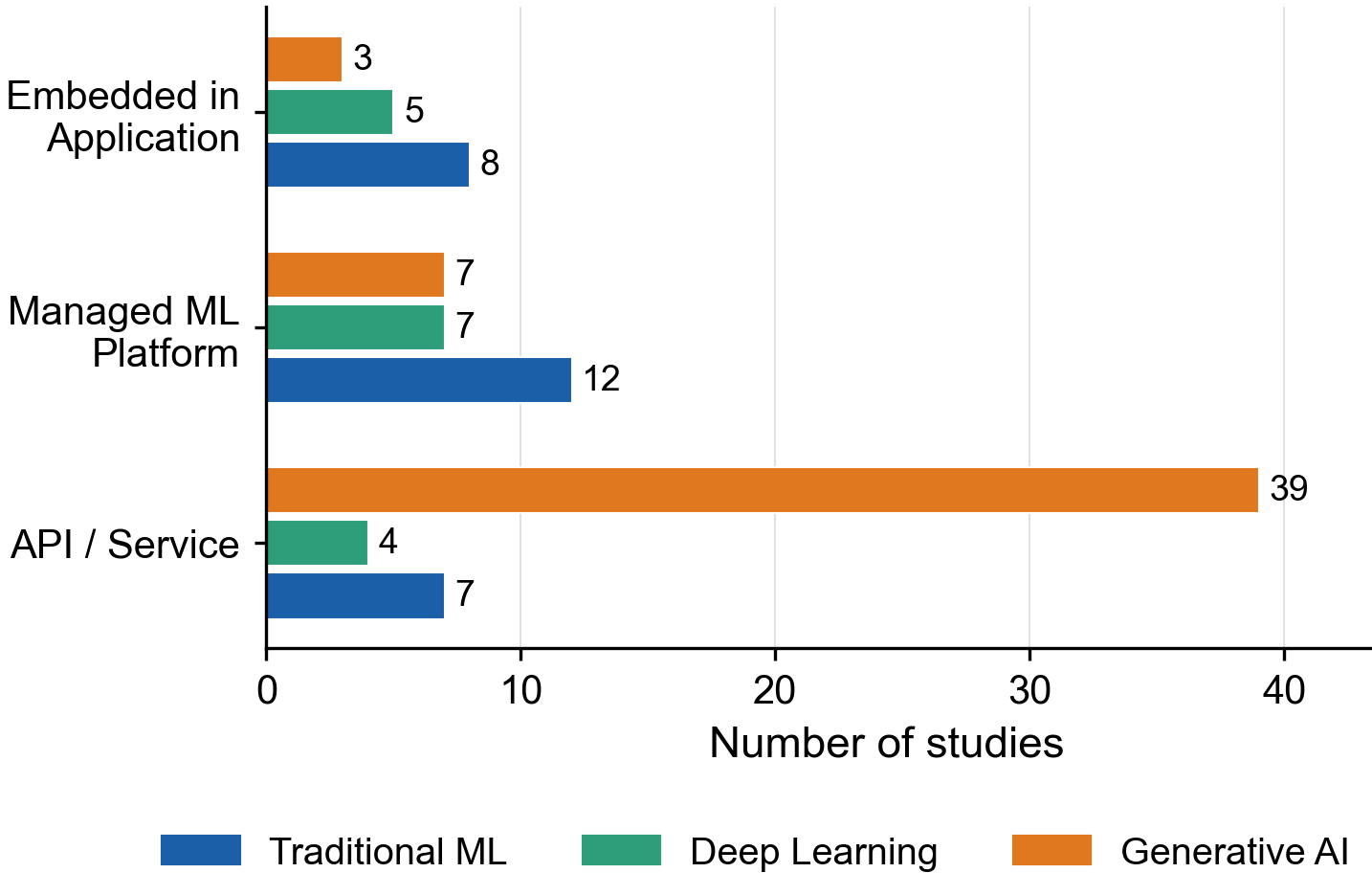}
        \caption{Serving Methods across ML Families}
        \label{fig:ML_serving}
    \end{subfigure}

    \caption{Operational Characteristics of Production ML Systems}
    \label{fig:deployment_and_serving}
\end{figure*}

\textbf{Deployment Environments. } Figure~\ref{fig:deployment_and_serving} presents the deployment environments and serving methods reported across the three ML families. Cloud and on-premise environments dominate the reported deployment settings, while edge monitoring is scarce. Generative AI systems are frequently reported in both cloud and on-premise environments, whereas traditional ML systems also span both settings. Cloud-based settings include managed monitoring services and hosted observability platforms \cite{sagemakermodel2022, alertiger2023, healthassurance2022, azurend, vertexai2026}, whereas on-premise settings include internally operated ML platforms and locally deployed monitoring components \cite{michelangelo2017, overton2019, nannyml2025, seldon2022, deployingmachine2018}. For example, Amazon SageMaker Model Monitor supports models deployed through managed cloud infrastructure \cite{sagemakermodel2022}, while Uber's Michelangelo integrates model serving and monitoring within an internally operated production platform \cite{michelangelo2017}. Edge monitoring is represented by only one source, which investigates runtime monitors for ML components embedded within safety-critical systems \cite{empiricalinvestigation2024}. The limited coverage indicates that current knowledge about production ML monitoring is concentrated on centrally operated infrastructure and provides little insight into monitoring under the resource, connectivity, and instrumentation constraints of edge environments.

\textbf{Serving Methods. } API- or service-based serving is the most frequently reported method, particularly for generative AI systems, for which it appears in 39 sources. These include LLM and agentic applications accessed through dedicated inference or application services \cite{llamafirewall2025, beyondblackbox2025, agentobservability2024}. Managed ML platforms are more prominent for traditional ML and deep learning systems \cite{sagemakermodel2022, vertexai2026, overton2019, scalingml2026, michelangelo2017, alertiger2023}, while models embedded directly within applications are reported less frequently across all three families \cite{pipelinemonitoring2023, endend2025, deepchecks2024, nannyml2025, empiricalinvestigation2024, deployingmachine2018}.  For example, LlamaFirewall operates as a guardrail around API-served agentic systems \cite{llamafirewall2025}. In contrast, the text-classification pipeline presented by Fu et al. integrates monitoring into the host application, where requests, predictions, and confidence scores are collected during system operation \cite{pipelinemonitoring2023}. 

\begin{rqanswer}{RQ1. What are the characteristics of production ML systems being monitored?}
\textbf{RQ1 Answer. }The literature covers diverse production ML systems, with traditional ML and generative AI most widely represented. Deep learning is typically examined alongside other ML families rather than independently. Generation, agent orchestration, classification, regression, and retrieval are the most frequently represented ML tasks, with gray literature providing most of the coverage of generative and agentic systems. Reported deployment settings are predominantly cloud and on-premise environments, with models commonly served through APIs or managed platforms. Edge deployments and application-embedded models receive comparatively limited coverage. \end{rqanswer}

\subsection{RQ2. How are production ML systems monitored?}
\subsubsection{RQ2 Taxonomy}
\begin{table}[!htbp]
\centering
\caption{RQ2 Taxonomy}
\label{tab:mo-mp-is-taxonomy}
\small
\begin{tabular}{@{}L{0.14\textwidth} L{0.25\textwidth} L{0.56\textwidth}@{}}
\toprule
\textbf{Taxonomy} & \textbf{Category} & \textbf{Description} \\
\midrule

\multirow{9}{=}{Monitored Object}
 & Preprocessing Layer & Data ingestion and feature engineering stages upstream of model inference \\
 & Inference Layer & The model serving path from input to prediction, including the model \\
 & Agentic Layer & Multi-step agent orchestration, tool use, and reasoning workflows built on top of a model \\
 & Serving Infrastructure & The compute, networking, and platform infrastructure hosting the deployed system \\
 & Application \& Business Layer & The downstream application and business context in which the system's outputs are used \\
\midrule

\multirow{21}{=}{Monitored Property}
 & Concept Drift & Change in the relationship between model inputs and target outcomes \\
 & Data Quality \& Integrity & Completeness, consistency, validity, and integrity of incoming data \\
 & Feature Drift & Distributional shift in input features relative to training or a reference window \\
 & Predictive Performance & Accuracy, error, or ranking quality of model predictions against outcomes \\
 & Prediction Drift & Distributional shift in model outputs, predicted scores, or predicted classes \\
 & Feature Attribution Drift & Change in which features drive predictions relative to a training baseline \\
 & Fairness \& Bias & Disparities in model behavior or performance across protected groups \\
 & LLM Generation Quality & Faithfulness, relevance, correctness, and coherence of generated text \\
 & RAG Retrieval Quality & Relevance, recall, and precision of retrieved context in retrieval-augmented generation \\
 & Safety & Harmful, adversarial, or policy-violating model behavior \\
 & Reasoning \& Alignment & Legibility and correctness of a model's reasoning relative to its goal \\
 & Agent Task Execution & Correctness, efficiency, and completion of multi-step agentic task execution \\
 & Operational Performance & Latency, throughput, error rate, cost, resource utilization, and infrastructure health of the serving system\\
 & Business \& Application Impact & Downstream business or user-facing outcomes attributable to the model \\
 & Compliance \& Governance & Auditability, policy adherence, and accountability for system behavior \\

\midrule

\multirow{14}{=}{Input Signal}
 & Feature Vectors & Raw feature values, input distributions, and feature statistics at inference time \\
 & Reference Dataset & Pre-computed training or recent production data snapshot used as a comparison baseline \\
 & Model Predictions & Model output scores, probabilities, or class predictions \\
 & Ground Truth Labels & Actual outcomes used for supervised performance evaluation \\
 & LLM Request/Response Pairs & Full LLM input prompt and generated completion \\
 & Embeddings & Dense vector representations of inputs or outputs \\
 & Execution Traces & Agent step logs, tool call records, and chain execution trees \\
 & LLM Reasoning Traces & Externalised chain-of-thought reasoning traces \\
 & System Telemetry & Raw infrastructure instrumentation: latency, compute utilisation, errors \\
 & Human Feedback \& Annotations & Human labels, ratings, or pairwise comparisons \\
 & User \& Business Activity & Raw user interaction events and business transactions \\

 \midrule
\multirow{3}{=}{Automation Level}
 & Manual & A person performs the monitoring action directly \\
 & Semi-automated & The monitoring action is partially automated, with human-in-the-loop \\
 & Fully automated & The monitoring action runs without human involvement \\

\midrule
\multirow{3}{=}{Monitoring Frequency}
 & Scheduled & Monitoring runs on a scheduled or periodic basis over accumulated data \\
 & Real-time & Monitoring runs continuously as data or predictions are produced \\
 & Hybrid & Combines batch and real-time monitoring within the same system \\
 
\bottomrule
\end{tabular}
\end{table}

The first RQ2 taxonomy defines the scope and operating configuration of production ML monitoring through five complementary dimensions: \textbf{monitored object}, \textbf{monitored property}, \textbf{input signal}, \textbf{automation level}, and \textbf{monitoring frequency}. Together, they distinguish where monitoring is applied, what concern it assesses, which observable evidence supports the assessment, and how the monitoring activity is executed. The dimensions are coded independently and may receive multiple values. This is necessary because a single approach may monitor several layers, combine prediction and telemetry signals, and perform continuous automated collection followed by scheduled human evaluation. Separating the dimensions prevents a metric, signal, or tool from being treated as a complete monitoring design.

The \textbf{monitored object} identifies the architectural location to which a monitoring claim applies. The preprocessing layer covers data ingestion and feature engineering upstream of inference, whereas the inference layer covers the path from prepared input to prediction, including the model. The agentic layer represents multi-step orchestration, tool use, and reasoning workflows built on top of a model. It is kept distinct because a satisfactory final response can conceal invalid tool calls, inefficient loops, or incomplete execution. Serving infrastructure covers the compute, networking, and platform services hosting the system. The application and business layer covers the downstream context in which outputs affect users, workflows, or organizational outcomes. These objects form an end-to-end system rather than isolated components; failures can originate in one layer and become visible in another~\cite{sculley2015hidden}.

The \textbf{monitored property} specifies the quality or behavior being assessed. Concept drift is the change in the relationship between model inputs and the corresponding target outcomes over time. Data quality and integrity and feature drift concern the validity and distribution of inputs. Predictive performance, prediction drift, feature-attribution drift, and fairness and bias concern learned behavior and its distribution across groups. Generative and agentic systems add LLM generation quality, RAG retrieval quality, safety, reasoning and alignment, and agent task execution. Operational performance captures latency, throughput, error rate, cost, resource utilization, and infrastructure health, measured at the relevant system layer; business and application impact captures downstream consequences; and compliance and governance captures auditability, policy adherence, and accountability. These categories distinguish a target property from its proxy. Prediction drift, for example, is observable without labels but does not by itself establish degraded predictive performance.

\textbf{Input signals} record the evidence available to the monitor: feature vectors, reference datasets, model predictions, ground-truth labels, LLM request/response pairs, embeddings, execution traces, LLM reasoning traces, system telemetry, human feedback and annotations, and user and business activity. Recording signals separately exposes the limits of the resulting claim, including delayed labels, incomplete traces, biased feedback, or unavailable business outcomes. Data-validation systems similarly separate captured statistics and schemas from inferred anomalies~\cite{tensorflowdata2020}.

Finally, \textbf{automation level} distinguishes manual actions, semi-automated actions involving human participation, and fully automated actions. \textbf{Monitoring frequency} distinguishes scheduled evaluation over accumulated data, real-time evaluation as events occur, and hybrid designs combining both. Neither dimension is a maturity ranking: appropriate configuration depends on evidence latency, failure severity, operational cost, and the degree of judgment required. Coding them per monitoring activity preserves mixed designs rather than assigning one configuration to an entire system.

\begin{table}[!htbp]
\centering
\caption{RQ2 Monitoring Techniques Taxonomy}
\label{tab:monitoring-taxonomy}
\small
\begin{adjustbox}{max height=0.85\textheight}
\begin{tabular}{@{}L{0.10\textwidth} L{0.28\textwidth} L{0.55\textwidth}@{}}
\toprule
\textbf{Monitoring Stage} & \textbf{Technique} & \textbf{Description} \\
\midrule

\multirow{7}{=}{Define}
 & Metric Selection & Deciding which metrics, signals, and monitoring objects to track \\
 & Schema Definition & Defining constraints, schemas, and rules that data or outputs must satisfy \\
 & Baseline Definition & Defining baseline distributions or reference datasets used for comparison \\
 & Threshold Configuration & Configuring thresholds, service-level objectives, and evaluation criteria used to judge system behavior \\
\midrule

\multirow{4}{=}{Observe}
 & System Instrumentation & Instrumenting pipelines and systems to collect monitoring signals \\
 & Distributed Trace Capture & Capturing execution traces across distributed system components \\
 & Request/Response Interception & Intercepting requests and responses in real time before or after inference \\
\midrule

\multirow{13}{=}{Detect}
 & Statistical Hypothesis Testing & Applying statistical significance tests to compare current and reference distributions \\
 & Distribution Distance Measurement & Quantifying the distance or divergence between current and reference distributions \\
 & Rule-Based Checking & Validating data or outputs against predefined rules, patterns, or business constraints \\
 & Threshold Detection & Flagging issues when computed metrics cross predefined thresholds \\
 & Classifier-Based Detection & Using a trained model to classify or flag anomalous behavior \\
 & LLM-Based Evaluation & Using a large language model as an automated judge of quality, safety, or alignment \\
 & Embedding Comparison & Comparing embedding representations to detect semantic drift \\
 & Feature Attribution Analysis & Computing feature attribution scores to identify influential inputs \\
 & Subgroup \& Slice-Level Detection & Detecting issues concentrated within specific data subgroups or slices \\
\midrule

\multirow{3}{=}{Notify}
 & Automated Alerting & Automatically pushing notifications to relevant stakeholders or systems \\
 & Reports \& Scores & Producing structured reports or scores for a person to review \\
 & Dashboard \& Visualization & Providing an interactive visual dashboard for exploring metrics \\
\midrule

\multirow{7}{=}{Analyze}
 & Log \& Trace Inspection & A person investigating traces, logs, or audit records to determine root cause \\
 & Subgroup \& Slice Investigation & A person investigating a specific data subgroup or slice in detail \\
 & Feature Attribution for Root Cause Analysis & Using computed attribution scores, interpreted by a person, to determine root cause \\
 & Human Review \& Annotation & A person labeling, annotating, or reviewing model outputs or monitoring findings \\
\midrule

\multirow{3}{=}{Respond}
 & Guardrail Enforcement & Automatically blocking, masking, or remediating policy-violating content \\
 & Automated Retraining & Automatically retraining or updating the model as part of its lifecycle \\
\bottomrule
\end{tabular}
\end{adjustbox}
\end{table}

The second RQ2 taxonomy classifies monitoring techniques according to the stage of the production ML monitoring process in which they perform their primary function: \textbf{Define}, \textbf{Observe}, \textbf{Detect}, \textbf{Notify}, \textbf{Analyze}, or \textbf{Respond}. This organization treats monitoring as a structured process that extends beyond runtime data collection and anomaly detection to include the activities required to configure monitoring, communicate findings, investigate problems, and initiate corrective action. The stage boundary is functional: define builds monitors, observation makes evidence available, detection evaluates that evidence, notification communicates a finding, analysis establishes meaning or cause, and response changes system behavior. This separation clarifies which parts of the monitoring process an approach implements and which it leaves to operators or external systems~\cite{domainspecific2023}.

\textbf{Define} establishes the monitoring specification. Metric selection chooses the properties, signals, and measures to track. Schema definition states the structural constraints or rules that data and outputs must satisfy. Baseline definition identifies the training distribution, reference dataset, validated period, or other comparator against which current behavior is judged. Threshold configuration sets decision boundaries, service-level objectives, or evaluation criteria. These techniques are separated because the same computed metric can produce different operational conclusions under different baselines and thresholds. Declarative monitoring specifications can make these decisions explicit, versioned, and reproducible~\cite{domainspecific2023}.

\textbf{Observe} covers the collection of monitoring evidence. System instrumentation emits data, model, application, and infrastructure signals. Distributed trace capture preserves causal paths across components, while request/response interception records or evaluates traffic around inference. Observation is not itself a judgement: instrumentation can expose a changed distribution without determining whether the change is harmful.

\textbf{Detect} contains techniques that turn collected signals into candidate findings. Statistical hypothesis testing assesses whether current and reference samples differ beyond a significance criterion, while distribution-distance measurement quantifies the magnitude of divergence. Rule-based checking validates predefined constraints, and threshold detection compares a computed metric with a configured boundary. Classifier-based detection uses a learned detector; LLM-based evaluation uses a language model to judge quality, safety, or alignment; and embedding comparison assesses semantic change in vector space. Feature-attribution analysis examines changes in influential inputs, while subgroup and slice-level detection searches for failures hidden by aggregate results. These techniques provide different forms of evidence and are therefore not interchangeable: detecting drift, for example, does not establish its cause or downstream consequence.

\textbf{Notify} communicates a finding through automated alerting, structured reports and scores, or a dashboard and visualization. The categories distinguish pushed notifications from artifacts that require active review. 

\textbf{Analyze} covers log and trace inspection, subgroup and slice investigation, feature attribution for root-cause analysis, and human review and annotation. Unlike automated detection, these techniques use contextual interpretation to determine cause, severity, or validity. 

Finally, \textbf{Respond} performs actions to minimize or mitigate the detected faults. Guardrail enforcement blocks, masks, or remediates unacceptable output, while automated retraining updates the ML model. A response technique is coded only when the approach changes or initiates a change to the system; generating an alert alone does not constitute remediation. 

\subsubsection{RQ2.1. What are the monitored objects and monitored properties of production ML systems?}
\begin{figure*}[!htbp]
    \centering

    \begin{subfigure}[t]{0.45\textwidth}
        \centering
        \includegraphics[width=\linewidth]{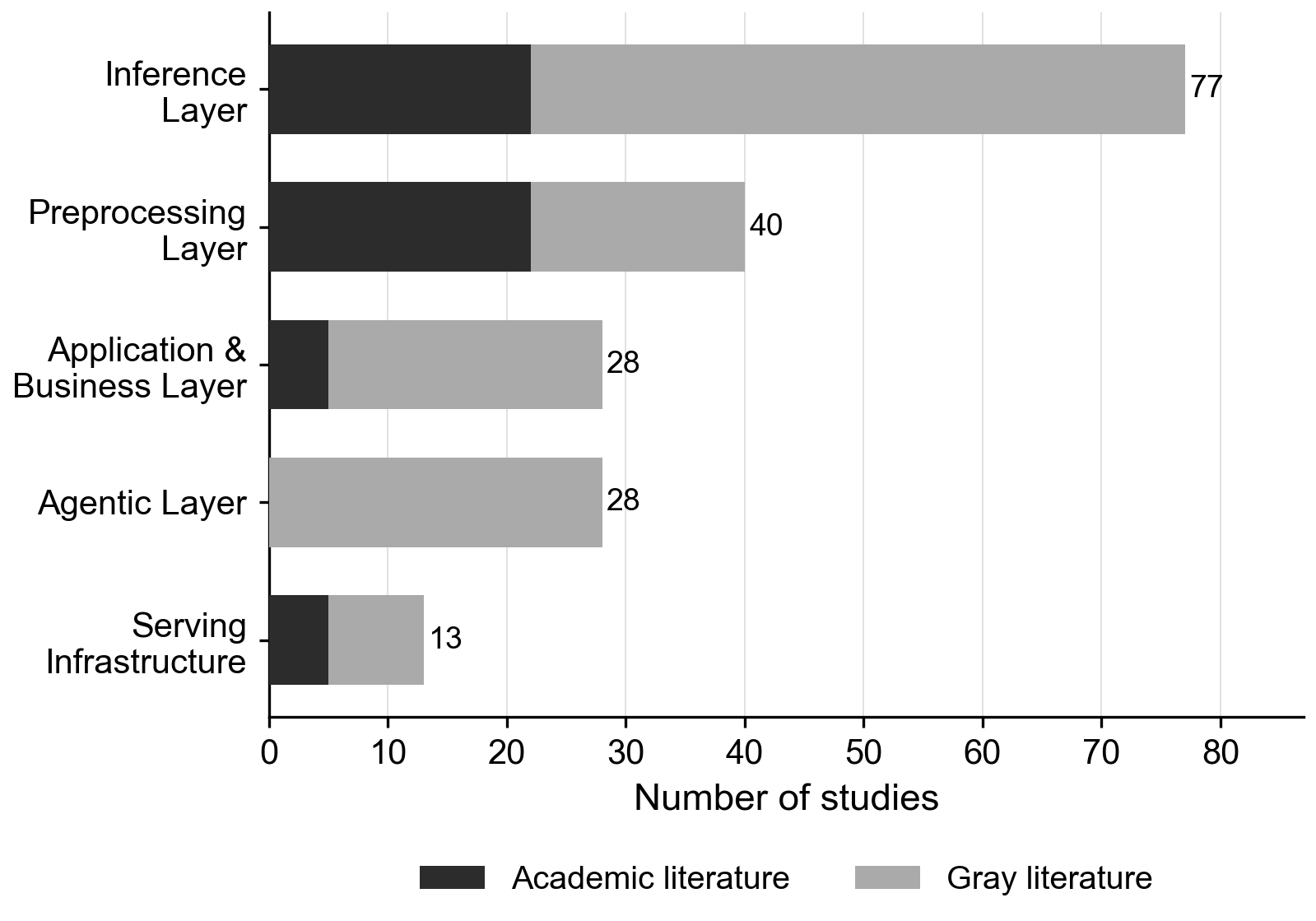}
        \caption{Monitored Objects}
        \label{fig:monitored_objects}
    \end{subfigure}
    \hfill
    \begin{subfigure}[t]{0.50\textwidth}
        \centering
        \includegraphics[width=\linewidth]{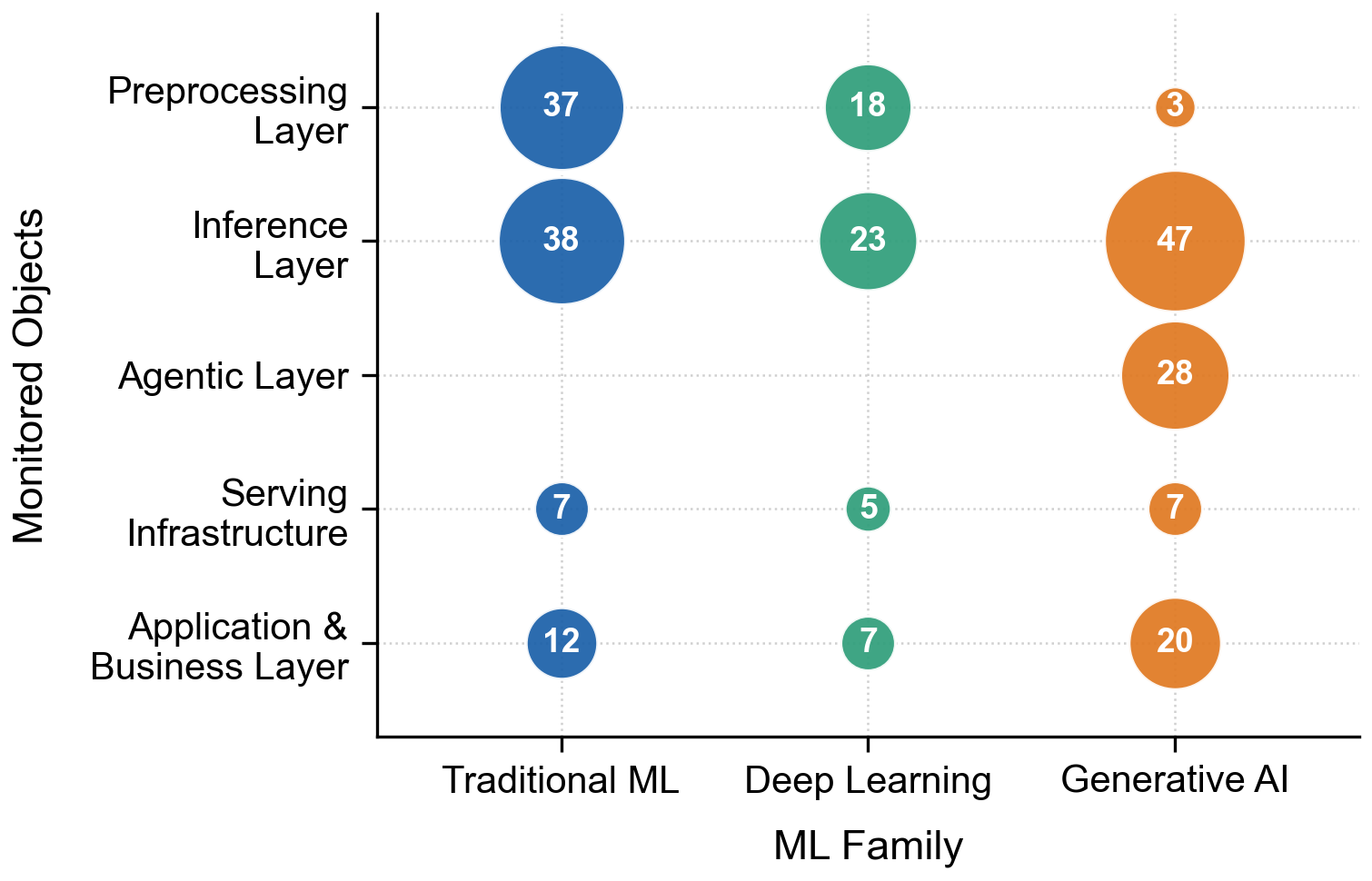}
        \caption{Monitored Objects across ML Families}
        \label{fig:objects_by_family}
    \end{subfigure}

    \caption{Monitored Objects of Production ML Systems}
    \label{fig:monitored_objects_overview}
\end{figure*}

\textbf{Monitored Objects. } Figure~\ref{fig:monitored_objects_overview} presents the monitored objects of the ML systems across literature types and ML families. The inference layer is the dominant focus of production ML monitoring \cite{mltestscore2017, superwise2026, timeretrain2025, constitutionalclassifiers2025, aitotal2021, witnessai2026, confidentnd}, followed by preprocessing \cite{advancedautomation2025, alertiger2023, automatinglarge2018, daqlmonitor2019, datasentinel2020, datavalidation2019}, application and business \cite{automatinglarge2018, deployingmachine2018, haveidea2024, loggingrequirement2025, measureimproved2025, successfulmachine2019}, and agentic layer \cite{langsmithnd, agentobservability2024, opikcomet2026, raindropnd, llmobservabilitynd}. Serving infrastructure \cite{coralogix2026, dynatracend, galileond, heliconend, honeycomb2026, langsmithnd} receives the least attention. Inference layer monitoring spans all three ML families \cite{evidentlynd, braintrust2026, monitoringfeature2021}, while preprocessing is concentrated primarily in traditional ML and deep learning systems \cite{datasentinel2020, tensorflowdata2020}. The agentic layer, by contrast, is exclusive to generative AI systems. Generative AI contributes substantially to agentic and application monitoring but provides comparatively limited coverage of serving infrastructure \cite{superwise2026, whylabsnd, arizeaind, heliconend}. Figure~\ref{fig:monitored_objects} also reveals a difference between the two bodies of literature. Academic studies are concentrated primarily on preprocessing and inference layers, reflecting an emphasis on data and model behavior \cite{alertiger2023, daqlmonitor2019, domainspecific2023}. Gray literature covers the entire agentic layer and provides most of the coverage for application and business, and serving infrastructure layers, reflecting its greater attention on business contexts and agentic systems \cite{raindropnd, sentry2026, splunknd, superwise2026}.

\begin{figure}[!htbp]
    \centering
    \includegraphics[width=0.9\linewidth]{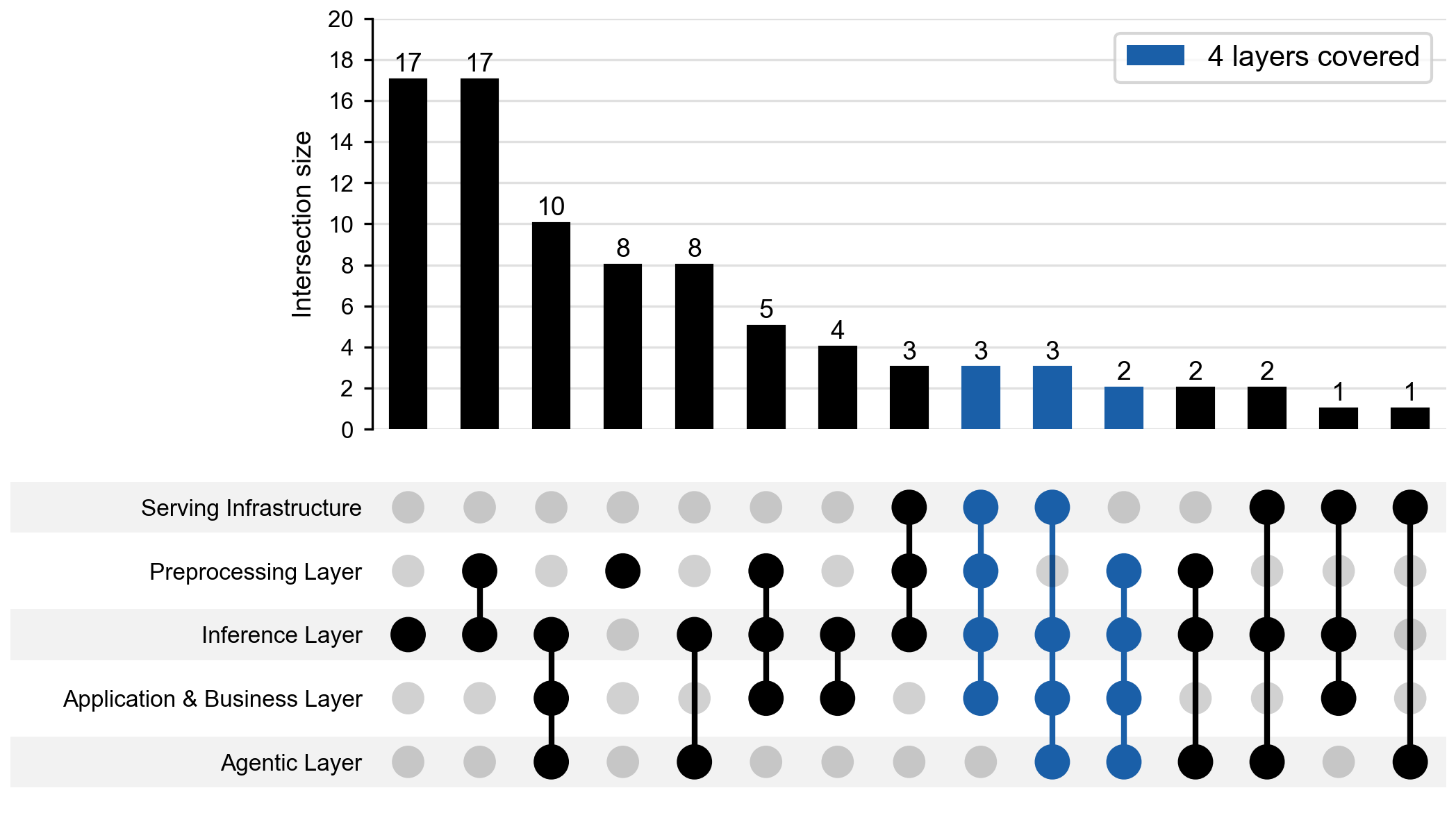}
    \caption{Monitored Objects Co-occurrence}
    \label{fig:MO_scope}
\end{figure}

Despite this breadth, none of the included sources monitor the complete production system across all five layers. Figure~\ref{fig:MO_scope} shows that monitoring scope varies considerably but is generally concentrated on a limited subset. Single-layer and two-layer monitoring configurations are common, with the inference layer frequently monitored either alone or together with preprocessing. Coverage becomes progressively less common as monitoring expands across additional layers, particularly when the agentic, serving infrastructure, and application and business layers are considered together. Most notably, only eight sources cover four layers, which is the highest observed coverage \cite{arizeaind, coralogix2026, superwise2026, newrelic2026, dynatracend, scalingml2026, monitoringobservability2025, measureimproved2025}. This indicates that production ML monitoring is generally organized around selected stages or concerns rather than providing end-to-end visibility across preprocessing, inference, agentic workflows, serving infrastructure, and downstream application and business outcomes \cite{understandingpractitioners2025, measureimproved2025, monitoringobservability2025}.

\begin{table}[!htbp]
\centering
\small
\caption{Mapping of Monitored Objects to Monitored Properties}
\label{tab:mo-mp-mapping}
\begin{tabular}{@{}L{0.25\textwidth} L{0.70\textwidth}@{}}
\toprule
\textbf{Monitored Object} & \textbf{Monitored Properties} \\
\midrule
Preprocessing Layer & Data Quality \& Integrity \textperiodcentered\ Feature Drift \\
Inference Layer & Concept Drift \textperiodcentered\ Predictive Performance \textperiodcentered\ Prediction Drift \textperiodcentered\ Feature Attribution Drift \textperiodcentered\ Fairness \& Bias \textperiodcentered\ LLM Generation Quality \textperiodcentered\ RAG Retrieval Quality \textperiodcentered\ Safety \textperiodcentered\ Reasoning \& Alignment \textperiodcentered\ Operational Performance\\
Agentic Layer & Agent Task Execution \textperiodcentered\ Operational Performance\\
Serving Infrastructure & Operational Performance \\
Application \& Business Layer & Business \& Application Impact \textperiodcentered\ Compliance \& Governance \textperiodcentered\ Operational Performance\\
\bottomrule
\end{tabular}
\end{table}

\begin{figure*}[!htbp]
    \centering

    \begin{subfigure}[t]{0.50\textwidth}
        \centering
        \includegraphics[width=\linewidth]{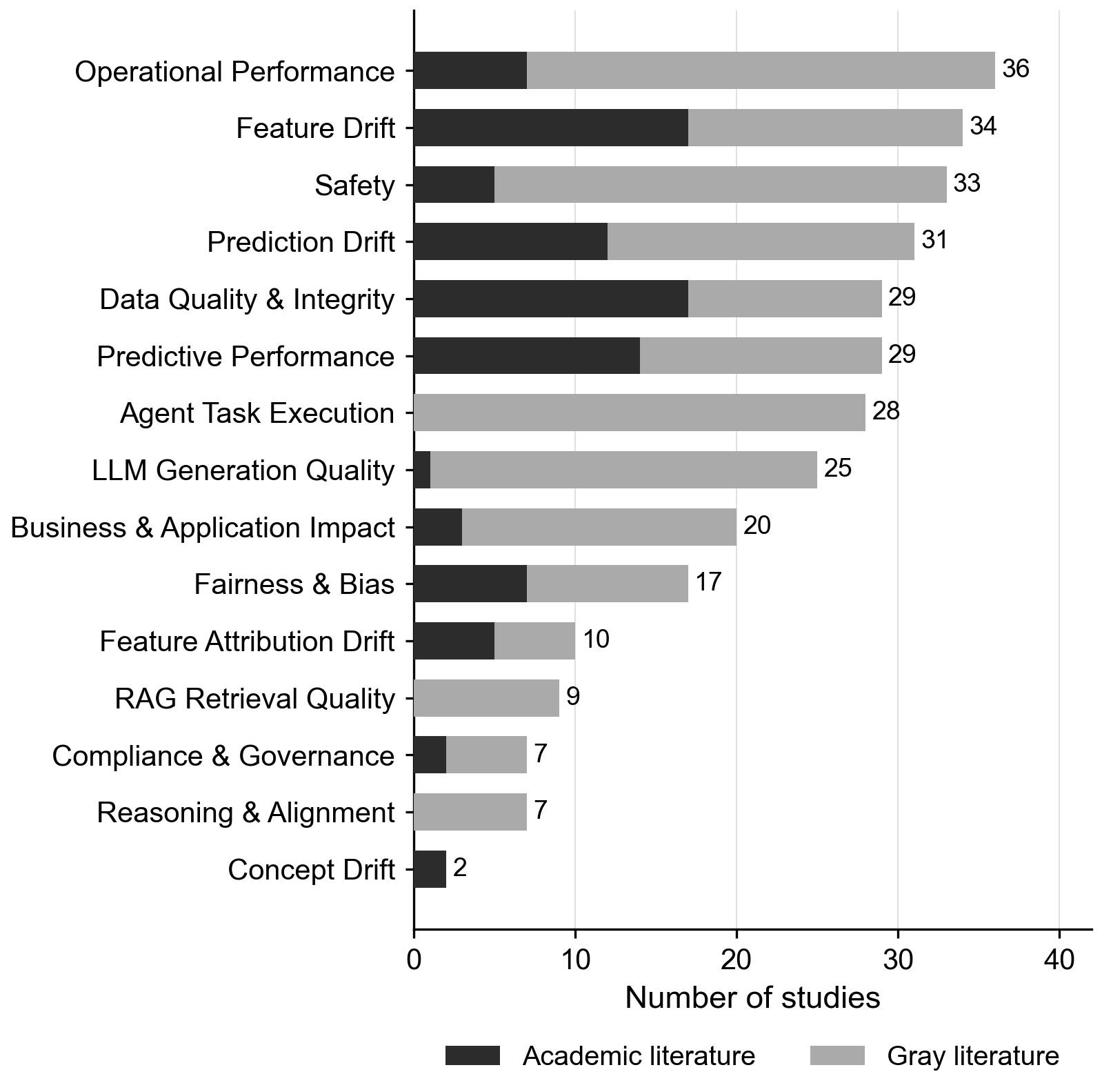}
        \caption{Monitored Properties}
        \label{fig:monitored_properties}
    \end{subfigure}
    \hfill
    \begin{subfigure}[t]{0.49\textwidth}
        \centering
        \includegraphics[width=\linewidth]{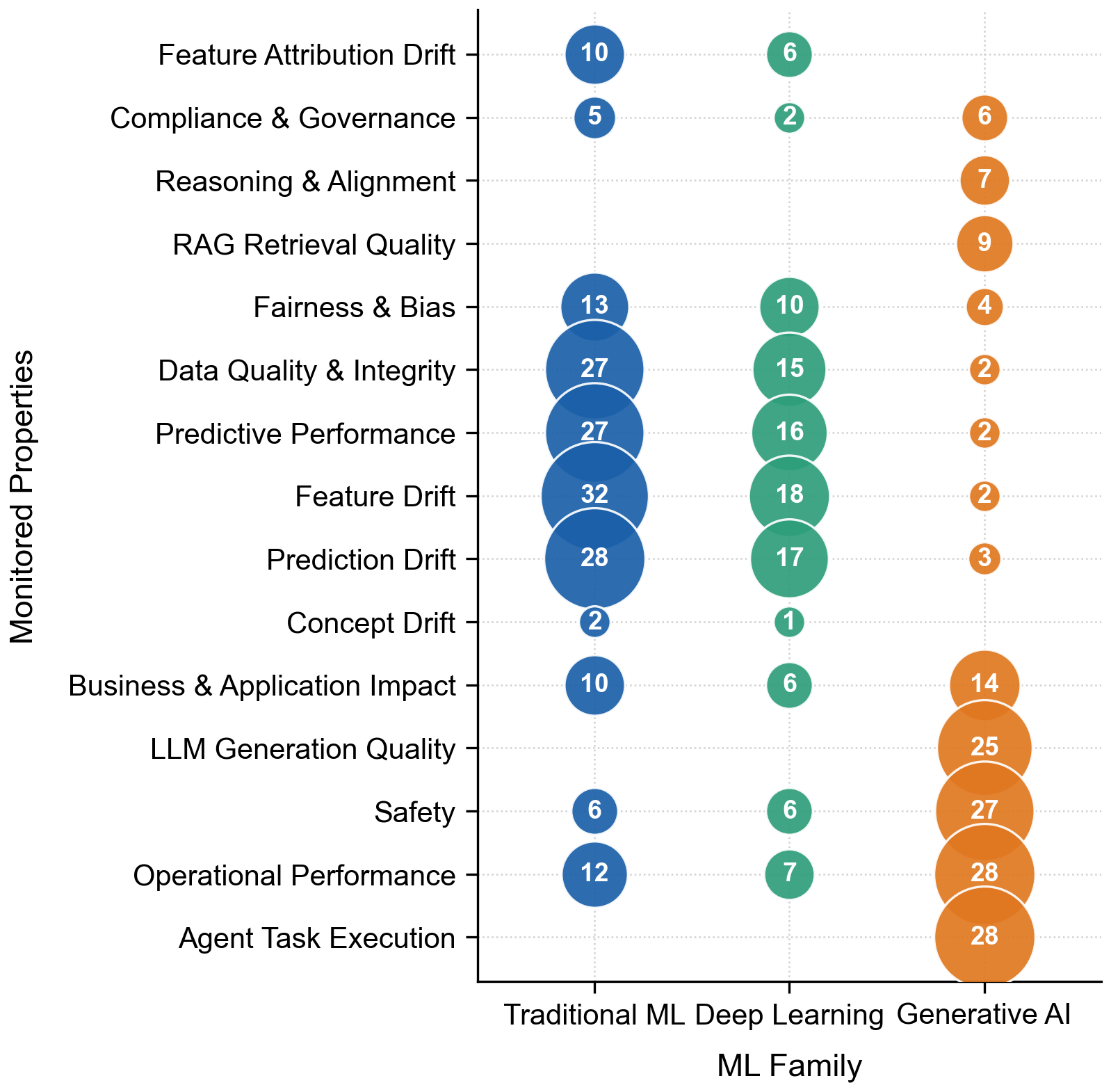}
        \caption{Monitored Properties across ML Families}
        \label{fig:properties_by_family}
    \end{subfigure}

    \caption{Monitored Properties of Production ML Systems}
    \label{fig:monitored_properties_overview}
\end{figure*}

\textbf{Monitored Properties. } Table~\ref{tab:mo-mp-mapping} organizes the monitored properties according to the object whose behavior they assess. The preprocessing layer is assessed through data quality and integrity and feature drift, both of which concern the condition and distribution of data before inference. The inference pipeline has the broadest monitoring scope, encompassing concept drift, predictive behavior, fairness, generation and retrieval quality, safety, and reasoning and alignment. The remaining properties concern behavior beyond individual model inference, such as agent task execution at the agentic layer and business impact and compliance within the downstream application and organizational context. Operational performance cuts across the inference, agentic, serving infrastructure, and application layers because latency, errors, throughput, and cost can be measured at any of these locations.

Figure~\ref{fig:monitored_properties_overview} shows the monitored properties of the ML systems across literature types and ML families. Operational performance \cite{coralogix2026, mltestscore2017, arizeaind}, feature drift \cite{domainspecific2023, alertiger2023, tensorflowdata2020}, safety \cite{llamafirewall2025, llamaguard2023, shieldgemma2024}, and prediction drift \cite{domainspecific2023, confidentnd, evidentlynd} are the most frequently monitored properties. These concerns reflect two major aspects of production ML monitoring: maintaining the operational health of the deployed system and identifying changes in production inputs, model outputs, or system behavior. Data quality and integrity \cite{tensorflowdata2020, alertiger2023, mltestscore2017} and predictive performance \cite{healthassurance2022, aitotal2021, nannyml2025} also receive substantial attention. Beyond these dominant concerns, business and application impact \cite{fiddlernd, galileond, monitoringobservability2025}, fairness and bias \cite{amazonsagemaker2021, dynatracend, censiusnd}, and feature-attribution drift \cite{vertexai2026, monitoringfeature2021, healthassurance2022} receive comparatively less attention. Coverage narrows further for RAG retrieval quality \cite{arizephoenixnd, confidentnd, dynatracend}, compliance and governance \cite{oneshield2025, superwise2026, loggingrequirement2025}, and reasoning and alignment~\cite{chainwhen2025, newrelic2026, monitoringreasoning2025}. Concept drift is the least frequently monitored property, appearing in only two sources  \cite{machinelearningdrift2019, timeretrain2025}.

The family-level distribution in Figure~\ref{fig:properties_by_family} reveals a clear difference in monitoring priorities. Traditional ML and deep learning sources primarily monitor production data and predictive behavior, including data quality, feature and prediction drift, predictive performance, fairness, and feature attribution. Generative AI monitoring leans towards generated content quality, execution trajectories, safety, retrieval quality, and operational performance. This difference is also reflected in the source types. Academic literature primarily addresses established data and model-centric properties, including data quality, drift, predictive performance, fairness, and feature attribution \cite{healthassurance2022, aitotal2021, daqlmonitor2019}. In contrast, gray literature provides all of the coverage of agent task execution, RAG retrieval quality, and reasoning and alignment, and most of the coverage of safety and LLM generation quality \cite{ evidentlynd, galileond, arizephoenixnd, witnessai2026}.

\subsubsection{RQ2.2. What are the tools, techniques, and input signals used for monitoring production ML systems?}

\begin{table}[!htbp]
\centering
\caption{Monitoring Tools for Production ML Systems}
\label{tab:tools}
\footnotesize
\begin{tabular}{
  l
  >{\raggedright\arraybackslash}p{3cm}
  >{\centering\arraybackslash}p{2.2cm}
  *{5}{c}
}
\toprule
\textbf{Category} & \textbf{Tool} & \textbf{ML Family} &
{\textbf{Preprocessing}} &
{\textbf{Inference}} &
{\textbf{Agentic}} &
{\textbf{Serving}} &
{\textbf{App.\ \& Business}} \\
\midrule

\multirow{23}{*}{OSS}
 & AlerTiger\cite{alertiger2023}                 & TML             & \checkmark & \checkmark &            & &            \\
 & Alibi Detect \cite{seldon2022}              & TML / DL        & \checkmark & \checkmark &            &            &            \\
 & Arize Phoenix \cite{arizephoenixnd}             & GenAI           &            & \checkmark & \checkmark &  &            \\
 & \mbox{Constitutional Classifiers \cite{constitutionalclassifiers2025}} & GenAI           &            & \checkmark &            &            &            \\
 & DaQL\cite{daqlmonitor2019}                       & TML             & \checkmark &            &            &            &            \\
 & Deepchecks  \cite{deepchecks2024}       & TML / DL        & \checkmark & \checkmark &            &            &            \\
 & Evidently AI \cite{evidentlynd}              & TML / GenAI     & \checkmark & \checkmark & \checkmark &            &            \\
 & Guardrails AI  \cite{guardrailsaind}            & GenAI           &            & \checkmark &            &            & \\
 & Helicone \cite{heliconend}                  & GenAI           &            &  \checkmark & \checkmark &  & \checkmark \\
 & LangChain \cite{llmtoolsnd}                & GenAI           &            & \checkmark & \checkmark &            & \checkmark \\
 & Langfuse \cite{agentobservability2024}                  & GenAI           &            & \checkmark & \checkmark &  &            \\
 & Llama Guard  \cite{llamaguard2023}              & GenAI           &            & \checkmark &            &            &            \\
 & LlamaFirewall  \cite{llamafirewall2025}            & GenAI           &            & \checkmark &            &            &  \\
 & Lunary  \cite{lunarnd}                   & GenAI           &            &  \checkmark          & \checkmark &  &            \\
 & \mbox{MLflow LLM Observability  \cite{mlflowllm2025}} & GenAI           &            &   \checkmark  & \checkmark &  &            \\
 & Mastra Observability \cite{mastra}      & GenAI           &            & \checkmark & \checkmark & &            \\
 & NannyML \cite{nannyml2025}                   & TML             & \checkmark & \checkmark &            &            &            \\
 & NeMo Guardrails  \cite{nemoguardrails2023}          & GenAI           &            & \checkmark &            &            &            \\
 & Opik by Comet \cite{opikcomet2026}             & GenAI           &            & \checkmark & \checkmark &  &  \\
 & ShieldGemma  \cite{shieldgemma2024}              & GenAI           &            & \checkmark &            &            &            \\
 & TFX Data Validation  \cite{baylor2017tfx}      & TML / DL        & \checkmark &            &            &            &            \\
 & Whylabs   \cite{whylabsnd}              & TML / GenAI     & \checkmark & \checkmark &  &  & \checkmark \\
\midrule

\multirow{5}{*}{Cloud}
 & \mbox{Azure ML Model Monitoring~\cite{azurend}} & TML / GenAI     & \checkmark & \checkmark &            &            &            \\
 & Google Vertex AI  \cite{vertexai2026}         & TML             & \checkmark & \checkmark &            &            &            \\
 & SageMaker Clarify \cite{amazonsagemaker2021}          & TML / DL        &            & \checkmark &            &            &            \\
 & \mbox{SageMaker Model Monitor \cite{sagemakermodel2022}}   & TML / DL        & \checkmark & \checkmark &            &  &            \\

\midrule

\multirow{23}{*}{SaaS}
 & Arize AI \cite{arizeaind}                      & TML / DL / GenAI & \checkmark & \checkmark & \checkmark &  & \checkmark \\
 & Braintrust \cite{braintrust2026}                    & GenAI            &            & \checkmark & \checkmark &  & \checkmark \\
 & Censius AI Observability  \cite{censiusnd}     & TML / DL         & \checkmark & \checkmark &            &            & \checkmark \\
 & Confident AI \cite{confidentnd}                  & GenAI            &            & \checkmark & \checkmark &  & \checkmark \\
 & Coralogix   \cite{coralogix2026}                   & GenAI            &            &  \checkmark          & \checkmark & \checkmark & \checkmark \\
 & Data Sentinel  \cite{datasentinel2020}                & TML              & \checkmark &            &            &            &            \\
 & Datadog AI Observability  \cite{llmobservability2nd}     & GenAI            &            & \checkmark & \checkmark & \checkmark &            \\
 & Dynatrace \cite{dynatracend}                     & GenAI            &            & \checkmark & \checkmark & \checkmark & \checkmark \\
 & Elastic LLM Observability  \cite{monitorstack2026}    & GenAI            &            & \checkmark & \checkmark &  &  \checkmark          \\
 & Fiddler \cite{fiddlernd}                       & TML / DL / GenAI & \checkmark & \checkmark &            &            & \checkmark \\
 & Galileo  \cite{galileond}                      & GenAI            &            & \checkmark & \checkmark &  & \checkmark \\
 & \mbox{Honeycomb AI Observability \cite{honeycomb2026}}    & GenAI            &            &  \checkmark  & \checkmark &  & \checkmark \\
 & LangSmith  \cite{langsmithnd}                    & GenAI            &            & \checkmark & \checkmark &  &            \\
 & Monte Carlo  \cite{mlcompleteGuide}                  & GenAI            & \checkmark & \checkmark & \checkmark &  &            \\
 & \mbox{New Relic AI Observability \cite{newrelic2026}}    & TML / GenAI      &            & \checkmark & \checkmark & \checkmark & \checkmark \\
 & \mbox{Observe LLM Observability \cite{observellm2025}}     & GenAI            &            &            & \checkmark & \checkmark &            \\
 & OneShield  \cite{oneshield2025}                    & GenAI            &            & \checkmark &            &            & \checkmark \\
 & \mbox{Radicalbit AI Monitoring~\cite{radicalbit2024}}     & TML / DL / GenAI & \checkmark & \checkmark &            &            &            \\
  & Raindrop    \cite{raindropnd}              & GenAI       &            & \checkmark & \checkmark &            & \checkmark \\
 & Sentry AI Observability   \cite{sentry2026}     & GenAI            &            &  \checkmark  & \checkmark &  & \checkmark           \\
 & \mbox{Splunk Observability Cloud  \cite{splunknd}}   & GenAI            &            & \checkmark & \checkmark & \checkmark &            \\
 & Superwise \cite{superwise2026}                 & TML / GenAI      & \checkmark & \checkmark & \checkmark &  & \checkmark \\
 & \mbox{TrueFoundry LLM Observability \cite{observabilityllm2026}}  & GenAI            &            & \checkmark & \checkmark &  & \checkmark \\
 & WitnessAI Observe \cite{witnessai2026}             & GenAI            &            & \checkmark &            &            & \checkmark \\
 
\midrule

\multirow{5}{*}{In-house}
 & AI Total \cite{aitotal2021}                 & TML         &            & \checkmark &            &            &            \\
 & \mbox{LinkedIn Health Assurance\cite{healthassurance2022}} & TML         & \checkmark & \checkmark &            &            &            \\
 & Michelangelo \cite{michelangelo2017}             & TML / DL    &            & \checkmark &            & &            \\
 & Overton \cite{overton2019}                  & DL          &            & \checkmark &            &            &            \\
\midrule

\multirow{6}{*}{Research}
 & CDSeer \cite{timeretrain2025}                      & TML / DL  &            & \checkmark &            &            &            \\
 & \mbox{CoT Monitorability Autorater\cite{pragmaticway2025}} & GenAI     &            & \checkmark &            &            &            \\
 & ML-On-Rails \cite{ensuringrobustness2025}                 & TML / DL  & \checkmark & \checkmark &            &            &            \\
 & Panoptes  \cite{domainspecific2023}                   & TML / DL  & \checkmark & \checkmark &            &            &            \\
 & TAMAS   \cite{beyondblackbox2025}                     & GenAI     &            & \checkmark & \checkmark &  &            \\
 & Tyra \cite{deployingmachine2018}                        & TML       & \checkmark & \checkmark &            &            & \checkmark \\
\bottomrule
\end{tabular}
\vspace{4pt}
\noindent\footnotesize\textit{TML} = Traditional Machine Learning; \textit{DL} = Deep Learning; \textit{GenAI} = Generative AI.
\\\noindent\footnotesize\textit{OSS} = Open Source Software; \textit{SaaS} = Software as a Service.
\end{table}

\textbf{Monitoring Tools. } Table~\ref{tab:tools} shows the range of monitoring tools available in the literature and their scope. The table contains 60 tool entries: 22 open-source tools, four cloud-provider services, 24 SaaS platforms, four in-house platforms, and six research prototypes. Open-source tools and SaaS platforms constitute the largest groups, indicating that production ML monitoring is supported by both reusable software and commercially managed offerings. The landscape also includes tools developed for particular organizational settings, such as AI Total \cite{aitotal2021} and Michelangelo \cite{michelangelo2017}, and research prototypes addressing specific monitoring problems, such as CDSeer \cite{timeretrain2025} and Panoptes \cite{domainspecific2023}.

Monitoring scope varies substantially within these tool categories. Among open-source tools, for example, TensorFlow Data Validation \cite{tensorflowdata2020} focuses on the preprocessing layer, whereas Evidently spans preprocessing, inference, and agentic layer monitoring \cite{evidentlynd}. SaaS offerings are similarly varied; for instance, Observe \cite{observellm2025} concentrates on the agentic layer and serving infrastructure, while Braintrust covers inference, agentic, and application and business layers \cite{braintrust2026}. These differences show that neither tool category implies comprehensive coverage. Among the tools listed in Table 8, Arize AI \cite{arizeaind}, Coralogix \cite{coralogix2026}, Dynatrace \cite{dynatracend}, New Relic AI Observability \cite{newrelic2026}, and Superwise \cite{superwise2026} provide the broadest coverage, each spanning four layers. Academic literature contributes primarily towards research prototypes and organization-specific implementations \cite{aitotal2021, healthassurance2022, timeretrain2025, ensuringrobustness2025}, while gray literature provides much of the coverage of externally available platforms \cite{heliconend, newrelic2026, nemoguardrails2023, nannyml2025}. 

\begin{figure}[!htbp]
    \centering
    \includegraphics[width=0.9\linewidth]{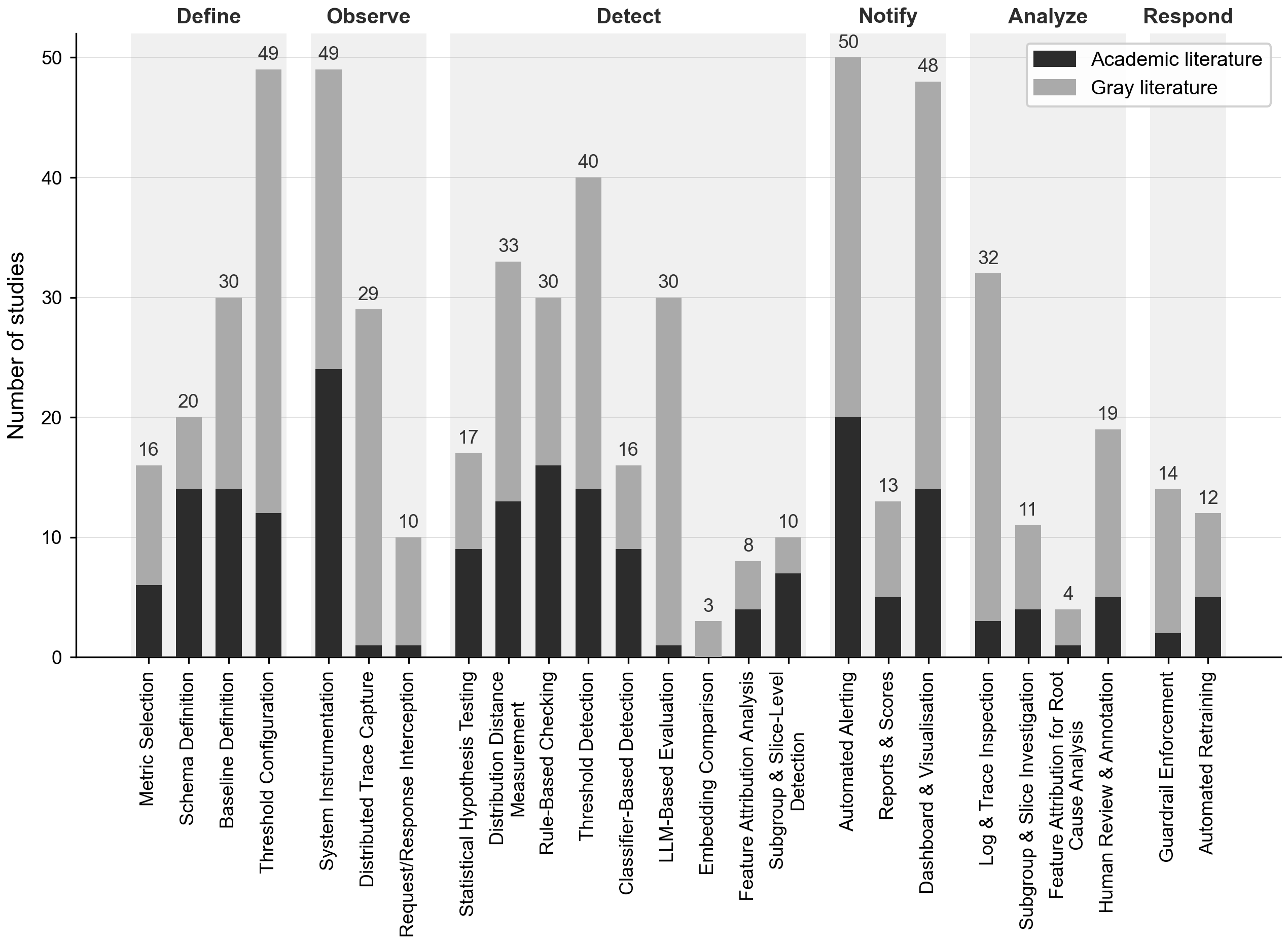}
    \caption{Monitoring Techniques}
    \label{fig:techniques}
\end{figure}

\textbf{Monitoring techniques. } Figure~\ref{fig:techniques} presents the techniques reported across the monitoring stages. Threshold configuration \cite{evidentlynd, daqlmonitor2019, tensorflowdata2020}, system instrumentation \cite{mltestscore2017, pipelinemonitoring2023, arizeaind}, automated alerting \cite{alertiger2023, sagemakermodel2022, splunknd}, and dashboard visualizations \cite{confidentnd, braintrust2026, healthassurance2022} receive the greatest coverage. Within detection, threshold-based checks \cite{dynatracend, fiddlernd, llmobservability2nd}, distribution distance measures \cite{domainspecific2023, tensorflowdata2020, michelangelo2017}, rule-based checking \cite{daqlmonitor2019, datasentinel2020, alertiger2023}, and LLM-based evaluation \cite{chainwhen2025, llamafirewall2025, llamaguard2023} are the most prominent techniques. Statistical testing \cite{seldon2022, modeldrificationweak2021, sagemakermodel2022} and trained classifiers \cite{aitotal2021, healthassurance2022,llamafirewall2025} are also represented, whereas embedding comparison \cite{arizeaind, evidentlynd}, feature attribution analysis \cite{vertexai2026, amazonsagemaker2021}, and subgroup-level detection \cite{aitotal2021, datasentinel2020} receive less attention. Beyond detection and notification stages, coverage becomes narrower. In the Analyze stage, log and trace inspection \cite{llmobservability2nd, braintrust2026, lunarnd} is relatively prominent, but subgroup investigation \cite{arizeaind, censiusnd, honeycomb2026} and feature-attribution-based root-cause analysis \cite{fiddlernd, whylabsnd, monitoringfeature2021} are much less common. In the Response stage, guardrail enforcement \cite{nemoguardrails2023, constitutionalclassifiers2025, llamafirewall2025} and automated retraining both receive less attention. 

The academic and gray literature comparison reveals both shared practices and distinct areas of emphasis. Statistical testing, instrumentation, rule-based checking, and alerting are represented in both bodies of literature \cite{advancedautomation2025, daqlmonitor2019, domainspecific2023}. Academic sources contribute particularly to schema definition and subgroup-level detection, while gray sources dominate distributed tracing, LLM-based evaluation, and log and trace inspection \cite{arizephoenixnd, monitorstack2026, sentry2026}. Overall, the literature provides more extensive coverage of identifying and surfacing problems than of explaining their causes and connecting findings to corrective action.

\begin{figure}[!htbp]
    \centering
    \includegraphics[width=0.9\linewidth]{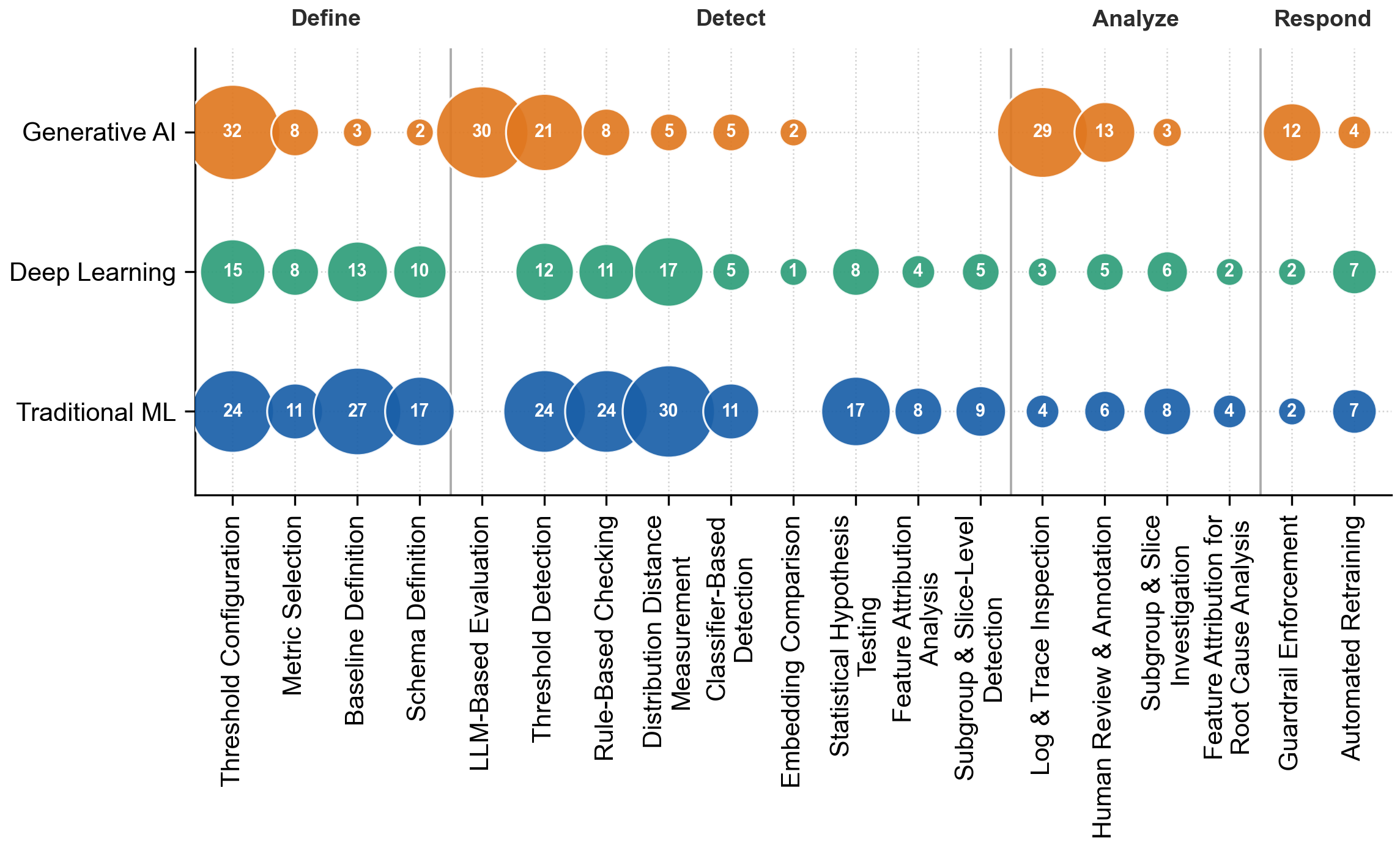}
    \caption{Monitoring Techniques across ML Families for Production ML Systems}
    \label{fig:detect_technique_family}
\end{figure}

The differences are also evident across ML families (Figure~\ref{fig:detect_technique_family}). At the definition stage, traditional ML and deep learning sources emphasize reference baselines and schemas \cite{amazonsagemaker2021, mltestscore2017, alertiger2023}, while threshold configuration is prominent across all three families \cite{evidentlynd, arizeaind, domainspecific2023, arizephoenixnd}. Detection in traditional ML and deep learning focuses mostly on distribution comparisons, statistical testing, and rule-based checks \cite{frameworkmonitor2019, successfulmachine2019, sagemakermodel2022}, whereas generative AI sources emphasize LLM-based evaluation \cite{chainwhen2025, mlcompleteGuide}. Threshold-based detection remains common across families, indicating that generative AI monitoring combines conventional metric-based checks with newer evaluators \cite{whylabsnd, daqlmonitor2019, lunarnd}. At the analysis stage, generative AI sources place greater emphasis on log and trace inspection \cite{superwise2026, llmobservabilitynd, langsmithnd} and human review \cite{llmtoolsnd, oneshield2025}, while subgroup investigation \cite{aitotal2021, evidentlynd} and feature-attribution-based root-cause analysis \cite{measureimproved2025, whylabsnd} are more closely associated with traditional ML and deep learning. Reported responses also differ; guardrail enforcement is concentrated in generative AI \cite{nemoguardrails2023, llamafirewall2025}, whereas automated retraining is more prominent in traditional ML and deep learning systems \cite{ measureimproved2025, pipelinemonitoring2023, timeretrain2025}. These patterns show that differences between ML families extend beyond detection to how monitoring criteria are established, findings are investigated, and corrective actions are taken.

\begin{figure}[!htbp]
    \centering
    \includegraphics[width=\linewidth]{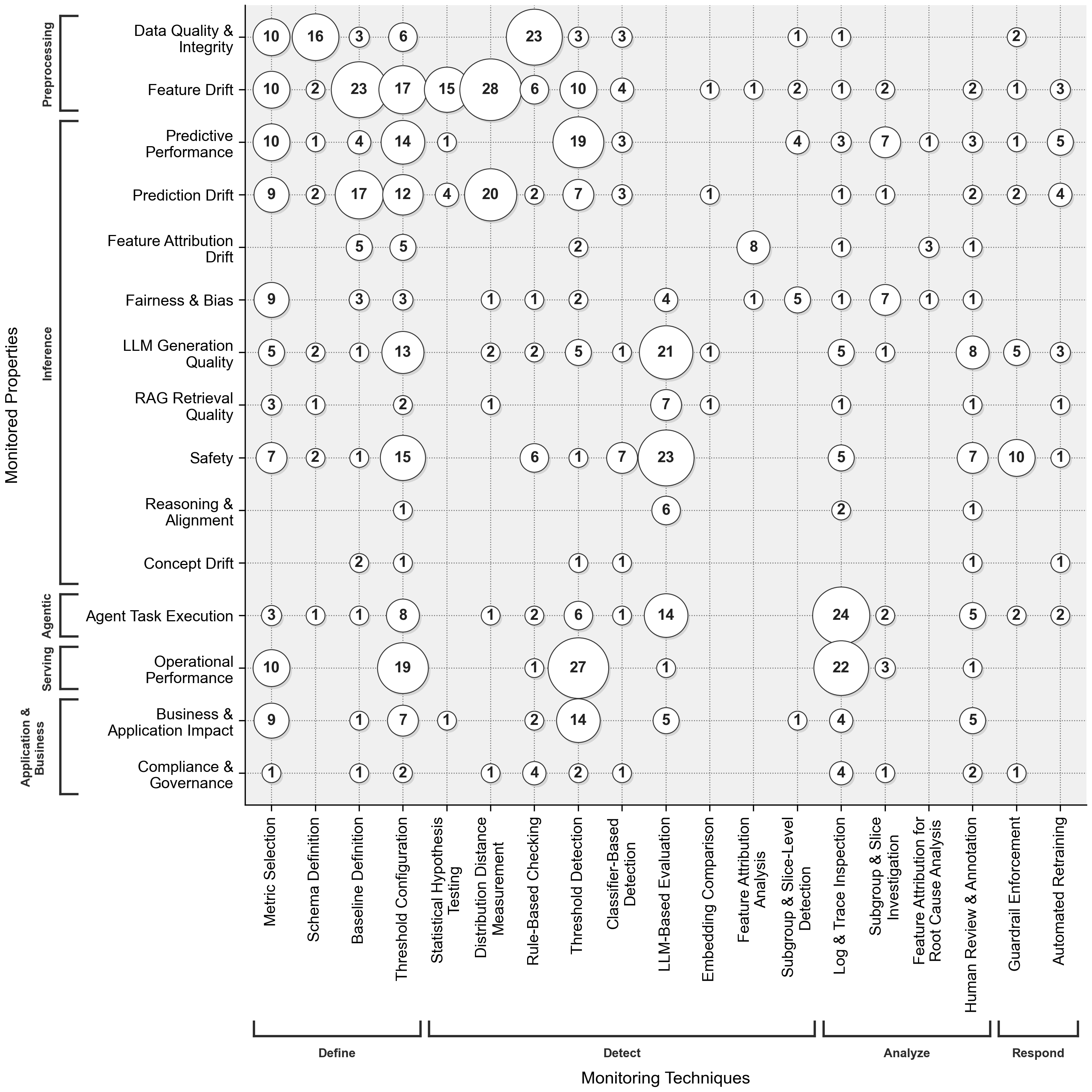} 
    \caption{Monitoring Techniques across Monitored Properties } 
    \label{fig:technique_MP}
\end{figure}

Figure~\ref{fig:technique_MP} shows that technique selection differs considerably by monitored property. Rule-based checking is concentrated on data quality and integrity monitoring \cite{daqlmonitor2019, datavalidation2019}, while distribution-distance measures mainly address feature and prediction drift \cite{evidentlynd, sagemakermodel2022}. Threshold-based detection has broader coverage, spanning predictive performance, operational performance, and business and application impact monitoring \cite{arizeaind, haveidea2024, mltestscore2017, galileond}. LLM-based evaluation is concentrated on generation quality, safety, retrieval quality, and reasoning and alignment \cite{chainwhen2025, nemoguardrails2023, arizeaind}. Safety monitoring, however, has a more varied set of techniques, including trained classifiers, rule-based checks, human review, and guardrail enforcement \cite{constitutionalclassifiers2025, aitotal2021, llamafirewall2025}. Compliance and governance instead use rule-based checks and log inspection \cite{oneshield2025, deployingmachine2018}. Overall, some properties are addressed through several complementary techniques, whereas others rely on a narrower range of monitoring methods.

\begin{figure}[!htbp]
    \centering
    \includegraphics[width=0.9\linewidth]{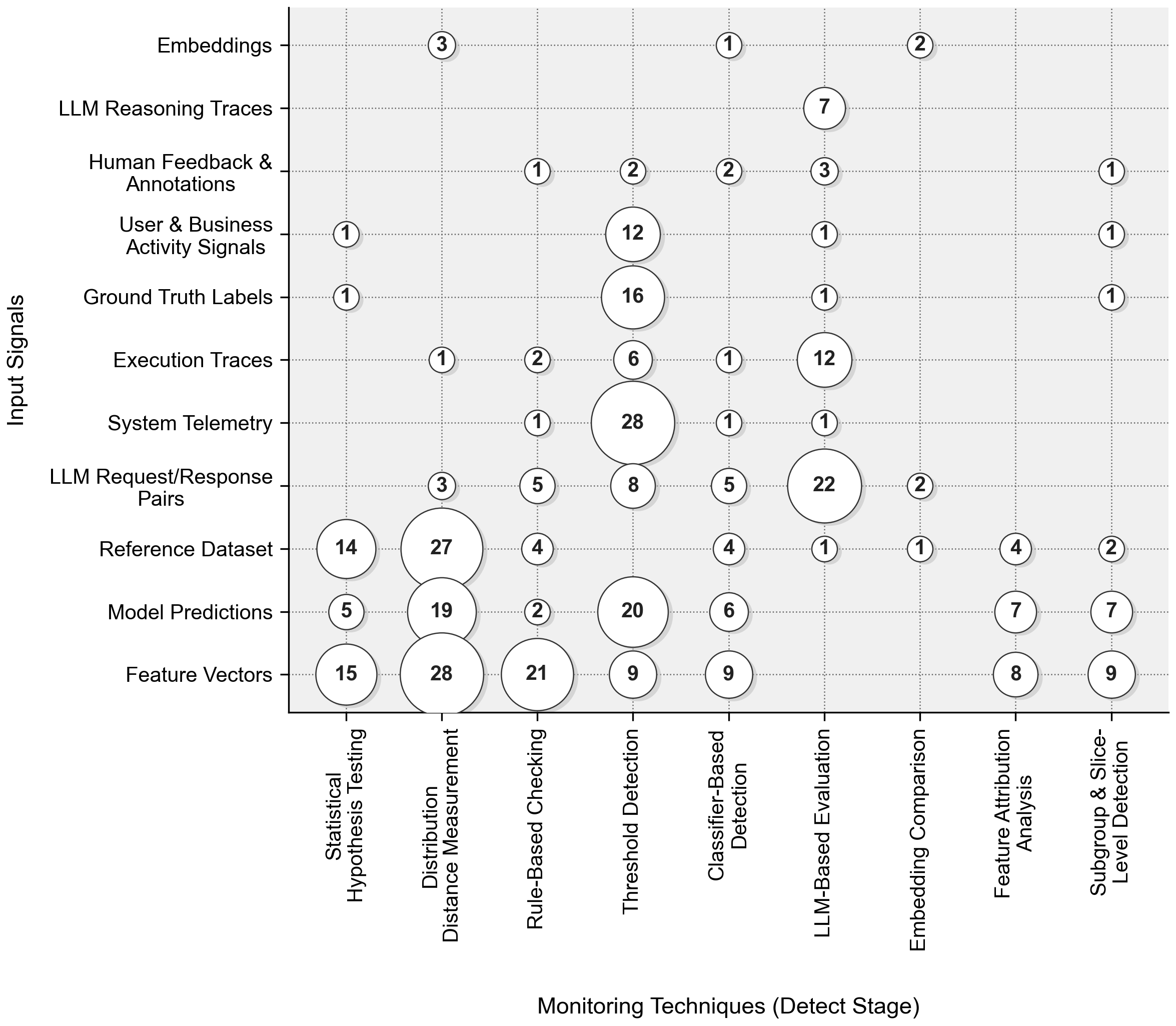}
    \caption{Monitoring Techniques across Input Signals}
    \label{fig:technique_IS}
\end{figure}

\textbf{Input signals.} Figure~\ref{fig:technique_IS} maps monitoring techniques from the detect stage to the data they use. Feature vectors \cite{sagemakermodel2022, alertiger2023, aitotal2021}, model predictions \cite{pipelinemonitoring2023, arizeaind, whylabsnd}, and reference datasets \cite{amazonsagemaker2021, mltestscore2017, seldon2022} are the input signals that support the widest range of techniques, particularly statistical testing and distribution comparisons, as well as rule-based, classifier-based, feature attribution-based, and subgroup-level detection. Ground-truth labels \cite{deployingmachine2018, azurend, newrelic2026} are more narrowly associated with threshold-based checks computed on predefined metrics. System telemetry \cite{langsmithnd, llmobservability2nd, mlcompleteGuide} and user and business activity \cite{haveidea2024, fiddlernd, honeycomb2026} signals are also predominantly associated with threshold-based detection. Generative AI-specific signals follow a different pattern; request–response pairs \cite{radicalbit2024, nemoguardrails2023, lunarnd, confidentnd} and execution traces \cite{arizephoenixnd, heliconend, coralogix2026} are strongly used by LLM-based evaluation, although they also support other detection techniques. LLM reasoning traces \cite{chainmonitorability2025, llamafirewall2025, llmtoolsnd, chainwhen2025} are used exclusively with LLM-based evaluation in the included sources, extending detection beyond final outputs to intermediate steps. Human feedback and annotations \cite{healthassurance2022, pragmaticway2025, llamafirewall2025} receive comparatively limited coverage as input signals, while embeddings \cite{evidentlynd, seldon2022, measuringdistributional2023} only support distribution and semantic comparisons. Overall, threshold-based detection is the most diverse in terms of supported input signals, whereas embedding comparison is the most limited.

\subsubsection{RQ2.3. What are the automation levels and frequencies of monitoring techniques for production ML systems?}

\begin{figure}[!htbp]
    \centering
    \includegraphics[width=0.9\linewidth]{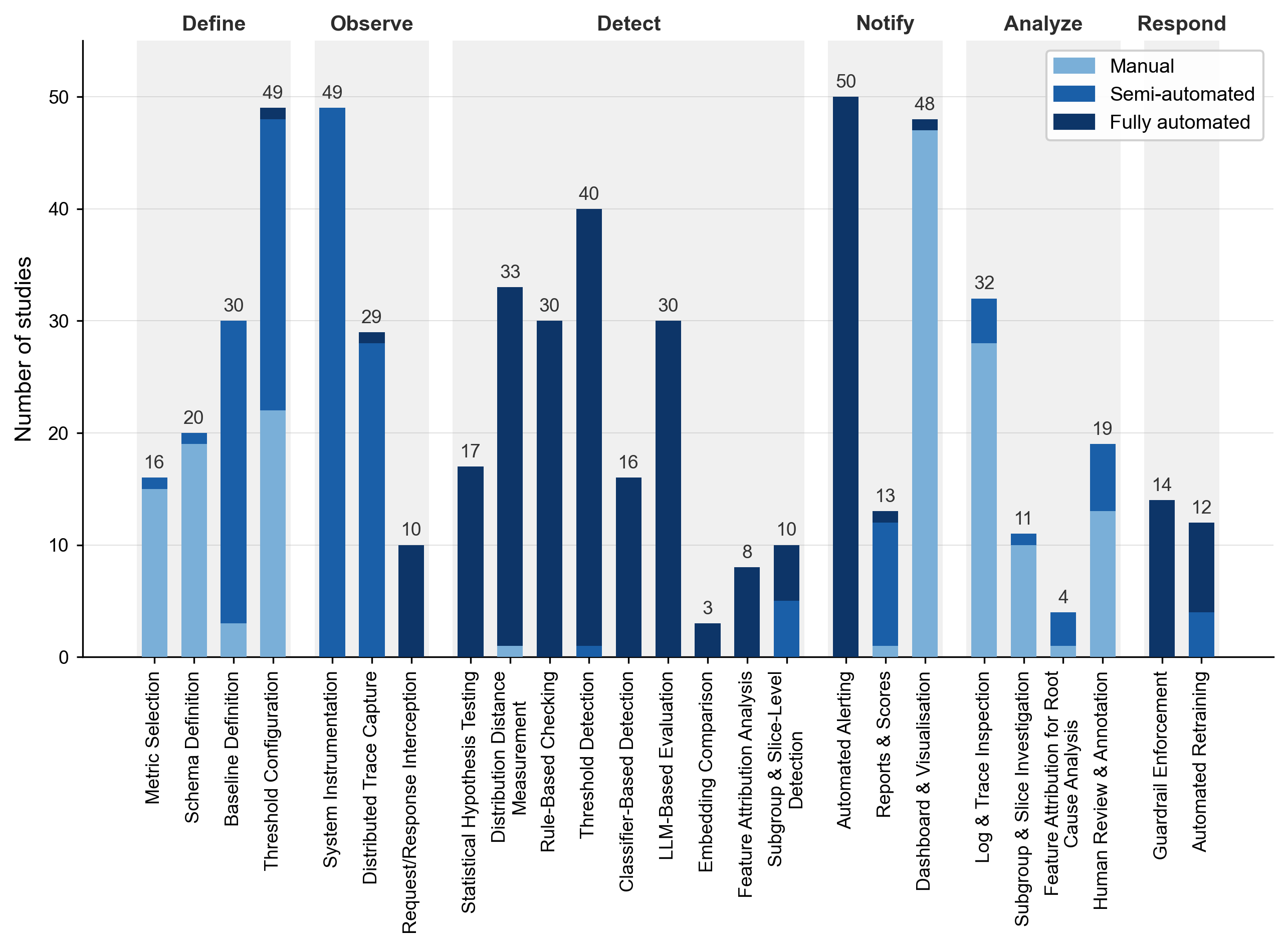}
    \caption{Automation Levels of Monitoring Techniques for Production ML Systems}
    \label{fig:technique_automation}
\end{figure}

\textbf{Automation levels.}  Figure~\ref{fig:technique_automation} shows the variation in automation across the six monitoring stages. At the Define stage, metric selection and schema definition are predominantly manual \cite{haveidea2024, successfulmachine2019, designprinciples2025, understandingpractitioners2025, datasentinel2020}, while baseline definition and threshold configuration often involve partial automation \cite{machinelearningdrift2019, daqlmonitor2019, measuringdistributional2023, scalingml2026}. The Observe stage is mainly semi-automated for instrumentation and distributed trace capture \cite{scalingml2026, michelangelo2017, arizephoenixnd}, whereas request–response interception is fully automated \cite{guardrailsaind, nemoguardrails2023, heliconend, witnessai2026}. At the Detect stage, most techniques execute automatically, including statistical tests, distribution comparisons, rule-based checks, and LLM-based evaluation \cite{domainspecific2023, deepchecks2024, seldon2022, whylabsnd}. At the Notify stage, alerting is fully automated, reports are semi-automated, and dashboard visualizations are mostly manual \cite{scalingml2026, deployingmachine2018, heliconend}. Human involvement remains prominent at the Analyze stage, where log and trace inspection, subgroup investigation, feature attribution for root cause analysis, and review are largely manual or semi-automated \cite{llmobservabilitynd, langsmithnd, haveidea2024}. At the Respond stage, guardrail enforcement is fully automated, while retraining includes both fully and semi-automated techniques \cite{nemoguardrails2023, intelligentmodel2024}.

\begin{figure*}[!htbp]
    \centering

    \begin{subfigure}[t]{0.38\textwidth}
        \centering
        \includegraphics[width=\linewidth]{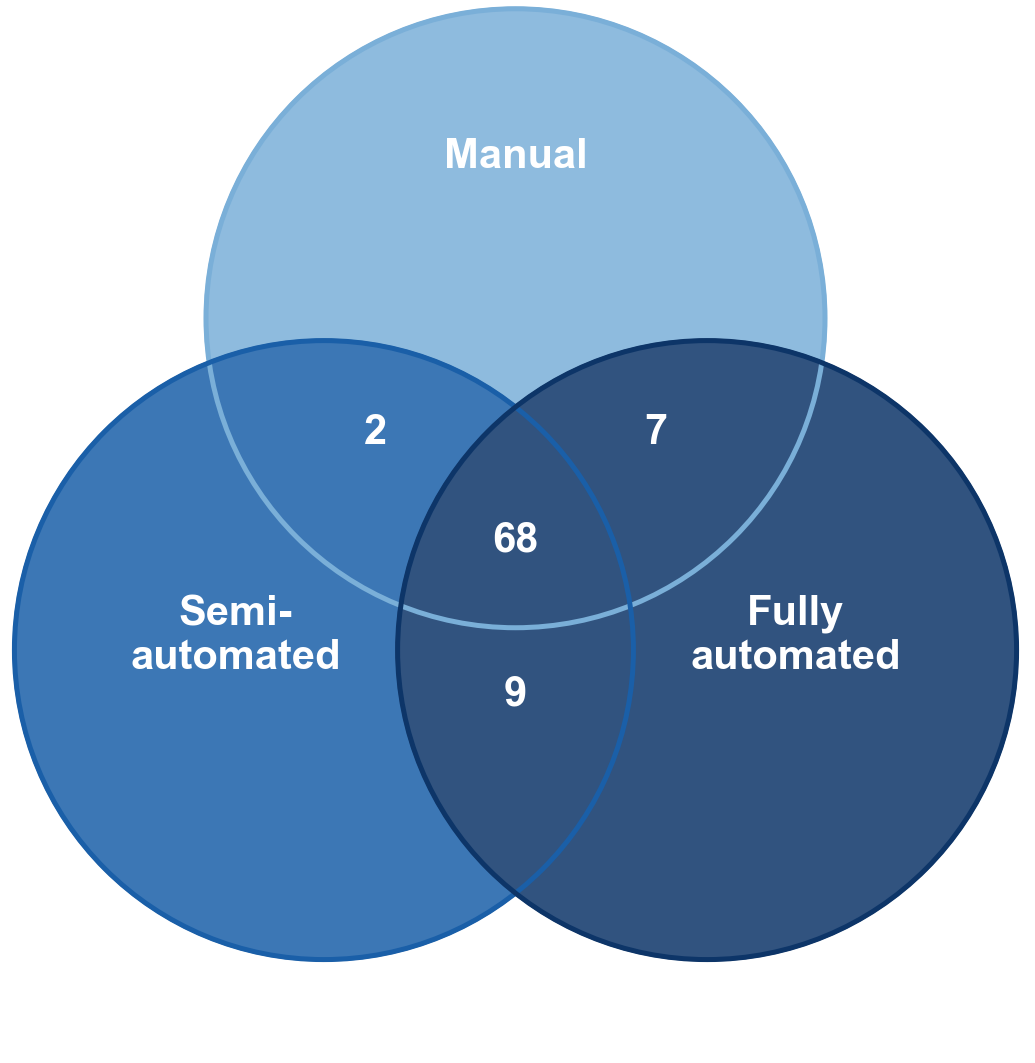}
        \caption{Automation Levels}
        \label{fig:automation_levels}
    \end{subfigure}
    \hfill
    \begin{subfigure}[t]{0.60\textwidth}
        \centering
        \includegraphics[width=\linewidth]{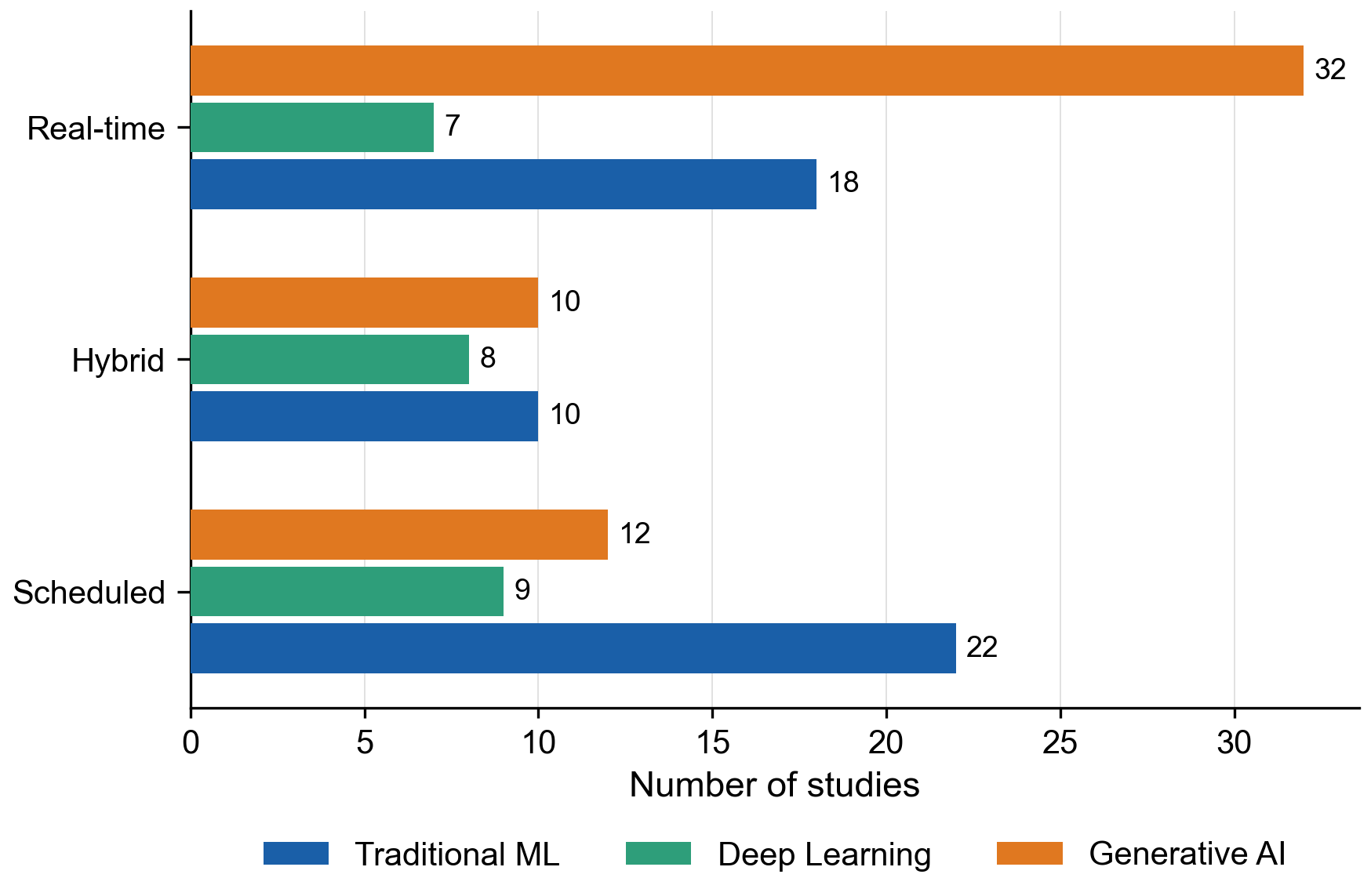}
        \caption{Monitoring Frequency across ML Families}
        \label{fig:frequency_family}
    \end{subfigure}

    \caption{Automation Levels and Monitoring Frequencies of Production ML Systems}
    \label{fig:deployment}
\end{figure*}

The Venn diagram in Figure~\ref{fig:automation_levels} highlights a pattern that most sources combine manual, semi-automated, and fully automated monitoring techniques. Production ML monitoring therefore involves a division of work between people and software; detection, alerting, and response techniques are mostly automated, while setting up monitors and analyzing faults still require substantial human involvement \cite{understandingpractitioners2025, haveidea2024, humancentricperspective2022}.

\textbf{Monitoring frequency. } Figure~\ref{fig:frequency_family} shows that scheduled, real-time, and hybrid monitoring occur across all three ML families, but with different emphasis. Traditional ML sources favor scheduled and real-time monitoring \cite{sagemakermodel2022, aitotal2021}, while deep learning sources are more evenly distributed across all three configurations \cite{datavalidation2019, mltestscore2017}. Generative AI sources report more real-time monitoring to capture ongoing interactions and enforce guardrails \cite{guardrailsaind, nemoguardrails2023, arizephoenixnd}. Hybrid configurations are represented across all families. 

\begin{rqanswer}{RQ2. How are production ML systems monitored?}
\textbf{RQ2 Answer. } Production ML systems are monitored through combinations of tools, input signals, and techniques applied across preprocessing, inference, agentic workflows, serving infrastructure, and application and business layers. Coverage is concentrated on inference, with no source monitoring all five system layers and only eight covering four layers. Open-source tools and SaaS platforms dominate the tooling landscape, complemented by cloud-provider services, in-house platforms, and research prototypes. 

Production ML systems are monitored through six stages: Define, Observe, Detect, Notify, Analyze, and Respond. These stages encompass selecting monitoring criteria and baselines, collecting runtime signals, detecting deviations, communicating findings, investigating causes, and taking corrective action. Traditional ML and deep learning approaches emphasize features, predictions, and reference datasets to assess data quality, drift, and predictive behavior. Generative AI approaches additionally evaluate LLM-generated content and execution traces using LLM-based methods, while also leveraging conventional metric-based checks. System telemetry and business activity signals mainly support threshold-based monitoring. Gray literature contributes much of the coverage of emerging generative and agentic monitoring practices. 

Automation is often stage-dependent; detection and alerting are largely automated, while configuration and investigation retain substantial human involvement. Scheduled monitoring is more prominent in traditional ML, real-time monitoring dominates generative AI, and hybrid configurations occur across all families. \end{rqanswer}

\subsection{RQ3. What evidence exists for the effectiveness of monitoring techniques and tools for production ML systems?}
\subsubsection{RQ3 Taxonomy}

\begin{table}[!htbp]
\centering
\begin{threeparttable}
\caption{RQ3 Taxonomy}
\label{tab:rq3-taxonomy}
\small
\begin{tabular}{@{}L{0.16\textwidth} L{0.24\textwidth} L{0.50\textwidth}@{}}
\toprule
\textbf{Taxonomy} & \textbf{Category} & \textbf{Description} \\
\midrule

\multirow{2}{=}{Evaluator Context}
 & Academia\tnote{$\dagger$} & The evaluation was conducted by academic researchers \\
 & Industry & The evaluation was conducted by industry practitioners \\
\midrule

\multirow{6}{=}{Evaluation Method}
 & Public dataset & Evaluated using a publicly available benchmark dataset \\
 & Proprietary dataset & Evaluated using an internal or non-public dataset \\
 & In production & Evaluated via live deployment in a real production system \\
 & User study & Evaluated by practitioners in a controlled lab setting \\
 & No evaluation, claims only & \mbox{Production effectiveness is asserted without a reported evaluation} \\
\midrule

\multirow{14}{=}{Evaluation Metric}
 & ML Performance & Evaluates the predictive accuracy or correctness of monitored ML models  \\
 & LLM Response Quality & Evaluates the quality, correctness, and usefulness of generated LLM responses and reasoning \\
 & Data Quality \& Distribution & Evaluates data integrity, statistical distribution changes, and drift-related characteristics \\
 & Safety, Fairness \& Security & Evaluates the ability to detect harmful, biased, adversarial, or non-compliant model behavior \\
 & Operational Performance & Evaluates the runtime efficiency, scalability, resource usage, and reliability of the monitoring approach \\
 & Business Impact & Evaluates improvements in organizational or business outcomes resulting from monitoring \\
 & Human Evaluation & Evaluates practitioners' perceptions of usability, usefulness, and effectiveness of the monitoring approach \\
 \midrule
\multirow{12}{=}{Evaluation Domain}
 & Research \& Development / Technology & General-purpose R\&D or technology-sector applications not tied to a specific industry \\
 & Multiple Domains & Evaluation spans more than one application domain \\
 & Finance \& Banking & Financial services, banking, or fintech applications \\
 & Manufacturing \& Industrial & Industrial process control or manufacturing applications \\
 & Telecommunications & Telecommunications network or service applications \\
 & Marketing \& Advertising & Marketing, advertising, or customer-targeting applications \\
 & Retail \& E-Commerce & Retail or online commerce applications \\
 & Healthcare & Clinical, medical, or healthcare applications \\
 & Automotive & Automotive or transportation applications \\
 & Cybersecurity & Security monitoring or threat-detection applications \\
 & Government \& Public Sector & Government, public administration, or civic applications \\
\midrule

\multirow{6}{=}{Evaluation Scale}
 & Dataset-based & Evaluated against a fixed dataset; no live user or practitioner count reported \\
 & Small Scale (1--9 users) & Evaluated with a small number of users or practitioners \\
 & Medium Scale (10--99 users) & Evaluated with a moderate number of users or practitioners \\
 & Large Scale (100--1{,}000 users) & Evaluated with a large number of users or practitioners \\
 & Huge Scale (1{,}000+ users) & Evaluated with a very large number of users or practitioners \\

\bottomrule
\end{tabular}
\begin{tablenotes}
\footnotesize
\item[$\dagger$] Academia evaluations in this corpus use proprietary or production data rather than public benchmarks, reflecting the review's scope: production ML monitoring evaluated with real practitioners or in real production contexts.
\end{tablenotes}
\end{threeparttable}
\end{table}

The RQ3 taxonomy characterises the evidence used to support claims that a production ML monitoring approach is effective. It separates \textbf{evaluation context}, \textbf{evaluation method}, \textbf{evaluation domain}, \textbf{evaluation scale}, and \textbf{evaluation metric}. These dimensions address different aspects of evidential strength and should not be collapsed into a single maturity label. A large benchmark experiment may provide reproducible technical evidence but little information about organizational use, whereas a small production case may provide strong ecological validity without supporting broad generalization. The taxonomy records these differences rather than imposing a universal ranking.

\textbf{Evaluator context} distinguishes academia from industry according to who conducted and reported the evaluation. Context is used to interpret the provenance and purpose of the evidence, not as a proxy for quality. Academic evaluations may emphasize controlled comparison and methodological transparency, while industry evaluations may expose real constraints involving scale, integration, privacy, delayed feedback, and incident response. The categories therefore remain separate from evaluation method: an academic team can use proprietary production data, and an industry team can evaluate a system on a public benchmark.

\textbf{Evaluation method} identifies how claims are tested. A public dataset provides an externally available and potentially reproducible benchmark; a proprietary dataset uses internal or non-public data that may better represent the deployment while limiting independent replication. In production denotes evaluation through a live operational deployment and is reserved for evidence collected from the running system rather than assertions that a method is production-ready. A lab user study evaluates practitioners under controlled conditions and is particularly suitable for usability, interpretability, and decision-support questions. No evaluation/claims only captures proposals or product descriptions that assert production effectiveness without reporting an empirical assessment. Explicitly retaining this last category prevents claimed deployment readiness from being silently interpreted as evaluated effectiveness.

\textbf{Evaluation metric} captures what effectiveness means: ML performance; LLM response quality; data quality and distribution; safety, fairness, and security; operational performance; business impact; or human evaluation. Multiple metrics may be assigned because a useful monitor can be accurate yet too costly, unusable, or disconnected from downstream outcomes. This multidimensional treatment follows the production-system principle that technical quality, operational reliability, and organizational usefulness are related but non-equivalent forms of evidence~\cite{sagemakermodel2022,alertiger2023,haveidea2024}.

\textbf{Evaluation domain} records the application context: research and development/technology, multiple domains, finance and banking, manufacturing and industrial settings, telecommunications, marketing and advertising, retail and e-commerce, healthcare, automotive, cybersecurity, and government and the public sector. Domain matters because acceptable error, label availability, regulation, data sensitivity, and remediation options vary substantially. Multiple domains are used only when an evaluation genuinely spans domains; general-purpose technical work is coded as research and development/technology rather than assigned an unsupported sector.

Finally, \textbf{Evaluation scale} distinguishes dataset-based assessment from evaluations involving small (1--9), medium (10--99), large (100--1,000), and huge (1,000+) populations of users or practitioners. Dataset-based is a separate category because record count is not comparable to the number of people participating in or affected by an evaluation. Scale describes reach, not methodological rigor.

\subsubsection{RQ3.1. How are monitoring approaches for production ML systems evaluated?}

\begin{table}[!htbp]
\centering
\caption{Evaluation Approaches of Included Sources}
\label{tab:eval_approach}
\small
\begin{tabular}{llccc}
\toprule
\textbf{Evaluator Context} & \textbf{Evaluation Method} & \textbf{Total} & \textbf{Academic Literature} & \textbf{Gray Literature} \\
\midrule
\multirow{4}{*}{Industry}
  & No evaluation — production claims only & 45 &  3 & 42 \\
  & Public dataset                         & 11 &  2 &  9 \\
  & In production                          &  6 &  5 &  1 \\
  & Proprietary dataset                    &  5 &  4 &  1 \\
\midrule
\multirow{3}{*}{Academia}
  & Proprietary dataset                    &  6 &  6 &  0 \\
  & In production                          &  3 &  3 &  0 \\
  & User study                             &  2 &  1 &  1 \\
\midrule
— & Not Applicable                           & 16 & 13 &  3 \\
\bottomrule
\end{tabular}
\end{table}
\begin{figure*}[!htbp]
    \centering

    \begin{subfigure}[t]{0.49\textwidth}
        \centering
        \includegraphics[width=\linewidth]{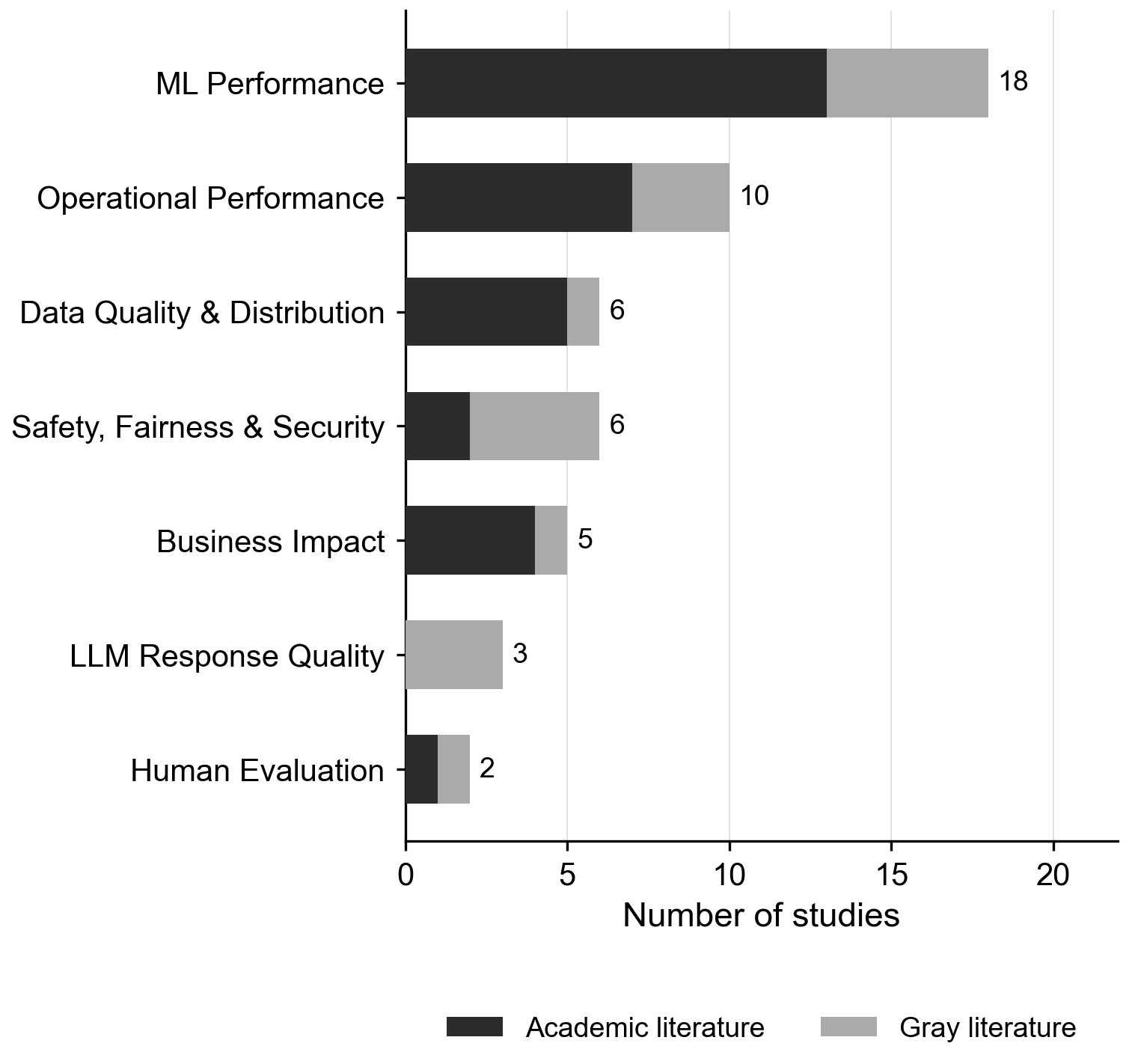}
        \caption{Evaluation Metric Categories}
        \label{fig:metric_categories}
    \end{subfigure}
    \hfill
    \begin{subfigure}[t]{0.49\textwidth}
        \centering
        \includegraphics[width=\linewidth]{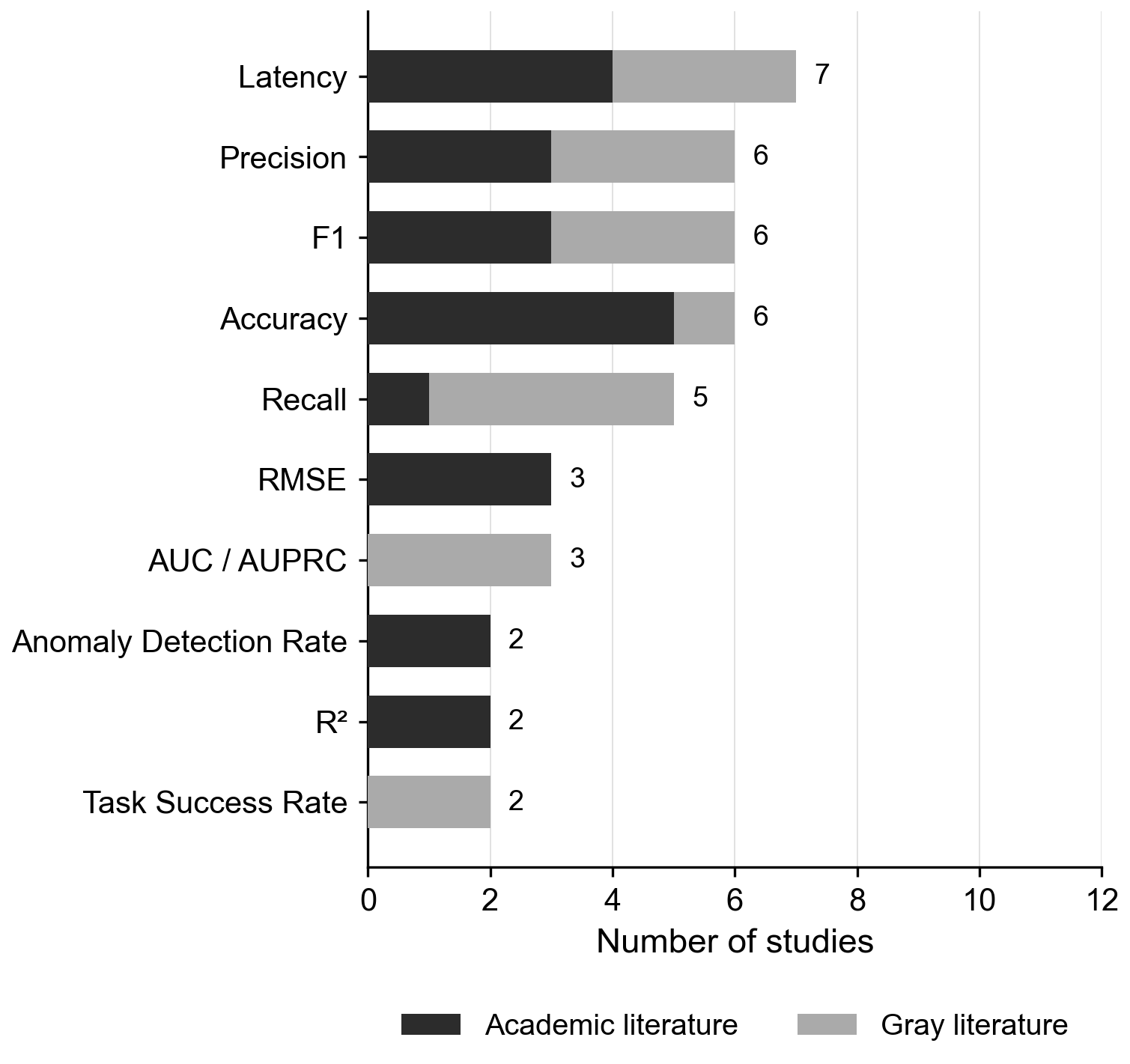}
        \caption{Top 10 Evaluation Metrics}
        \label{fig:top_10_metrics}
    \end{subfigure}

    \caption{Metrics for Evaluating Monitoring Approaches for Production ML Systems}
    \label{fig:metrics}
\end{figure*}

Table~\ref{tab:eval_approach} separates evaluator context from publication type. Evaluator context identifies the academic or industry setting of those reporting the approach or evaluation, including cases where no evaluation is supplied; academic and gray sources identify the publication channel. These axes are independent: an industry provider can publish an academic paper, and an academic provider can report in a preprint. The evaluations reported in the corpus remain limited. Production usage claims without empirical evaluation are the most common form of evidence, which comes mainly from monitoring tool websites that describe production capabilities and real-world usage without reporting a formal evaluation \cite{confidentnd, heliconend, splunknd, arizeaind, sentry2026}. Reported evaluations rely mainly on public or proprietary datasets, which test whether monitors detect known failures \cite{machinelearningdrift2019, timeretrain2025}, \cite{pipelinemonitoring2023, shieldgemma2024, oneshield2025}. Live production evaluations are less common \cite{datasentinel2020, constitutionalclassifiers2025, deployingmachine2018, alertiger2023}, and user studies are rare \cite{domainspecific2023, beyondblackbox2025}. 16 practitioner-oriented sources are classified as not applicable because they report monitoring practices and challenges rather than presenting an approach to be evaluated.

The evidence also differs between academic and gray literature. Academic publications provide most of the empirical and live production evaluations \cite{amazonsagemaker2021, sagemakermodel2022, deployingmachine2018, datavalidation2019}, whereas gray literature accounts for most unevaluated production claims \cite{scalingml2026, langsmithnd, michelangelo2017, evidentlynd}. Although generative AI monitoring is widely represented, most sources make production claims without evaluation. For example, Coralogix \cite{coralogix2026} claims use by over 4,000 teams, and Superwise \cite{superwise2026} reports monitoring more than 10 billion inferences. Sentry provides an example, reporting use by over 600 Anthropic engineers, a 10--15\% productivity increase, and 20--30\% faster incident resolution \cite{sentry2026}.

Figure \ref{fig:metrics} shows that evaluations focus primarily on ML performance \cite{machinelearningdrift2019, alertiger2023}, operational performance \cite{successfulmachine2019, llamafirewall2025, oneshield2025}, and data quality and distribution \cite{amazonsagemaker2021, automatinglarge2018}. ML-performance evaluations use measures such as accuracy, F1, precision, recall, and RMSE, while data-focused evaluations rely mainly on drift measures and distribution distances. Latency is the most frequently reported operational metric in both literatures. However, these metrics provide different kinds of evidence depending on what they measure. Metrics such as detection rate directly evaluate the monitoring approach, whereas changes in model performance or business outcomes after a monitor initiates a response provide indirect evidence of effectiveness. Safety, fairness, and security evaluations use measures such as harmful content detection accuracy \cite{llamafirewall2025, monitoringreasoning2025}, while generative AI evaluations additionally use legibility and coverage scores \cite{pragmaticway2025}. In comparison, human and business outcomes are rarely evaluated. Human evaluation is based mainly on user feedback, while the reported business metrics are domain-specific and vary across sources \cite{datasentinel2020, successfulmachine2019}. Overall, existing evaluations provide considerably more evidence about technical detection and operational performance than about the effects of monitoring on practitioners or business outcomes.

\subsubsection{RQ3.2. What are the evaluation scales and domains of production ML systems being monitored?}
Among sources reporting an evaluation domain, research and development is the most common \cite{llamafirewall2025, llamaguard2023}. Some sources evaluate monitoring across multiple application domains rather than focusing on a single sector \cite{overton2019, oneshield2025, pipelinemonitoring2023}. Less common sector-specific evaluations cover telecommunications \cite{sagemakermodel2022, domainspecific2023}, manufacturing \cite{daqlmonitor2019}, finance \cite{frameworkmonitor2019, scalingml2026, sagemakermodel2022}, marketing \cite{scalingml2026}, retail \cite{sagemakermodel2022}, healthcare \cite{scalingml2026}, government \cite{deployingmachine2018}, and automotive applications \cite{sagemakermodel2022}, and are reported primarily in academic literature. 

Evaluation scale is reported differently for dataset-based and live evaluations. Dataset-based studies assess an approach using a fixed collection of records and do not report a user or practitioner population \cite{llamafirewall2025, llamaguard2023}. For evaluations reporting operational scale, huge-scale (1,000+ users) deployments are the most common and appeared in 6  sources that are all academic publications \cite{automatinglarge2018, datasentinel2020}. Large (100-1000 users) and medium-scale (10-99 users) evaluations occur in both literature types, while only one small-scale (1-9 users) evaluation is reported \cite{alertiger2023, sagemakermodel2022, domainspecific2023, constitutionalclassifiers2025, beyondblackbox2025}.

\begin{rqanswer}{RQ3. What evidence exists for the effectiveness of monitoring techniques and tools for production ML systems?}
\textbf{RQ3 Answer. } Evidence of effectiveness is limited and varied. Empirical evaluations rely mainly on public or proprietary datasets, while live-production evaluations and user studies are less common. Academic publications provide most of the empirical and production evidence, whereas gray literature—particularly monitoring-tool websites—accounts for most production claims without formal evaluation. Evaluations focus primarily on operational performance, ML performance, and data quality and distribution. Generative AI evaluations additionally measure response quality and safety, often through LLM evaluators. Human and business outcomes receive comparatively limited assessment. Reported evaluations are concentrated in research and development or span multiple domains, with less evidence from specific sectors. Evaluation domain and scale are also frequently unspecified, although a small number of studies report large or huge operational deployments.
\end{rqanswer}

\subsection{RQ4. What are the challenges in monitoring production ML systems, to what extent do existing monitoring
approaches address them, and what are the open problems?}
\subsubsection{RQ4 Taxonomy}
\begin{table}[!htbp]
\centering
\caption{RQ4 Monitoring Challenges Taxonomy}
\label{tab:challenges}
\small
\begin{adjustbox}{max height=0.99\textheight}
\begin{tabular}{@{}L{0.40\textwidth} L{0.55\textwidth}@{}}
\toprule
\textbf{Challenge} & \textbf{Description} \\
\midrule

Absence of monitoring standards &
No well-accepted standards or guidelines exist for monitoring ML systems \\

Alert quality \& threshold configuration &
Misconfigured thresholds cause alert fatigue and desensitize engineers \\

Business impact attribution &
Technical metrics do not reliably predict business outcomes; attribution to KPIs is difficult \\

Context-specific monitoring customizations &
Generic monitors underperform in domain-specific deployments, forcing bespoke redevelopment for each system \\

Data pipeline integrity \& quality at scale &
Data corruption, anomalies, and schema drift silently propagate through distributed pipelines \\

Deciding between monitoring tradeoffs &
No principled way to balance monitoring objectives \\

Distinguishing \& measuring drift &
Seasonal variation, batch effects, and noise are indistinguishable from real drift \\

Ground truth unavailability for monitoring &
Production labels arrive late, never, or are removed by corrective action \\

Human-in-the-loop dependency &
Closing the loop from detection to remediation requires manual effort even where partial automation exists \\

Model opacity \& behavioral unpredictability &
Non-deterministic black-box models with unknown failure modes \\

Monitoring cost \& inference overhead &
Financial, compute, and human-resource costs of monitoring \\

Monitoring output interpretability \& actionability &
Outputs that cannot be interpreted or translated into concrete remediation actions \\

Monitoring tool maintainability \& extensibility &
Monitoring tools accumulate technical debt and become difficult to extend as systems evolve \\

Organizational \& skill gaps in monitoring &
Unclear responsibilities, siloed ownership, and skill gaps lead to fragmented monitoring with no shared accountability \\

Privacy \& data governance in monitoring &
Capturing and routing data to hosted services violates privacy policies \\

Responsible AI monitoring gaps &
Fairness and ethical metrics are rarely monitored \\

Root cause isolation across pipeline components &
Cross-component entanglement makes it difficult to locate faults \\

Scalability of monitoring infrastructure \& methods &
Approaches that work on small scale fail at enterprise scale under high data volume \\

Tooling integration barriers &
Monitoring tools work in isolation and lack interfaces for integration with existing ML pipelines and technology stacks \\

\bottomrule
\end{tabular}
\end{adjustbox}
\end{table}

The RQ4 taxonomy defines a flat set of 19 challenges that can obstruct the monitoring of production ML systems. Each category represents a distinct problem reported in practitioner-oriented or approach-focused literature, rather than a monitoring technique, monitored property, or lifecycle stage. The categories are not mutually exclusive: one incident or monitoring approach may expose several challenges simultaneously. The flat structure is intentional because many challenges span multiple stages of monitoring. Their later mapping to Define, Detect, Notify, Analyze, and cross-cutting concerns provides a secondary analytical view, but those stages are not parent categories in this taxonomy.

Several categories concern how monitoring objectives are specified. \textbf{Absence of monitoring standards} captures the lack of accepted guidance for choosing and implementing monitoring practices. \textbf{Deciding between monitoring tradeoffs} represents choices such as coverage versus cost or sensitivity versus false alarms. \textbf{Alert quality and threshold configuration} concerns thresholds that generate noisy, repeated, or poorly prioritized alerts. \textbf{Context-specific monitoring customizations} captures the need to adapt generic monitors to a particular domain, model, risk profile, or workflow. \textbf{Business impact attribution} addresses the difficulty of linking technical metrics to user or organizational outcomes when seasonality, interventions, and other confounders are present.

A second group concerns the production evidence available for detection. \textbf{Data pipeline integrity and quality at scale} covers corruption, schema change, and anomalies that propagate through distributed pipelines. \textbf{Distinguishing and measuring drift} concerns separating consequential distribution change from normal variation, noise, or batch effects. \textbf{Ground truth unavailability for monitoring} captures labels that are delayed, missing, selectively observed, or changed by corrective action. \textbf{Model opacity and behavioral unpredictability} covers black-box or non-deterministic behavior and failure modes that cannot be fully specified in advance. These categories are distinct because valid data do not guarantee stable model behavior, and detectable drift does not necessarily demonstrate a loss of performance.

Other categories describe the gap between detection and remediation. \textbf{Root cause isolation across pipeline components} concerns locating a fault within interdependent data, model, infrastructure, and application components. \textbf{Monitoring output interpretability and actionability} captures results that do not explain their operational significance or support a concrete response. \textbf{Human-in-the-loop dependency} denotes the continuing need for manual review, contextual judgement, approval, or repair even when collection and detection are automated. Together, these categories distinguish observing a symptom from understanding and resolving the underlying problem.

The remaining challenges concern the sustainability and governance of monitoring. \textbf{Monitoring cost \& inference overhead} includes financial, computational, storage, latency, and human-attention costs. \textbf{Scalability of monitoring infrastructure and methods} concerns whether monitoring remains effective as traffic, data volume, and model fleets grow. \textbf{Tooling integration barriers} and \textbf{monitoring tool maintainability and extensibility} respectively capture difficulty connecting monitors to heterogeneous production stacks and evolving them without accumulating technical debt. \textbf{Organizational and skill gaps in monitoring} covers fragmented ownership, siloed expertise, and unclear accountability. \textbf{Privacy and data governance in monitoring} concerns lawful and policy-compliant collection, retention, access, and transfer of production evidence. Finally, \textbf{responsible AI monitoring gaps} captures inadequate coverage of fairness, ethics, safety, and broader societal effects. Including these categories treats monitoring as a socio-technical capability whose effectiveness depends on organizational processes and governance as well as detectors~\cite{haveidea2024,mlenabledsystems2024,monitoringobservability2025}.


\subsubsection{RQ4.1. What research methods and practitioner populations underpin these findings?}
\begin{table}[!htbp]
\centering
\caption{Sources of Reported ML Monitoring Challenges}
\label{tab:challenge_sources}
\small
\begin{tabular}{llrr}
\toprule
\textbf{Study Orientation} & \textbf{Challenge Source} & \textbf{Number of Sources} & \textbf{Number of Challenge Mentions} \\
\midrule
Approach/Tool & Research Paper & 38 &  109 \\
Approach/Tool & Tool Website   & 39 &  63 \\
Practitioner  & Interview      & 8 &  44 \\
Practitioner  & Survey         & 7 &  22 \\
Practitioner  & Focus Group    & 2 &  11 \\
Practitioner  & Workshop       & 1 &   9 \\
Approach/Tool & Blog           & 1 &   2 \\
\bottomrule
\end{tabular}
\end{table}

\begin{table}[!htbp]
\centering
\caption{Summary of Practitioner-Oriented Studies}
\label{tab:practitioner-studies}
\small
\begin{adjustbox}{max width=\textwidth}
\begin{tabular}{@{}L{0.14\textwidth} c c L{0.28\textwidth} L{0.36\textwidth}@{}}
\toprule
\textbf{Method} & \textbf{Studies} & \textbf{Total Participants} & \textbf{Domains} & \textbf{Participant Roles} \\
\midrule

Interview  \cite{lifecyclemodels2021, haveidea2024, humancentricperspective2022, measureimproved2025, collaborationchallenges2022, misbehaviourfault2021, empiricalinvestigation2024, sustainability2026}     & 8 & 126 & Finance, Healthcare, Retail, Telecom, Education, Government, Automotive, Multiple & ML Engineers, Software Engineers, Ops/MLOps, Managers, Business Stakeholders, Domain Experts, QA, Systems Engineers \\
\midrule
Survey \cite{comparativeanalysis2022, understandingpractitioners2025, mlenabledsystems2024, loggingrequirement2025, empiricalinvestigation2024, sustainability2026, ensuringrobustness2025}         & 7 & 720 & Finance, Healthcare, Retail, Telecom, Education, Government, Automotive, Multiple & ML Engineers, Software Engineers, Ops/MLOps, Managers, Business Stakeholders, QA, Systems Engineers \\
\midrule
Focus Group \cite{designprinciples2025, monitoringobservability2025}    & 2 & 67  & Finance, Retail                                                                   & ML Engineers, Software Engineers, Ops/MLOps, Managers, Business Stakeholders \\
\midrule
Workshop  \cite{challengesmonitoring2026}      & 1 & 30  & Not reported                                                                      & ML Engineers, Software Engineers, Ops/MLOps \\

\bottomrule
\end{tabular}
\end{adjustbox}
\end{table}

Table~\ref{tab:challenge_sources} shows that the challenge synthesis utilizes both practitioner-oriented and approach/tool-oriented sources. Research papers describing approaches contribute the largest number of challenge mentions, followed by tool websites. Table~\ref{tab:practitioner-studies} shows further details of practitioner-oriented sources which report ML monitoring challenges. Interviews are the most frequently used method, while surveys account for the largest participant coverage. These methods provide complementary perspectives; interviews examine monitoring experiences in context, whereas surveys capture practices and concerns across broader practitioner populations. Focus groups and workshops are less frequently represented. 

The participant populations span both technical and organizational roles. ML engineers and software engineers are represented across all methodological categories, alongside operations and MLOps practitioners, managers, business stakeholders, quality-assurance professionals, systems engineers, and domain experts. The reported domains include finance, healthcare, retail, telecommunications, education, government, and automotive applications. Interviews and surveys provide the broadest domain coverage, while focus groups are in specific settings. 


\subsubsection{RQ4.2. What are the challenges in monitoring production ML systems?}

\begin{table}[!htbp]
\centering
\caption{Mapping of Monitoring Challenges to Monitoring Stages}
\label{tab:challenges-steps}
\small
\begin{adjustbox}{}
\begin{threeparttable}
\begin{tabular}{@{}L{0.18\textwidth} L{0.60\textwidth} r@{}}
\toprule
\textbf{Monitoring Stage} & \textbf{Challenge} & \textbf{Number of Sources} \\
\midrule

\multirow{5}{=}{Define}
 & Deciding between monitoring tradeoffs                                        & 19 \\
 & Alert quality \& threshold configuration\tnote{$\star$}           & 15 \\
 & Context-specific monitoring customizations        & 13 \\
 & Business impact attribution                       & 10 \\
 & Absence of monitoring standards                                   &  9 \\
\midrule

\multirow{4}{=}{Detect}
 & Model opacity \& behavioral unpredictability\tnote{$\dagger$}    & 17 \\
 & Ground truth unavailability for monitoring                        & 15 \\
 & Distinguishing \& measuring drift                    &  9 \\
 & Data pipeline integrity \& quality at scale                       &  7 \\
\midrule

\multirow{1}{=}{Notify}
 & Alert quality \& threshold configuration\tnote{$\star$}           & 15 \\
\midrule

\multirow{4}{=}{Analyze}
 & Root cause isolation across pipeline components              & 31 \\
 & Human-in-the-loop dependency                                      & 16 \\
 & Model opacity \& behavioral unpredictability\tnote{$\dagger$}    & 17 \\
 & Monitoring output interpretability \& actionability               &  8 \\
\midrule

\multirow{7}{=}{Cross-cutting}
 & Organizational \& skill gaps in monitoring                        & 30 \\
 & Scalability of monitoring infrastructure \& methods               & 16 \\
 & Privacy \& data governance in monitoring                          & 10 \\
 & Monitoring cost \& inference overhead                             &  8 \\
 & Tooling integration barriers                                      &  7 \\
 & Monitoring tool maintainability \& extensibility                  &  7 \\
 & Responsible AI monitoring gaps                                    &  4 \\

\bottomrule
\end{tabular}
\begin{tablenotes}
\footnotesize
\item[$\star$] Appears in Define and Notify stages.
\item[$\dagger$] Appears in Detect and Analyze stages.
\end{tablenotes}
\end{threeparttable}
\end{adjustbox}
\end{table}

Table~\ref{tab:challenges-steps} shows that monitoring challenges extend across the entire monitoring process and its organizational context. Root-cause isolation across pipeline components and organizational and skill gaps are the most frequently reported challenges. They are followed by difficulties in balancing monitoring tradeoffs, understanding opaque model behavior, managing human involvement, and scaling monitoring. 

At the Define stage, deciding between monitoring tradeoffs is the most prominent challenge \cite{nemoguardrails2023, deployingmachine2018, newrelic2026, haveidea2024, amazonsagemaker2021}. Sources also report difficulties configuring thresholds \cite{understandingpractitioners2025, humancentricperspective2022, haveidea2024, mltestscore2017}, adapting monitoring to specific contexts \cite{misbehaviourfault2021, timeretrain2025, newrelic2026}, linking technical metrics to business outcomes \cite{monitoringobservability2025}, and working without accepted monitoring standards \cite{sustainability2026, humancentricperspective2022, challengesmonitoring2026}. These findings indicate that establishing effective monitoring is itself a substantial problem. Practitioner-oriented studies similarly identify metric selection, custom tooling, and initial configuration as difficulties in operationalizing monitoring \cite{mlenabledsystems2024, understandingpractitioners2025}. Alert quality and threshold configuration also appear at the Notify stage, highlighting that decisions made during setup affect whether subsequent alerts are useful or contribute to alert fatigue.

At the Detect stage, model opacity and behavioral unpredictability \cite{chainmonitorability2025, beyondblackbox2025} and unavailable ground truth \cite{haveidea2024, monitoringobservability2025, mlenabledsystems2024} receive the greatest attention. These challenges limit the ability to determine whether the observed behavior represents an actual failure, particularly when ground truth is delayed or unknown. Distinguishing meaningful drift from normal variation \cite{humancentricperspective2022, measuringdistributional2023, deployingmachine2018} and maintaining data pipeline integrity \cite{deployingmachine2018, daqlmonitor2019} are also reported, although less frequently. Together, these findings highlight difficulties in collecting monitoring evidence and also establishing what that evidence implies about system behavior \cite{haveidea2024, mlenabledsystems2024}.

The Analyze stage contains the most frequently reported challenge: root cause isolation across pipeline components \cite{haveidea2024, humancentricperspective2022, overton2019, monitoringobservability2025}. Human-in-the-loop dependency \cite{monitoringobservability2025, understandingpractitioners2025, pipelinemonitoring2023} and model opacity \cite{chainmonitorability2025, beyondblackbox2025} further complicate investigation, while monitoring outputs may also be difficult to interpret or fail to provide clear guidance on how to respond \cite{datavalidation2019, haveidea2024, alertiger2023}. The key difficulty here is therefore moving from an observed symptom to an explanation of where and why the problem occurred. Practitioners frequently mention a continued reliance on manual investigation and a need for domain expertise to resolve runtime issues \cite{understandingpractitioners2025}. This complements the RQ2 finding that automated detection and notification receive more extensive coverage than diagnosis and corrective action.

Among cross-cutting challenges, organizational and skill gaps are particularly prominent \cite{monitoringobservability2025, sagemakermodel2022, aitotal2021, mltestscore2017, collaborationchallenges2022}, followed by scalability \cite{datasentinel2020, experiencesadopting2021, understandingpractitioners2025}. Privacy and data governance \cite{challengesmonitoring2026, braintrust2026, observabilityllm2026, mastra}, monitoring cost \cite{challengesmonitoring2026, constitutionalclassifiers2025, measuringdistributional2023}, tooling integration \cite{mastra, experiencesadopting2021, understandingpractitioners2025, datasentinel2020}, and maintainability \cite{experiencesadopting2021, llamafirewall2025, oneshield2025} also constrain the operation of monitoring across stages. These concerns show that monitoring depends on ownership, expertise, resources, and integration with existing workflows, rather than on detector capabilities alone. Responsible AI monitoring gaps receive comparatively fewer mentions \cite{loggingrequirement2025, amazonsagemaker2021, sustainability2026}. Overall, the findings characterize production ML monitoring as both a technical and organizational activity, with the greatest reported difficulties concerning diagnosis, coordination, and the practical decisions needed to make monitoring useful.

\subsubsection{RQ4.3. Which monitoring challenges are addressed by existing approaches and what are the open problems?}\label{RQ4.3}
Table~\ref{tab:challenge-benefit-coverage} shows how well existing monitoring approaches cover each reported challenge. The coverage column reports the number of sources that identify each challenge and the number of sources that address it. These counts are independent. An approach was coded as \textit{addressed} when it targeted one or more aspects of a challenge, as summarized in the aspects addressed column. Open problems were derived by identifying aspects not addressed by the corresponding approaches. The comparison shows which challenges existing approaches focus on and where gaps remain.

\newcolumntype{L}[1]{>{\RaggedRight\arraybackslash}p{#1}}
\setlength{\LTcapwidth}{\textwidth}
\begingroup
\setlength{\tabcolsep}{2pt}
\renewcommand{\arraystretch}{1.0}
\small
\begin{longtable}{@{}
    L{0.14\textwidth}
    L{0.10\textwidth}
    L{0.30\textwidth}
    L{0.45\textwidth}
@{}}

\caption{Challenges Addressed in Studies and Open Problems}
\label{tab:challenge-benefit-coverage}\\

\toprule
\textbf{Challenge} &
\textbf{Coverage} &
\textbf{Aspects addressed } &
\textbf{Open problems} \\
\midrule
\endfirsthead

\multicolumn{4}{l}{\small\textit{Table~\ref{tab:challenge-benefit-coverage} continued}}\\[2pt]
\toprule
\textbf{Challenge} &
\textbf{Coverage} &
\textbf{Aspects addressed } &
\textbf{Open problems} \\
\midrule
\endhead

\midrule
\multicolumn{4}{r}{\small\textit{Continued on next page}}\\
\endfoot

\bottomrule
\endlastfoot

Absence of monitoring standards &
\mbox{\textit{Reported:} 9} 
\mbox{\textit{Addressed:} 2}

&
Production-readiness checklists and process frameworks provide some standardization of monitoring workflows  &
\textbf{Monitoring metrics:} consensus on what to monitor and how to measure it;
\textbf{Monitor specification:} systematic definition of monitoring scope and requirements;
\textbf{External standards:} alignment with regulatory, professional, and community guidance.
\\
\midrule

Alert quality \& threshold configuration &
\mbox{\textit{Reported:} 15} 
\mbox{\textit{Addressed:} 19}
 &
Statistical drift detectors support threshold-based alerts, with some approaches incorporating adaptive or dynamic thresholds and standardized alert schemas  &
\textbf{Alert fatigue:} reduction of repeated and false-positive alerts;
\textbf{Threshold optimization:} principled sensitivity--false-alarm balancing;
\textbf{Alert deduplication:} consolidation of correlated alerts arising from the same failure.
\\
\midrule

Business impact attribution &
\mbox{\textit{Reported:} 10} 
\mbox{\textit{Addressed:} 3}
 &
Dashboards and lightweight frameworks connect technical monitoring metrics with business KPIs and revenue-related signals. &
\textbf{Causal attribution:} separating model effects from market shifts, seasonality, and interventions;
\textbf{Delayed feedback:} accounting for lags between model actions and business outcomes;
\textbf{Confounding:} controlling alternative explanations for observed business changes.
\\
\midrule

Context-specific customization &
\mbox{\textit{Reported:} 13} 
\mbox{\textit{Addressed:} 6}

 &
Domain-adapted evaluation frameworks and custom metric libraries support monitoring in specialized contexts such as healthcare, finance, and NLP. &
\textbf{Monitor transfer:} adaptation of monitors across domains;
\textbf{Reusable templates:} common monitoring configurations for related contexts;
\textbf{Domain adaptation:} generalizable adaptation frameworks beyond individual domain-specific evaluators.
\\
\midrule

Data pipeline integrity \& quality at scale &
\mbox{\textit{Reported:} 7} 
\mbox{\textit{Addressed:} 10}
 &
Schema validation, data-quality checks, anomaly detection, declarative constraints, and automated integrity tests support production data-pipeline monitoring. &
\textbf{Retroactive changes:} recomputation of monitoring results after upstream corrections;
\textbf{Schema ownership:} responsibility for failures across organizational boundaries;
\textbf{Training--serving skew:} divergence between offline and online feature transformations.
\\
\midrule

Distinguishing \& measuring drift &
\mbox{\textit{Reported:} 9} 
\mbox{\textit{Addressed:} 18} &
Drift detection is extensively addressed through statistical tests, stability indices, divergence measures, and windowed distribution comparisons. &
\textbf{Drift disambiguation:} distinguishing genuine shifts from seasonality, batch effects, and noise;
\textbf{Semantic/multimodal drift:} drift in embeddings, multimodal inputs, and LLM outputs;
\textbf{Statistical calibration:} controlling excessive sensitivity at production scale;
\textbf{Impact-aware detection:} linking detected shifts to model and business outcomes.
\\
\midrule

Ground truth unavailability for monitoring &
\mbox{\textit{Reported:} 15} 
\mbox{\textit{Addressed:} 8} &
Proxy metrics, behavioral signals, and surrogate indicators enable performance monitoring when outcome labels are unavailable or delayed. &
\textbf{Proxy reliability:} uncertainty and confidence in proxy estimates relative to ground truth;
\textbf{Monitoring under intervention:} accounting for distributional changes introduced by retraining and other interventions.
\\
\midrule

Human-in-the-loop dependency &
\mbox{\textit{Reported:} 16}
\mbox{\textit{Addressed:} 11} &
Automated remediation pipelines handle predefined failures, while runtime guardrails enforce safety constraints without manual approval. &
\textbf{Threshold configuration:} systematic or automated configuration;
\textbf{Agent trace review:} automatic identification of violations across multi-step trajectories;
\textbf{Data repair:} automated inspection and correction of problematic production data.
\\
\midrule

Model opacity \& behavioral unpredictability &
\mbox{\textit{Reported:} 17}
\mbox{\textit{Addressed:} 39} &
LLM and agent observability uses prompt--response logging, trace inspection, and token-level analysis; chain-of-thought monitoring exposes reasoning behavior; attribution tools support post-hoc failure analysis. &
\textbf{Traditional ML interpretability:} explanations of individual production outputs;
\textbf{Production uncertainty:} quantification and communication of prediction uncertainty;
\textbf{Novel failure modes:} detection of previously unseen production behaviors.
\\
\midrule

Monitoring cost \& inference overhead  &
\mbox{\textit{Reported:} 8}
\mbox{\textit{Addressed:} 14} &
Distilled evaluators, adaptive sampling, lightweight monitors, and proxy metrics reduce the computational cost of production ML monitoring. &
\textbf{Reasoning-monitoring cost:} latency overhead from inspecting reasoning traces;
\textbf{Embedding-based monitoring:} computational cost of embedding production traffic;
\textbf{Cost--quality tradeoffs:} balancing sampling, coverage, and evaluation thoroughness against cost.
\\
\midrule

Deciding between monitoring tradeoffs &
\mbox{\textit{Reported:} 19}
\mbox{\textit{Addressed:} 9}
&
Process frameworks provide general monitoring guidance, while production-evaluation feedback loops provide empirical evidence for refining monitoring decisions. &
\textbf{Coverage--cost:} principled signal selection under resource constraints;
\textbf{Precision--recall:} balancing missed failures and false alarms;
\textbf{Sampling:} determining appropriate production traffic coverage.
\\
\midrule

Organizational \& skill gaps  in monitoring &
\mbox{\textit{Reported:} 30}
\mbox{\textit{Addressed:} 10} &
Low-code dashboards, simplified monitoring interfaces, documentation, and process frameworks reduce monitoring expertise requirements and reliance on tacit knowledge. &
\textbf{Cross-team ownership:} allocation of monitoring responsibilities and handoffs;
\textbf{Monitoring governance:} ownership of requirements, escalation, and actions;
\textbf{Monitoring maturity:} progressive establishment and improvement of monitoring practices.
\\
\midrule

Monitoring output interpretability \& actionability&
\mbox{\textit{Reported:} 8}
\mbox{\textit{Addressed: }30} &
Attribution tools support actionable diagnosis; business-metric mappings improve stakeholder relevance; automated remediation and guardrails act on detected problems; evaluation feedback loops provide evidence for subsequent monitoring decisions. &
\textbf{Automated-action explanation:} rationale for remediation and guardrail actions;
\textbf{Real-time explanation:} practitioner-readable interpretation of monitoring outputs;
\textbf{Alert prioritization:} distinguishing urgent alerts from those that can be deferred.
\\
\midrule

Privacy \& data governance in monitoring &
\mbox{\textit{Reported:} 10} 
\mbox{\textit{Addressed: }6} &
Compliance logging, access controls, and regulatory audit trails support governance of captured monitoring data. &
\textbf{Privacy-aware collection:} privacy-preserving capture of prompts, responses, and production data;
\textbf{Privacy-preserving drift detection:} detecting shifts without exposing individual records;
\textbf{PII-safe logging:} safe handling of sensitive information in monitoring pipelines.
\\
\midrule

Responsible AI monitoring gaps &
\mbox{\textit{Reported:} 4} 
\mbox{\textit{Addressed: }17} &
Fairness metrics, demographic slicing, bias detection, content-safety evaluation, and reasoning monitoring support responsible AI properties. &
\textbf{Audit trails:} standardized and interoperable responsible-AI audit records;
\textbf{Societal impacts:} downstream feedback loops, structural exclusion, and long-term harms;
\textbf{Practitioner adoption:} barriers to adoption despite available monitoring capabilities.
\\
\midrule

Root cause isolation across pipeline components  &
\mbox{\textit{Reported:} 31} 
\mbox{\textit{Addressed: }12} &
Component attribution, correlation, and drill-down tools support post-hoc failure diagnosis. &
\textbf{Multi-component diagnosis:} joint diagnosis across data, model, and infrastructure;
\textbf{Agent-specific diagnosis:} failure tracing across multi-step agentic pipelines;
\textbf{Automated diagnosis:} automatic identification and recommendation of likely root causes.
\\
\midrule

Scalability of monitoring infrastructure \& methods &
\mbox{\textit{Reported:} 16} 
\mbox{\textit{Addressed: }11} &
Distributed architectures, streaming metric pipelines, and horizontally scalable infrastructure support monitoring at increasing production scale. &
\textbf{Large model fleets:} scalable management of model-specific thresholds and baselines;
\textbf{Distributed ownership:} coordination of monitoring responsibilities across teams;
\textbf{Cost-effective scaling:} economic tradeoffs as monitoring and model fleets expand.
\\
\midrule

Monitoring tool maintainability \& extensibility  &
\mbox{\textit{Reported:} 7} 
\mbox{\textit{Addressed: }11} &
Unified platforms reduce reliance on independently maintained point tools, while standardized processes can reduce monitoring-related maintenance complexity. &
\textbf{Monitor versioning:} migration of thresholds, baselines, and configurations as models evolve;
\textbf{Monitor testing:} systematic validation of monitoring implementations;
\textbf{Monitor deprecation:} safe retirement of obsolete monitors without coverage gaps.
\\
\midrule

Tooling integration barriers &
\mbox{\textit{Reported:} 7} 
\mbox{\textit{Addressed: }11} &
Unified platforms consolidate monitoring capabilities, while standardized schemas reduce ad hoc integration across heterogeneous tools. &
\textbf{Interoperability:} open interfaces across monitoring tools;
\textbf{Framework integration:} standardized interfaces across ML frameworks;
\textbf{Legacy migration:} migration of existing monitoring pipelines to newer architectures.
\\

\end{longtable}
\endgroup

Challenges that receive the highest support from existing approaches, include monitoring cost and inference overhead \cite{alertiger2023, domainspecific2023, datavalidation2019, fiddlernd, overton2019}, distinguishing and measuring drift \cite{timeretrain2025, endend2025, healthassurance2022, arizeaind}, monitoring output interpretability and actionability \cite{arizephoenixnd, tensorflowdata2020, dynatracend, llamafirewall2025, braintrust2026}, and responsible AI monitoring gaps \cite{amazonsagemaker2021, chainmonitorability2025, pragmaticway2025, oneshield2025, galileond}. However, the remaining open problems show that coverage is uneven across several aspects of these challenges. Drift approaches predominantly address structured data, leaving semantic and multimodal drift, statistical calibration at scale, and impact-aware detection as open problems. For monitoring cost, unresolved problems include the additional computing time and resources required to monitor reasoning traces and embeddings, and the absence of structured cost--quality trade-offs. Open problems in monitoring output interpretability and actionability include real-time explanation and alert prioritization, while responsible AI gaps include standardized audit trails, societal impacts, and practitioner adoption. 

Other challenges are addressed by existing approaches, but their support is limited to particular aspects. These include ground-truth unavailability \cite{timeretrain2025, monitoringfeature2021, machinelearningdrift2019}, alert quality and threshold configuration \cite{alertiger2023, azurend, scalingml2026, sagemakermodel2022}, scalability of monitoring infrastructure and methods \cite{datasentinel2020, datavalidation2019, michelangelo2017, whylabsnd}, tooling integration barriers \cite{arizeaind, coralogix2026, fiddlernd, whylabsnd, newrelic2026}, monitoring-tool maintainability and extensibility \cite{dynatracend, splunknd, whylabsnd, arizeaind, fiddlernd}, model opacity and behavioral unpredictability \cite{raindropnd, chainmonitorability2025, overton2019, splunknd, dynatracend}, and data-pipeline integrity and quality at scale \cite{daqlmonitor2019, datasentinel2020, datavalidation2019, deepchecks2024}. The identified open problems include validating proxy measures in the absence of labels, reducing repeated and false-positive alerts, and detecting previously unseen model behavior. Other gaps concern monitoring large model fleets, providing interoperable interfaces between tools, and testing, versioning, and retiring monitors as systems evolve. For data-pipeline integrity, unresolved issues include retroactive data changes, cross-team schema ownership, and training--serving skew. Coverage of model opacity is concentrated more heavily on LLM and agentic systems, while production uncertainty and output explanations for traditional ML receive less attention.

The largest gaps are found where frequently reported challenges receive minimal support from existing approaches. These include the absence of monitoring standards \cite{mltestscore2017, experiencesadopting2021}, decisions between monitoring trade-offs \cite{experiencesadopting2021, mltestscore2017}, context-specific monitoring customization \cite{arizeaind, arizephoenixnd, whylabsnd, galileond}, business-impact attribution \cite{healthassurance2022, azurend, alertiger2023}, human-in-the-loop dependency \cite{pipelinemonitoring2023, llamafirewall2025, llamaguard2023, oneshield2025}, organizational and skill gaps \cite{overton2019, domainspecific2023, coralogix2026, splunknd}, privacy and data governance in monitoring \cite{witnessai2026, dynatracend, superwise2026, fiddlernd}, and root-cause isolation across pipeline components \cite{raindropnd, dynatracend, scalingml2026, fiddlernd}. The open problems include defining common monitoring metrics and specifications, balancing coverage against cost and alert quality, and transferring monitoring configurations across domains. Other gaps concern causal attribution of delayed business outcomes, privacy-aware data collection and logging, cross-team ownership, and reducing manual effort in threshold configuration, trace review, and data repair.

\begin{rqanswer}{RQ4. What are the challenges in monitoring production ML systems, to what extent do existing
monitoring approaches address them, and what are the open problems?}
\textbf{RQ4 Answer. }The findings draw on practitioner studies using several empirical methods, together with challenges reported in approach and tool-focused sources. The participants span technical and organizational roles across multiple application domains. The most frequently reported challenges concern isolating root causes, coordinating monitoring across teams, making monitoring design decisions, and managing opaque or unpredictable model behavior. Existing approaches are concentrated on established technical concerns such as drift detection and monitoring cost, although many address only particular aspects of these challenges. Comparatively fewer sources report approaches supporting organizational, contextual, and diagnostic challenges. Root-cause isolation presents the clearest mismatch: it is the most frequently reported challenge, yet only a few sources provide direct support for diagnosing faults across interacting pipeline components. The remaining open problems include selecting and configuring appropriate monitors, improving alert quality, diagnosing failures across interacting system components, maintaining monitors as systems evolve, and integrating monitoring with organizational responsibilities, privacy, and governance.
\end{rqanswer}

\section{Discussion}\label{discussion}
This section examines the key findings and discusses their implications.

\subsection{The Challenge of Designing Effective Monitoring}
Even when suitable monitoring capabilities exist, deciding how to use them remains difficult. The literature offers many tools, signals, and detection techniques. However, it provides limited guidance on translating organizational and application-specific requirements into operational monitors \cite{haveidea2024, understandingpractitioners2025, mlenabledsystems2024}. It also offers limited guidance on selecting appropriate properties, signals, and thresholds, determining which evidence is sufficient for decision-making, and maintaining monitors as systems, data, and operational requirements change. What should be monitored depends on the application domain, available evidence, consequences of failure, and organizational objectives. A generic configuration may detect technical changes while overlooking harms to affected users, failures that prevent users from completing their tasks, or changes in service and business outcomes against the objectives set by system owners \cite{leest2025tea}. The alert quality issue demonstrates how these design decisions affect monitoring in practice. Although alerting is widely automated, practitioners continue to report false and poorly prioritized alerts resulting from unsuitable metrics and thresholds \cite{understandingpractitioners2025, humancentricperspective2022, haveidea2024}. Increasing the number or sensitivity of monitors often increases the burden of manual investigation without improving response. Effective monitoring requires explicit trade-offs among monitored properties, false alarms, latency, cost, and human attention. These prioritization decisions are also evident in responsible AI monitoring. Although few sources report it as a challenge, existing approaches focus mainly on fairness and safety, while other responsible AI concerns receive limited attention. This may reflect narrow monitoring scope rather than maturity. The field has developed a broad set of monitoring capabilities, but comparatively limited guidance on how to combine them into a strategy that fits the context and supports operational and business goals.

\subsection{Production Monitoring Detects Problems More Readily Than It Resolves Them}
A clear pattern across the results is that support decreases as monitoring moves from detection towards diagnosis and response. A wide range of automated techniques can identify potential problems and communicate them through alerts, reports, and dashboards. Once a problem has been identified, however, its interpretation relies largely on manual or semi-automated inspection of logs, traces, subgroups, and feature attributions. This issue is also evident in RQ4, where root-cause isolation across pipeline components is one of the most reported challenges, yet only a few sources address it \cite{haveidea2024, humancentricperspective2022, monitoringobservability2025}. The limited coverage of the Response stage further increases this gap. Reported responses consist mainly of guardrail enforcement in generative AI and automated retraining in traditional ML and deep learning \cite{nemoguardrails2023, llamafirewall2025, timeretrain2025}. These mechanisms usually execute predefined automated actions rather than selecting a response based on an established root cause. Failures outside those predefined conditions continue to require human interpretation and intervention. Production ML monitoring is therefore imbalanced and only partially automated, with systems considerably better at detecting and communicating symptoms than at diagnosing, explaining, and resolving their underlying causes. 

\subsection{Fragmented Monitoring Limits End-to-End Diagnosis}
Fragmented monitoring is a key factor that makes fault diagnosis in ML systems difficult. Most approaches observe selected parts of the production ML system rather than connecting evidence across preprocessing, inference, agentic, infrastructure, and application layers. None of the 94 included sources cover all five monitored layers, and only eight cover four layers. Even the broadest approaches therefore omit part of the production ML system, and empirically evaluated evidence of end-to-end monitoring remains scarce. The consequences of this fragmentation become visible during diagnosis. A change observed in model outputs may originate in production data, preprocessing, infrastructure, or application logic; similarly, an acceptable agent response may conceal retrieval or tool-execution faults. Layer-specific monitoring can reveal individual faults, but it provides limited support for tracing faults across these dependencies. The same disconnect extends to application decisions. Drift, latency, and error rates are widely monitored, while their effects on users and business objectives receive less attention. As a result, monitoring may establish that a system has changed without showing where the change originated or whether it had a meaningful consequence.

\subsection{Monitoring Frequently Relies on Indirect Evidence}
Ground-truth labels may arrive late or not at all, while human feedback and business outcomes are difficult to collect continuously. Monitoring therefore depends heavily on proxy metrics such as distribution distances, threshold violations, evaluator scores, latency, and error rates. These proxies make continuous monitoring possible, but they also constrain the claims that can be made from its findings. Drift monitoring highlights this limitation. Feature and prediction drift receive substantial attention, while concept drift, the changing relationship between inputs and target outcomes, is monitored in only two sources \cite{machinelearningdrift2019, timeretrain2025}. A change in the feature distribution does not necessarily indicate reduced model performance, just as a stable distribution does not guarantee that the model remains correct. LLM evaluator scores introduce a similar uncertainty; they allow response quality and safety to be assessed without reference answers, but remain proxies for user or expert judgment. Existing approaches are therefore better at finding visible changes than at showing whether those changes are actually harmful or important.

\subsection{Evidence for Production Monitoring Remains Limited, Particularly for Generative AI}
Only nine of the included sources evaluated their approaches and tools in live production. Most evidence comes from offline datasets, controlled settings, or descriptive claims, leaving limited understanding of how these approaches perform under changing workloads, delayed feedback, and interacting production failures. The emergence of generative AI has expanded monitoring from individual predictions to prompts, generated content, retrieval, reasoning, tool calls, and agent execution. LLM-based evaluation and trace inspection offer ways to assess these new forms of ML behavior, while conventional thresholds and operational metrics continue to monitor cost, latency, and failures. Generative AI monitoring represents a broader shift from observing a model to observing a multi-component application and its execution. However, the corpus provides limited evaluation evidence for these practices. Of the 50 sources coded as covering generative AI, 46 (92\%) are gray literature and 33 (66\%) are tool websites. Some platforms claim substantial adoption and operational scale \cite{coralogix2026, superwise2026, sentry2026}, but these claims do not establish that their monitoring techniques detect production problems accurately or improve decisions. This is particularly important for LLM evaluators, whose reliability across models, domains, changing production conditions, and agreement with users and experts remains weakly evaluated.

\subsection{Monitoring Is an Evolving Socio-Technical Capability}
People remain responsible for selecting ML system monitoring requirements, configuring monitors, interpreting findings, investigating failures, and deciding how to respond. These activities cross technical and organizational boundaries, making ownership, expertise, and coordination essential to monitoring effectiveness \cite{collaborationchallenges2022, humancentricperspective2022, understandingpractitioners2025, monitoringobservability2025}. A monitoring platform may provide the necessary instrumentation, but organizations must still determine which outcomes matter, who investigates findings, and who has authority to act. These roles and responsibilities must also be maintained with the monitored system. As models, data, applications, and policies change, existing baselines, thresholds, evaluators, guardrails, and alert rules may no longer represent acceptable behavior. Production ML monitoring should therefore be treated not as a component added once after deployment, but as a continuing capability maintained alongside the technical system and its organizational context.

\section{Future Recommendations}\label{future}
The findings identify several areas in which production ML monitoring requires improvement. Future work should improve how monitors are designed, how evidence is connected and interpreted, how systems respond to findings, and how monitoring is evaluated and maintained. We propose six concrete directions for researchers, practitioners, and tool developers.

\subsection{Structure and Automate Monitor Design}
Creating ML monitors currently requires substantial manual effort, while default configurations may not reflect the runtime issues and objectives of a particular application \cite{amershi2019software,paleyes2022challenges}. Future work should develop methods that generate an initial monitoring configuration from a structured description of the ML system, including its architecture, task, deployment context, failure consequences, available data, and operational and business objectives. The generated configuration could recommend monitored objects and properties, required signals, suitable metrics, initial thresholds, monitoring frequency, alert priority, and response procedures. Each recommendation should remain traceable to the requirement. Historical operating data could inform provisional thresholds only after checking its relevance to the current model, data pipeline, operating conditions, and seasonal patterns. Thresholds should be validated on a later time window and reviewed when these conditions change; past distributions should not be assumed to remain representative. Confirmed and dismissed alerts could inform recalibration, alongside sampled unflagged cases to identify missed failures. Recalibration should also be checked against task outcomes or expert judgment, since a stable distribution or an adapted threshold does not establish correct model behavior. However, this process should remain semi-automated: system owners and domain experts should validate the generated configuration, particularly its trade-offs and high-risk responses. Such methods could be implemented as a domain-specific language, monitor configuration schemas, reusable system profiles, or recommendation tools and evaluated against manual configuration using requirement coverage, configuration effort, alert usefulness, and alignment with operational and business outcomes \cite{ali2010goalcontext,naveed2024responsiblemonitoring, monitoringobservability2025, leest2025tea}.

\subsection{Develop Cross-Layer Root-Cause Analysis}
Root-cause isolation is the most frequently reported challenge, yet existing support remains focused on checking individual components and comparing events after they happen \cite{haveidea2024, humancentricperspective2022, understandingpractitioners2025}. Future approaches should connect evidence across preprocessing, inference, agentic workflows, serving infrastructure, and application layers. This requires shared identifiers that follow a request through data transformations, model calls, retrieval operations, tool use, and downstream outcomes \cite{wu2020microrca,lee2023eadro}. A runtime dependency graph could then combine these observations and use their temporal order, dependencies, and anomaly evidence to rank likely fault locations. Fault injection should form a central part of evaluating these monitoring approaches \cite{wu2020microrca,lee2023eadro}. Studies should deliberately introduce data corruption, preprocessing errors, model degradation, infrastructure failures, retrieval and reranking faults, incorrect tool use, and compound failures spanning multiple layers, then assess whether the monitoring system detects, correlates, and explains the resulting signals. This would test whether monitoring can identify the originating component rather than merely raise alerts for downstream symptoms. Evaluations should report root-cause localization accuracy, top-$k$ ranking accuracy, detection and diagnosis time, alert precision, and the manual effort required to move from an alert to an actionable monitoring response.

\subsection{Assess the Validity of Proxy Signals Against Ground Truth}
Production monitoring often relies on drift measures, thresholds, and evaluator scores because ground truth, human feedback, and business outcomes are delayed or unavailable \cite{rabanser2019failing}. Future work should assess whether these proxies reliably indicate meaningful degradation in real-world contexts. Proxy monitoring should be combined with sampled direct evaluation. A selection of flagged and unflagged cases could be linked retrospectively to delayed labels, expert review, user feedback, incident outcomes, or business measures when this evidence becomes available. When no form of feedback is available, evaluation should instead rely on triangulation among multiple metrics. This evidence should be used to estimate when a proxy is reliable, identify the failures it misses, and recalibrate it as the system changes. Active sampling could prioritize uncertain cases, disagreements between monitors, and high-risk decisions, while monitors should be able to abstain or request human review when confidence is insufficient. For generative AI, LLM evaluator and guardrail judgments should be compared with human assessments across multiple models, prompts, domains, and time periods \cite{zheng2023judging}. Evaluations should report agreement, calibration, false-positive and false-negative rates, and association with downstream outcomes rather than only the proxy's own score.

\subsection{Close the Gap Between Diagnosis and Response}
Current response mechanisms are concentrated mainly on guardrail enforcement and automated retraining \cite{nemoguardrails2023, llamafirewall2025, timeretrain2025}. Future work should develop a catalog that maps diagnosed causes and their severity to appropriate corrective actions \cite{garlan2004rainbow}. Depending on the source of failure, these actions could include isolating a data source, rolling back a deployment, switching to a fallback model, rerouting traffic, repairing a pipeline, requesting human review, or initiating retraining. This would assist on-call engineers and provide a starting point for automated responses specific to the issue. Monitoring tools should also support interactive debugging, allowing engineers to trace failures across system layers, test hypotheses, and inspect relevant signals. AI-assisted root-cause analysis could synthesize logs, traces, alerts, model outputs, data changes, deployment events, and past incidents to identify likely causes and recommend corrective actions \cite{coralogix2026, splunknd, monitorstack2026}.
New response policies should progress from recommendations for operators to human-approved execution, and finally to automation once their correctness has been demonstrated. The responses should also retain explanations, audit records, and rollback mechanisms. Evaluation can continue after an action is taken. Relevant measures include whether the original problem was resolved, recovery time, recurrence, inappropriate responses, and unintended effects elsewhere in the system. This would extend monitoring evaluation beyond successful detection to whether the complete monitoring process produces an effective recovery.

\subsection{Improve Production-Oriented Evaluation}
The effectiveness of many monitoring approaches remains weakly established, particularly for generative AI, where tool websites frequently report production adoption without formal evaluation. Future studies should use production or production-like workloads that represent temporal changes, delayed feedback, seasonal variation, interacting component failures, and changes caused by interventions such as retraining \cite{mltestscore2017,sculley2015hidden,paleyes2022challenges}. Such data could be obtained through industry partnerships, consortia, synthetic datasets, simulated workloads, and fault-injection experiments. Shared benchmarks could provide separate scenarios for traditional ML and generative AI systems, covering the full monitoring process from observation to response. Evaluation of these monitoring approaches should extend beyond detection accuracy to include detection delay, alert precision, root-cause localization, response success, operational overhead, human effort, and business outcomes. Studies should also report the deployment environment, domain, operational scale, monitored objects and properties, available signals, threshold-selection procedure, and degree of human involvement. For generative AI, academic studies should examine whether evaluators, guardrails, and reasoning monitors remain reliable across different models and changing production conditions \cite{zheng2023judging}. These practices would improve comparison and provide evidence of production readiness.

\subsection{Strengthen Collaboration, Ownership, and Maintenance}
Technical improvements will have limited effect if monitoring responsibilities and collaboration processes remain unclear. Monitoring involves data, ML, software, operations, domain, and business teams, which may use different tools and have access to different evidence \cite{collaborationchallenges2022,amershi2019software,humancentricperspective2022,understandingpractitioners2025}. Organizations should clarify who is responsible for each monitor, who investigates its alerts, and who can update its configuration or approve a response. For incidents that involve multiple parts of the system, teams should agree on who coordinates the investigation. Shared dashboards, incident records, and response records can support coordination and reduce repeated transfers of work between teams \cite{collaborationchallenges2022,amershi2019software}. Periodic reviews can help ensure that monitoring remains relevant as systems and requirements change. Monitoring configurations must also be maintained as the production system changes. Updates to models, pipelines, prompts, tools, or policies should trigger a review of the associated baselines, thresholds, signals, and alerts. Monitors should be tested against historical incidents or injected faults before deployment. Structured templates for monitor configurations, shared terminology, and documented investigation procedures could further reduce configuration effort and dependence on domain experts.

\section{Threats to Validity}\label{threats}
This section presents the potential threats to the validity of our MLR and the mitigation strategies applied.

\subsection{Internal Validity}
Selection bias is the main threat to the internal validity of our study. To mitigate this threat, we followed the multivocal literature review guidelines proposed by Garousi et al. \cite{garousi2019guidelines}. The review protocol was developed by the first and second authors and reviewed by the remaining authors before the formal search was conducted. The search string and study selection criteria were iteratively refined by assessing the relevance of the results retrieved from the four academic databases (IEEE Xplore, ACM Digital Library, Scopus, and Web of Science) and two gray literature sources (ArXiv and Google Search Engine). Another threat related to the search string is that the ten seed papers used to derive the search keywords were identified through exploratory searches and the authors’ domain knowledge, which may have introduced bias into the terminology included in the search string. To reduce this threat, the search terms were iteratively refined by incorporating synonyms, abbreviations, and related terminology, and the final search string was validated against all seed papers. We also restricted machine learning terms to the title field and monitoring and production-related terms to the title and abstract to keep the number of retrieved studies manageable. This restriction may have excluded relevant studies in which these terms appear only in the full text; however, we mitigated this threat through complementary snowballing.

Screening was conducted in two stages, as illustrated in Figure~\ref{fig:search_results}. Studies were first screened based on their titles and abstracts, followed by full-text review. The first and second authors independently applied the selection criteria to each study, with disagreements resolved through discussion with the remaining authors until consensus was reached. Although no formal inter-rater agreement was calculated, independent screening and consensus-based resolution reduced the risk of individual interpretation affecting the selection results. To reduce the risk of missing relevant studies during the automated search, we complemented it with backward and forward snowballing following the guidelines of Wohlin \cite{wohlin2014guidelines}, which identified 34 additional studies. However, because snowballing was initiated from studies identified through the automated search, relevant studies outside their citation networks may still have been missed, particularly those from disconnected research areas or recent studies with limited citation links. This threat was mitigated by the broad coverage of academic and gray literature sources in the automated search, although it cannot be fully eliminated. Lastly, during data extraction, the first and second authors extracted data from separate sets of studies. Differences in interpretation between the two extractors therefore remain a potential threat. We mitigated this threat by aligning extraction definitions during the pilot rounds and resolving ambiguous cases through discussion.

\subsection{Construct Validity}
The terminology used to describe monitoring concepts varied across the included studies, posing a threat to consistent interpretation and classification. We mitigated this threat by discussing ambiguous cases among the authors until consensus was reached. Another threat concerns the subjectivity involved in developing our taxonomies. Although we followed established guidelines for taxonomy development in software engineering \cite{ralph2018toward}, the categories were derived through an inductive and interpretive coding process, and alternative categorizations may therefore be possible. We mitigated this threat by grounding the categories in the extracted data and discussing category definitions and boundaries among the authors until consensus was reached.

Another threat concerns variation in the quality of the included sources, particularly gray literature. We did not conduct a formal quality assessment or exclude sources based on quality because our objective was to capture the breadth of production monitoring knowledge, including emerging tools and practices that have not yet been formally evaluated. We mitigated this threat by distinguishing academic and gray literature throughout the analysis, identifying the types of included sources, and separately examining whether monitoring approaches were empirically evaluated or supported only by production claims. Finally, our findings rely on information reported by the primary studies. We did not independently verify the monitoring tools, techniques, evaluation results, or challenges reported in the included studies. While this reliance on reported evidence is inherent to an MLR and cannot be fully eliminated, we aimed to interpret and categorize the reported findings as accurately and consistently as possible.

\subsection{Conclusion Validity}
Threats to conclusion validity arise from the interpretation and synthesis of the extracted data. Descriptive and cross-dimensional analyses were used to identify patterns and relationships within the extracted data; however, the observed patterns depend on the distribution and coverage of studies in the reviewed literature. Taxonomy development and qualitative synthesis relied on inductive coding, which introduces subjectivity in the formation of categories and themes. 

To mitigate these threats, coding decisions, category definitions, and cross-dimensional interpretations were iteratively refined by the first and second authors, with ambiguities and differences in interpretation resolved through discussion. Findings across the different analyses were examined together during synthesis to ensure that the resulting interpretations were consistent with the extracted evidence. Furthermore, conclusions were limited to patterns supported by the extracted data, reducing the risk of overgeneralization beyond the reviewed literature.

\subsection{External Validity}
Our goal was to provide a comprehensive overview of current practices for monitoring production ML systems by reviewing both formal and gray literature. We conducted a systematic search following established guidelines \cite{garousi2019guidelines} and applied the predefined inclusion and exclusion criteria presented in Table~\ref{tab:selection_criteria}. However, restricting the review to English-language studies with accessible full texts of at least four pages may have excluded relevant studies. Gray literature sources whose terms of use restricted research use or required special permission were also excluded (EX7). Furthermore, for feasibility, Google Search results were limited to the first 200 results, following the recommendation of Garousi et al. \cite{garousi2019guidelines}.

Despite the systematic search process, we cannot guarantee that all relevant studies were identified, which may limit the generalizability of our findings. We mitigated this threat by complementing the automated search with backward and forward snowballing. To reduce publication bias and capture a broader range of evidence, studies were not excluded based on publication date or study quality. Finally, our findings represent the state of the literature at the time of the search. Given the rapid evolution of production ML monitoring, particularly for LLM and agent-based systems, the identified practices, challenges, and research gaps may evolve as new technologies and practices emerge.

\section{Conclusion}\label{conclusion}
This MLR provides a comprehensive synthesis of the literature on monitoring production ML systems. From 7,186 initially retrieved sources, a rigorous screening and snowballing process identified 94 relevant sources, comprising 37 academic publications and 57 gray literature sources. We analyzed the characteristics of monitored systems; the monitored objects, properties, signals, techniques, tools, and stages; the evidence of monitoring effectiveness; monitoring challenges; and open problems.

Our findings show that production ML monitoring provides broad support for observing runtime behavior, detecting deviations, and communicating findings, but remains considerably weaker in diagnosis and response. The main unresolved problems are: (1) monitor selection and configuration require manual effort and are weakly connected to business objectives; (2) fragmented monitoring across system layers limits end-to-end fault diagnosis; (3) root-cause isolation continues to rely heavily on manual investigation; (4) drift measures, thresholds, and LLM evaluator scores are often used without sufficient evidence that they indicate actual failures; (5) automated monitoring responses are limited to guardrail enforcement and automated retraining, rather than actions selected according to diagnosed fault; (6) evidence of effectiveness remains limited, particularly in live-production and generative AI settings; and (7) collaboration, ownership, expertise, and monitor maintenance receive less support than technical detection concerns.

To address these problems, we propose six directions for future work. Monitor design should be automated while keeping a clear link from business goals to system requirements, signals, thresholds, alerts, and responses. Cross-layer root-cause analysis should connect runtime evidence using shared identifiers and dependency graphs, and it should be evaluated through systematic fault injections. Proxy signals should be checked against delayed labels, expert judgments, user feedback, incidents, and business outcomes. Catalogs should link diagnosed causes and severity levels to suitable corrective actions, moving from operator recommendations toward validated automation. Evaluations should use production or similar synthetic data to measure diagnosis, response, operational overhead, human effort, and application outcomes. Finally, monitoring should be supported by clear collaboration, ownership, and maintenance practices across the teams that operate ML systems.

These recommendations shift monitoring production ML systems from isolated detection toward an integrated process connecting system objectives, runtime evidence, diagnosis, and corrective action. By combining academic and gray knowledge and identifying existing gaps, this review provides a direction for future monitoring approaches that are easier to configure, more effective at explaining and resolving failures, more rigorously evaluated, and better integrated with operational and organizational practice.

\section{Acknowledgments}
Naveed is supported by a Faculty of IT Post-graduate scholarship. Grundy and Haggag are supported by ARC Laureate Fellowship FL190100035. Haggag is also supported by a National Intelligence Post-doctoral Fellowship. This work is also
partly supported by ARC Discovery Project DP200100020.

\bibliographystyle{ACM-Reference-Format}
\bibliography{sample-base}



\end{document}